\documentclass[twocolumn,amsfonts,amssymb,amsmath,tightenlines,nofootinbib]{revtex4-2}

\usepackage{hyperref}
\usepackage{mathrsfs}
\usepackage{txfonts}
\usepackage{braket}
\usepackage{bm}

\newcommand\dm{\mathrm d}
\newcommand\im{\mathrm i}
\newcommand\rme{\mathrm e}
\newcommand\he{\hat{\bm e}}
\newcommand\hbk{\skew{-3}\hat\Bbbk}
\newcommand\del{\bm\nabla}
\newcommand\dotp{\bm\cdot}
\newcommand\stack[2]{\genfrac{}{}{0pt}{}{#1}{#2}}

\newcommand\mnras{Mon.~Not.~R.~Astron.~Soc.}

\defcitealias{DLMF}{DLMF}
\defcitealias{EdZ92}{E-dZ}

\numberwithin{equation}{section}
\begin{document}

\title{Potentials of axisymmetric razor-thin disks}

\author{J.~\surname{An}}
\noaffiliation

\begin{abstract}
We investigate the gravitational potentials generated by axisymmetric, razor-thin disks. Within certain limitations, the potential on one side of the disk is shown to be equivalent to the potential produced by a linear mass distribution along the axis perpendicular to the disk. We first establish the connection between two mass distributions that generate the same potential. We then consider all disk surface density profiles that produce the potential equivalent to those generated by linear mass distributions, specifically those defined by the elementary beta distribution and its relatives on the interval $[0,1]$, $[1,\infty)$ or $[0,\infty)$. These families of models are important because the potentials in all cases are given by, at most, a single real quadrature of elementary functions of the coordinates, and furthermore, many cases result in closed-form expressions. The potential of many realistic disks may be constructed from some combinations of these disk models.
\end{abstract}

\label{firstpage}
\maketitle
\section{Introduction}

Despite their elementary nature, calculating the gravitational potentials of completely flattened mass distributions has proven to be a significant challenge. Although integrating the Newtonian potential of mass elements is, in principle, straightforward, one immediately discovers that such calculations typically involve multiple integrals or integral transformations with the elliptic integral kernels (which effectively add another layer of integration) even for simple, circularly symmetric surface density profiles. If one were to construct a family of orbits around such a disk, the computational demand quickly becomes impractical. Consequently, many settle for a compromise, modeling the disk with profiles that generate simpler potentials. Still, a striking dearth of known exact expressions for the potentials generated by disks persists, even when the surface mass density is circularly symmetric.

This paper provides a new physical interpretation of the method for finding the potential of a circularly symmetric, razor-thin disk, originally attributed to \cite{EdZ92}. Subsequently, this method is extended to its full generality by means of the Mellin transform. We then construct a large array of circular disk models with exact analytical expressions for their potentials, which may be considered as a generalization of the model set first proposed by \cite{Qi92}. Ultimately, the current paper serves as the opening chapter of a compendium of the ``analytic'' disk potentials, providing both the requisite mathematical machinery and a resource for modeling various highly flattened astronomical objects.

\vfill
\section{Preliminary}
\subsection{The Mestel disk}
\label{sec:mestel}

Let us consider the potential generated by a semi-infinite uniform string of mass. If it coincides with the negative-$z$ part of the $R=0$ axis, the gravitational potential $\Phi$ it generates in cylindrical polar coordinates $(R,\phi,z)$ can be written down as 
\begin{equation}\label{eq:psi}
\Phi(R,z)=-G\Lambda\int_0^\infty\!\frac{\dm h}{\!\sqrt{R^2+(z+h)^2}},
\end{equation}
where $\Lambda$ is the linear mass density of the string. Although one finds that equation (\ref{eq:psi}) does not converge, the resulting gravitational acceleration $\bm\varg=-\del\Phi$ is nevertheless well-defined:
\begin{equation}\label{eq:ga}
\frac{-\bm\varg}{G\Lambda}
=\int_0^\infty\!\frac{R\,\he_R+(z+h)\,\hbk}
{\left[R^2+(z+h)^2\right]^{3/2}}\dm h
=\left(1-\frac{z}r\right)\frac{\he_R}R+\frac\hbk r,
\end{equation}
where $\he_R=\del R$ and $\hbk=\del z$ are the unit vectors in the increasing $R$ and $z$ directions, and $r^2=R^2+z^2$. In principle, the explicit expression for $\Phi(R,z)$ may be obtained by integrating equation (\ref{eq:ga}) back. Instead let us consider the equipotential surfaces first. If $\he_r=\del r$ is the radial unit vector, we find $r\he_r=R\he_R+z\hbk$ and so
\begin{equation}\label{eq:gp}
\frac{\del\Phi}{G\Lambda}=\frac{r-z}{R^2}(\he_r+\hbk)
=\frac{\del(r+z)}{r+z},
\end{equation}
that is, the equipotential surfaces are also the set of surfaces with constant $r+z$. The meridional cross-sections of these surfaces are convex-upward confocal parabolae with the focus at the origin and the $R=0$ axis being its symmetry axis (see \cite{An13}, sect.~5.1). Hence the potential due to a semi-infinite uniform string of mass is stratified over the family of the confocal paraboloids of revolution. The potential itself is a function of $s=r+z$ alone, $\Phi(R,z)=f(s)$, and so $\del\Phi=f'(s)(\del s)=f'(s)(\he_r+\hbk)$. Then from equation (\ref{eq:gp}), $f'(s)=G\Lambda/s$ and $f(s)=G\Lambda\ln s+\text{constant}$. Consequently
%
\begin{equation}\label{eq:pot}
\Phi(R,z)=G\Lambda\ln\left(\frac{z+\!\sqrt{R^2+z^2}}{2a}\right)
\end{equation}
up to an arbitrary zero-point $a$. Formally if the unintegrable singularity was removed in equation (\ref{eq:psi}) by introducing an infinite offset, one obtains the same result:
\begin{equation}\label{eq:potshi}
\frac{\Phi(R,z)}{G\Lambda}
=\int_0^\infty\!\dm h\,
\left(\frac1{a+h}-\frac1{\!\sqrt{R^2+(z+h)^2}}\right).
\end{equation}

Equation (\ref{eq:pot}) satisfies the Laplace equation ($\nabla^2\Phi=0$) everywhere except the negative-$z$ part of the $R=0$ axis, where it is singular. The same function therefore corresponds to the potential due to some localized mass distributions in the source-free part of the space, subject to the boundary conditions. In particular, an axisymmetric harmonic potential $\Phi(R,z)$ in the upper half-space ($z>0$) and the reflection-symmetric function in the lower half-space results from an infinitesimally thin disk on the mid-plane ($z=0$) whose surface mass density $\Sigma(R)$ is given by $2\pi G\Sigma(R)=\lim_{z\downarrow0^+}\hbk\dotp\del\Phi$. Since $\hbk\dotp\bm\nabla\Phi=\partial\Phi/\partial z=G\Lambda/r$ for $\Phi(R,z)$ in equation (\ref{eq:pot}), the razor-thin disk with the surface density given by
\begin{equation}
\Sigma(R)=\frac{\varv_0^2}{2\pi GR}
\end{equation}
therefore generates the potential behaving like
\begin{equation}\label{eq:mestel}
\Phi(R,z)=\varv_0^2\ln\left(\frac{|z|+\!\sqrt{R^2+z^2}}{2a}\right)
\end{equation}
where $a$ is again an arbitrary integration constant. Note that the circular speed is given by $\varv_\mathrm c^2(R)=\lim_{z\downarrow0^+}R(\partial\Phi/\partial R)=G\Lambda=\varv_0^2$, and so this model is equivalent to the scale-free Mestel disk \cite{Me63}, which is characterized by a flat mid-plane rotation curve. In other words, the equipotential surfaces of the scale-free Mestel disk are stratified over the set of confocal paraboloids of revolution whose foci are at the center of the disk. This fact contrasts with the Kepler potential due to a point mass, which is stratified over the set of concentric spherical surfaces, as well as the plane-parallel potential due to an infinitely-extended uniform plate (see \cite{Ro92}, sect.~22). Furthermore, the potential and the gravity field of the scale-free Mestel disk on one side of the disk are indistinguishable from those due to the uniform string of mass extending from the center infinitely to the opposite side in the perpendicular direction to the disk. This last fact is again compared with the Kuzmin disk \cite{Ku53,Ku56} whose potential on one side of the disk reproduces the potential of a point mass located at a position on the symmetry axis on the opposite side.

A similar procedure for obtaining the Kuzmin disk potential from the Kepler potential may be applied to the scale-free Mestel disk potential. That is to say, consider displacing the potential of the scale-free Mestel disk farther along the symmetry axis and reflecting it off with respect to the mid-plane. This is equivalent to reflecting equation (\ref{eq:pot}) with respect to the plane $z=c>0$, which is now interpreted as the new plane of the disk. The surface density and the mid-plane rotation curve of the corresponding disk are then
%
\begin{equation}\label{eq:cmd}
\Sigma(R)=\frac{\varv_\infty^2}{2\pi G\!\sqrt{R^2+c^2}}
;\quad
\varv_\mathrm c^2(R)=
\varv_\infty^2\left(1-\frac c{\!\sqrt{R^2+c^2}}\right),
\end{equation}
where $\varv_\infty^2$ is the scale constant setting the asymptotic rotation speed due to the disk. This is the Mestel disk modified by replacing its $R^{-1}$ central cusp with a flat-topped core, $\Sigma(0)=(2\pi Gc)^{-1}\varv_\infty^2$, of a ``core radius'' $c$. This ``cored Mestel disk'' is also recognized as the $n=0$ member of the \citet{To63}'s disk family \citep{EdZ92}. The explicit expression of the potential due to the cored Mestel disk is given by equation (\ref{eq:mestel}) with $|z|\to|z|+c$, 
\begin{equation}\label{eq:cmdp}
\Phi(R,z)
=\varv_\infty^2\ln\left(\frac{|z|+c+\!\sqrt{R^2+(|z|+c)^2}}{2a}\right).
\end{equation}
Obviously the $c=0$ limit recovers the potential of the scale-free Mestel disk \citep{Ly89} in equation (\ref{eq:mestel}). While the potential diverges as $r\to\infty$, the central value $\Phi(0,0)$ is finite, which may be set to zero by choosing $a=c$. The equipotential surfaces of the cored Mestel disk in the upper half-space are stratified over the set of confocal paraboloids of revolution but the common foci of the paraboloids are displaced along the symmetry axis to the point $z=-c$ on the opposite side of the disk.

\subsection{The Mestel-difference disk}
\label{sec:mdd}

The paraboloid of revolution is the limiting case of a prolate spheroid as one of its foci is sent to the point at infinity. By contrast, the sphere is a prolate spheroid whose two foci coincide. In other words, the equipotential surfaces of the Kuzmin and Mestel disks are composed of the special class of prolate spheroids. Generalizing these, we would like to find the disks that generate the potential stratified over more general classes of the confocal prolate spheroids.

If the symmetry axis of the prolate spheroid is on the $R=0$ axis, the meridional cross-section of the spheroid becomes an ellipse whose long axis is also on the $R=0$ axis. Since an ellipse is a locus such that the sum of the distances to the two foci is constant (and the line joining the two foci is on its major axis), a member of the family of the confocal prolate spheroids whose foci are located $z=-c_1$ and $z=-c_2$ on the $R=0$ axis with $c_1>c_2\ge0$ is identified with the sum of the distances to each focus $r_1+r_2$ being constant, where $r_1^2=R^2+(z+c_1)^2$ and $r_2^2=R^2+(z+c_2)^2$. Suppose that these spheroids constitute the equipotential surfaces for $z>0$ due to an infinitesimally thin disk placed in the $z=0$ plane. The potential for $z>0$ is a function of $s=r_1+r_2$ that is also a solution of the Poisson equation. However, since the space with $z\ne0$ is empty, $\Phi=f(r_1+r_2)$ should actually satisfy 
\begin{equation*}
\nabla^2\Phi
=\frac1{r_1r_2}
\frac\dm{\dm s}\left\{\bigl[s^2-(c_1-c_2)^2\bigr]\frac{\dm f}{\dm s}\right\}=0,
\end{equation*}
which reduces to an ordinary differential equation on $f(s)$, the general solution of which is found to be ($s>c_1-c_2>0$)
\begin{equation*}
f(s)
=-C\tanh^{-1}\!\left(\frac{c_1-c_2}s\right)
\end{equation*}
with an integration constant $C$. Since $f(s)\simeq-(c_1-c_2)C/s$ as $s\to\infty$, the choice of the potential (with $c_1>c_2\ge0$)
\begin{equation}\label{eq:pmd}
\Phi=-\frac{2GM_\mathrm t}{c_1-c_2}
\tanh^{-1}\!\left(\frac{c_1-c_2}{r_1+r_2}\right)
\end{equation}
where $r_j^2=R^2+(|z|+c_j)^2$, exhibits the behavior, $\Phi\sim-GM_\mathrm t/r$ as $r\to\infty$ consistent with a localized mass distribution. Equation (\ref{eq:pmd}) is found to be generated by the razor-thin disk in the $z=0$ plane with
\begin{equation}\label{eq:mdsd}
2\pi G\Sigma(R)=\lim_{z\downarrow0^+}\frac{\partial\Phi}{\partial z}
=\frac{GM_\mathrm t}{c_1-c_2}\left(\frac1{R_2}-\frac1{R_1}\right)
\end{equation}
where $R_j^2=R^2+c_j^2$. Provided $c_1>c_2\ge0$, we find $\Sigma(R)>0$ for all $R\ge0$ and $4\pi\Sigma\simeq(c_1+c_2)M_\mathrm tR^{-3}$ as $R\to\infty$, implying a finite total mass, $M_\mathrm t=2\pi\int_0^\infty\dm R\,R\Sigma(R)$. At the center, the surface density and the potential well depth are both finite with $\Sigma(0)=M_\mathrm t/(2\pi c_1c_2)$ and $|\Phi(0,0)|=GM_\mathrm t\ln(c_1/c_2)/(c_1-c_2)$ unless $c_2=0$. If $c_1>c_2=0$ however,
%
\begin{equation*}
\Sigma(R)
=\frac{M_\mathrm t}{2\pi c_1}\left(\frac1R-\frac1{R_1}\right),
\end{equation*}
exhibits an $R^{-1}$ singularity and the potential diverges logarithmically as $R\to0$, like those of the scale-free Mestel disk.

In fact, equation (\ref{eq:mdsd}) is simply the difference between two Mestel disks with the same asymptotic circular speed but different core radii (with $c_2=0$ for the scale-free Mestel disk), while the $c_1\to\infty$ limit is the Mestel disk (of infinite mass and indeterminate $M_\mathrm t/c_1$) with the core radius of $c_2\ge0$. The circular speed for this ``Mestel-difference disk'' is found to be
\begin{equation*}
\varv_\mathrm c^2(R)
=\frac{GM_\mathrm t}{c_1-c_2}\left(\frac{c_1}{R_1}-\frac{c_2}{R_2}\right),
\end{equation*}
which is the difference of $\varv_\mathrm c^2(R)$ in equation (\ref{eq:cmd}). Given the finite total mass ($c_1\ne0$), this rotation curve falls off like $\varv_\mathrm c^2(R)\simeq GM_\mathrm t/R$ as $R\to\infty$, while the maximum occurs at $R$ where $R^2=c_1^{4/3}c_2^{2/3}+c_1^{2/3}c_2^{4/3}=c_1^{2/3}c_2^{2/3}(c_1^{2/3}+c_2^{2/3})$ with
\begin{equation*}
\frac{\varv_{\mathrm c,\max}^2}{GM_\mathrm t}
=\frac{c_1+c_2}{\bigl[c_1^{4/3}+(c_1c_2)^{2/3}+c_2^{4/3}\bigr]^{3/2}}.
\end{equation*}
The rotation curve for $c_1>c_2>0$ behaves like $\varv_\mathrm c^2(R)\sim R^2$ as $R\to0$, but the circular speed for $c_1>c_2=0$ monotonically decreases from the finite maximum at the center $\varv_{\mathrm c,\max}^2=\varv_\mathrm c^2(0)=GM_\mathrm t/c_1$. The result for the $c_1\to\infty$ limit (the Mestel disk) is found in equation (\ref{eq:cmd}), for which $\varv_\mathrm c^2(R)$ increases monotonically (unless $c_2=0$ too) and tends to a finite limiting value (asymptotically flat).

In the $c=c_1=c_2>0$ limit maintaining $M_\mathrm t$ constant, it is easy to observe that the Mestel-difference disk reduces to
\begin{equation*}
\Sigma(R)=\frac{M_\mathrm t}{2\pi}\frac c{(R^2+c^2)^{3/2}};\quad
\varv_\mathrm c^2(R)=\frac{GM_\mathrm tR^2}{(R^2+c^2)^{3/2}},
\end{equation*}
which is actually the Kuzmin disk, the potential of which follows the same $c=c_1=c_2$ limit of equation (\ref{eq:pmd}); that is,
\begin{equation}\label{eq:kdp}
\Phi(R,z)=-\frac{GM_\mathrm t}{\!\sqrt{R^2+(|z|+c)^2}}.
\end{equation}
All of the limiting behaviors of the Kuzmin disk are consistent with it being a member of the Mestel-difference disks.

\section{The Evans--de Zeeuw method}

As the potential of the Mestel disk on one side is identical to that generated by a semi-infinite string on the opposite side, the potential of the Mestel-difference disk (which is stratified over the confocal prolate spheroids) is also indistinguishable from the difference of the potentials generated by two semi-infinite strings of the same linear mass density. That is to say, the potential of the Mestel-difference disk on one side is the same as that due to a string of uniform linear mass density of a finite length, $c_1-c_2$. In fact, suppose that the potential is given by $\Phi=-C\tanh^{-1}[(c_1-c_2)/(r_1+r_2)]$ in the whole space, except for $z\in[-c_1,-c_2]$ and $R=0$, for which $r_1+r_2=c_1-c_2$ and so $\Phi$ becomes singular. Integrating the Poisson equation over the region where the potential is singular, we find $\Phi$ is consistent with the boundary condition, $2G\Lambda(z)=\lim_{R\to0}R\he_R\dotp\del\Phi=C$ for $z\in[-c_1,-c_2]$ (so $c_2+z\le 0\le c_1+z$) with $\Lambda(z)$ being the constant linear mass density. The two ends of the string equivalent to the Mestel-difference disk are located at $z=-c_1,-c_2$ on the $R=0$ axis, and so the total mass of the equivalent string is also the same as that of the corresponding Mestel-difference disk $M_\mathrm t=(c_1-c_2)\Lambda$.

Generalizing this, consider the pair of a razor-thin axisymmetric disk and a linear mass distribution along the symmetry axis on the opposite side that generates the identical potential on one side of the disk. The potential is expressible as a simple integral quadrature of the linear mass density $\Lambda$:
\begin{equation}\label{eq:genpot}
\Phi(R,z)=-\int_I\frac{G\Lambda(h)\,\dm h}{\!\sqrt{R^2+(|z|+h)^2}}
\end{equation}
where $h\in I\subset[0,\infty)$, whereas the surface density of the corresponding disk and the resulting rotation curve are:
\begin{equation}\label{eq:sigv}\begin{split}
\Sigma(R)
&=\frac1{2\pi G}\lim_{z\downarrow0^+}\frac{\partial\Phi}{\partial z}
=\frac1{2\pi}\int_I\frac{h\Lambda(h)\,\dm h}{(R^2+h^2)^{3/2}};
\\\varv_\mathrm c^2(R)
&=\lim_{z\downarrow0^+}R\frac{\partial\Phi}{\partial R}
=GR^2\int_I\frac{\Lambda(h)\,\dm h}{(R^2+h^2)^{3/2}}.
\end{split}\end{equation}
This approach is basically equivalent to the method proposed by \cite{EdZ92}, who considered decomposing the surface mass density of the disk to a weighted integral over the set of the Kuzmin disks with varying scale lengths. What has been shown is that we may assign a physical interpretation on the weight function of \cite{EdZ92} (henceforth \citetalias{EdZ92}) as the linear mass density of the string that can generate the same potential as the circularly symmetric razor-thin disk under consideration.

\subsection{How to get $\Lambda$ from $\Sigma$}
\label{sec:inv}

The main difficulty involved with this technique is that finding the weight $\Lambda$ from the known surface density is non-trivial, whereas obtaining the potential and surface density from a given $\Lambda$ is straightforward. Here we consider a few inversion methods, although none is entirely satisfactory.

First we extend the support of the integral in equation (\ref{eq:sigv}) to $I\to[0,\infty)$ by assuming $\Lambda(h)=0$ for $h\notin I$. This allows us to consider the integral transformation over the standard interval $[0,\infty)$. Also note that the transformation for $\Lambda\to\Sigma$ is formally equivalent to that for $h^{-1}\Lambda\to R^{-2}\varv_\mathrm c^2$. Hence the provided inversion method for $\Sigma\to\Lambda$ is equally applicable to $\varv_\mathrm c^2\to\Lambda$ after obvious adjustments.

\subsubsection{The Laplace transform}

Let us observe that
\begin{equation*}
\int_0^\infty\!\dm x\,\rme^{-x(R^2+h^2)}\!\sqrt x
=\frac{\!\sqrt\pi}{2(R^2+h^2)^{3/2}}.
\end{equation*}
Thus equation (\ref{eq:sigv}) may be reducible to
\begin{equation*}\begin{split}
\Sigma(R)&=\frac1{\pi^{3/2}}\int_0^\infty\!\dm h\,h\Lambda(h)
\int_0^\infty\!\dm x\,\rme^{-x(R^2+h^2)}\!\sqrt x
\\&=\frac1{2\pi^{3/2}}\int_0^\infty\!\dm x\,\rme^{-xR^2}\!\sqrt x
\int_0^\infty\!\dm(h^2)\,\rme^{-xh^2}\Lambda(h).
\end{split}\end{equation*}
This is recognized as an iterated Laplace transform;
\begin{equation*}
2\pi^\frac32\Sigma(R)
=\underset{x\to R^2}{\mathcal L}\!\left\lbrace\!\sqrt x
\underset{h^2\to x}{\mathcal L}\!\left[\Lambda(h)\right]\right\},
\end{equation*}
and so we can at least formally invert this for $\Lambda$:
\begin{equation*}
\Lambda(h)=2\pi^\frac32
\underset{x\to h^2}{\mathcal L^{-1}}\!\left\lbrace x^{-\frac12}
\underset{R^2\to x}{\mathcal L^{-1}}\!\left[\Sigma(R)\right]\right\rbrace.
\end{equation*}
Without an explicit algorithm for the inverse Laplace transform, this formula is purely pedagogical. Although the inverse transformations are expressible explicitly via the Bromwich integrals, the resulting formula can be reduced to the direct inversion formula to be presented later, which is better understood through alternative approaches.

\subsubsection{The Abel--Stieltjes transform}

The right-hand sides of equations (\ref{eq:sigv}) are recognized as the generalized Stieltjes transform (GST) $h^2\to R^2$ \citep[chap.~XIV]{Er54}. Although the inversion formulae for the GST are known \citep{Wi38,Sc05} (see also Appendix~\ref{app:d_inv}), it is instructive to approach this through the canonical Stieltjes transform first:
\begin{equation}\label{eq:stiel}
\mathscr S[g(t);x]=\int_0^\infty\frac{g(t)\,\dm t}{t+x}=f(x).
\end{equation}
which is invertible through the Stieltjes--Perron formula:
%
\begin{equation}\label{eq:spf}\begin{split}
g(t)&=\lim_{\epsilon\downarrow0^+}
\frac{f(-t-\im\epsilon)-f(-t+\im\epsilon)}{2\pi\im}\quad
\text{for $t\in(0,\infty)$}
\\&=\frac{f(t\rme^{-\im\pi})-f(t\rme^{\im\pi})}{2\pi\im}
=\frac1\pi\bigl\lvert\Im[f(-t)]\bigr\rvert
\end{split}\end{equation}
assuming that the analytic continuation of $f(x)$ has a branch cut along the negative real line. The last formula with $\Im$ being the imaginary part is valid assuming $f(\bar x)=\overline{f(x)}$, which follows $g(t)\in\mathbb R$ for $t\in(0,\infty)$.

Next we observe that the GST of any order is reducible to the canonical Stieltjes transform utilizing fractional calculus. In particular, consider the Abel transform -- which is equivalent to the Weyl integral of order one-half -- of equation (\ref{eq:sigv}):
\begin{equation*}\begin{split}
\mathcal A^-(r)&=\int_r^\infty\!\frac{R\Sigma(R)\,\dm R}{\!\sqrt{R^2-r^2}}
\\&=\frac1{2\pi}\!\int_0^\infty\!\dm h\,h\Lambda(h)
\int_r^\infty\!\frac{R\,\dm R}{\!\sqrt{(R^2-r^2)(R^2+h^2)^3}}
\\&=\frac1{2\pi}\int_0^\infty\frac{h\Lambda(h)\,\dm h}{r^2+h^2}
=\frac{\mathscr S[\Lambda(\!\sqrt{h^2});r^2]}{4\pi},
\end{split}\end{equation*}
which is in the form of the Stieltjes transform ($h^2\to r^2$). So
\begin{equation}\label{eq:invint}\begin{split}
\Lambda(h)&=\frac2\im\left[
\mathcal A^-(h\rme^{-\im\frac\pi2})
-\mathcal A^-(h\rme^{\im\frac\pi2})\right]
\\&=\frac2\im\!\int\limits_{h\rme^{-\im\pi/2}}^{h\rme^{\im\pi/2}}
\!\frac{R\Sigma(R)\,\dm R}{\!\sqrt{R^2+h^2}},
\end{split}\end{equation}
where the last integral is along the complex contour starting and ending at the points $R=h\rme^{\pm\im\pi/2}=\pm\im h$ on the imaginary axis. Unless $\Sigma(R)$ falls off faster than $\sim R^{-1}$ as $R\to\infty$, the Abel transform $\mathcal A^-$ does not converge, but we may consider another Abel transform (equivalent to the Riemann--Liouville integral of order one-half):
\begin{equation*}
\mathcal A^+(r)=\int_0^r\!\frac{R\Sigma(R)\,\dm R}{\!\sqrt{r^2-R^2}}
=\frac r{2\pi}\!\int_0^\infty\frac{\Lambda(h)\,\dm h}{r^2+h^2}.
\end{equation*}
The resulting transformation $h^{-1}\Lambda(h)\to4\pi r^{-1}\mathcal A^+(r)$ is also recognized as the Stieltjes transform ($h^2\to r^2$), and equation (\ref{eq:spf}) thus leads to
\begin{equation*}
\Lambda(h)=2\im\left[\rme^{-\im\frac\pi2}\mathcal A^+(h\rme^{\im\frac\pi2})
-\rme^{\im\frac\pi2}\mathcal A^+(h\rme^{-\im\frac\pi2})\right],
\end{equation*}
which results in the same integral as equation (\ref{eq:invint}).

\subsubsection{The generalized Stieltjes transform}

The last integral in equation (\ref{eq:invint}) is also consistent with the general inversion formula of the GST \citep{Sc05}:
\begin{equation}\label{eq:stielt_g}
f(x)=\int_0^\infty\!\frac{g(t)\,\dm t}{(t+x)^{1+\eta}}
\end{equation}
which admits the inversion via a contour integral: if $\eta>0$,
\begin{equation}\label{eq:sif}
g(t)=\frac{\eta}{2\pi\im}\!
\int\limits_{-t}^{(0+)}\dm z\,(t+z)^{\eta-1}f(z).
\end{equation}
Here we have used the notation of \cite{WW27} for the path of integration; that is, the contour starts and ends at $z=-t$ (on the opposite side of the branch cut) while it encircles the point $z=0$ in the positive sense.

Since equation (\ref{eq:sigv}) is the GST of order $1+\eta=\frac32$, applying the integral inversion formula (i.e.\ eq.~\ref{eq:sif}) results in
\begin{equation*}
\Lambda(h)=\frac1\im\!
\int\limits_{-h^2}^{(0+)}
\frac{\Sigma(\rho^{1/2})\,\dm\rho}{\!\sqrt{\rho+h^2}}
=\frac2\im\!\int\limits_{-\im h}^{\im h}
\!\frac{R\Sigma(R)\,\dm R}{\!\sqrt{R^2+h^2}},
\end{equation*}
which recovers the same formula as equation (\ref{eq:invint}). Here the integration path of the last $R$-integral starts at $R=-\im h$, cuts (upward) the positive half of the real-$R$ axis, and ends at $R=\im h$. Thanks to the Cauchy theorem, the specific choice of the integration path is immaterial, provided that $\Sigma(R)$ is extended analytically to the right half-space of the complex number (i.e.\ $R\in\mathbb C$ and $\Re[R]>0$). In other words, if the \emph{analytic continuation} of $\Sigma(R)$ or $\varv_\mathrm c^2(R)$ is known for the \emph{complex} domain, the \citetalias{EdZ92} weight, provided that it actually exists, can be recovered via a complex quadrature.

For instance, if $\Sigma(R)=C(R^2+a^2)^{-p}$, then (here $R=\im y$)
\begin{equation}\label{eq:lamh}
\frac{\Lambda(h)}C
=2\im\!\int_{-h}^h\!\frac{y\,\dm y}{(a^2-y^2)^p\!\sqrt{h^2-y^2}}.
\end{equation}
This vanishes if $a>h>0$, for the integrand with $y\in[-h,h]$ is a real-valued even function of $y$. If $h>a>0$ on the other hand, the integrand has the branch points at $y=\pm a\in[-h,h]$. With proper branch cuts, equation (\ref{eq:lamh}) for $h>a>0$ is reducible to
\begin{equation}\label{eq:kmt_w}\begin{split}
\frac{\Lambda(h)}C
&=\int_a^h\!\frac{4\sin(p\pi)\,y\,\dm y}{(y^2-a^2)^p\!\sqrt{h^2-y^2}}
\\&=\frac{2\pi^{3/2}(h^2-a^2)^{1/2-p}}{\Gamma(p)\Gamma(3/2-p)},
\end{split}\end{equation}
where $\Gamma(x)$ is the Gamma function. That is to say, if we use $\Lambda\propto(h^2-a^2)^{1/2-p}\Theta(h^2-a^2)$ -- where $\Theta(x)$ is the Heaviside unit step -- as the weight, the GST in equation (\ref{eq:sigv}) then results in $\Sigma\propto(R^2+a^2)^{-p}$ (for $0<p<\frac32$), which can be confirmed through direct calculations. See Sect.~\ref{sec:mkt} for more details of this family of the models.

\section{The Mellin transform approach}
\label{sec:melt}

While the inversion formula of equation (\ref{eq:invint}) completes the formal machinery to apply the method of \cite{EdZ92}, its actual utility is still limited. This is not only because the inversion requires the analytic continuation in the complex domain but also because an arbitrary surface density is not necessarily the GST of a proper non-negative weight. In fact, the surface density $\Sigma(R)$ in equation (\ref{eq:sigv}) with any non-negative $\Lambda(h)$ cannot decay faster than $R^{-3}$ as $R\to\infty$ (in particular, if $\int_I\dm hh\Lambda(h)$ converges, then $\Sigma\sim R^{-3}$). In other words, any surface density decaying faster than $R^{-3}$ (including all finite disks) cannot be expressed as the GST in equation (\ref{eq:sigv}) with a proper weight.

Nevertheless we can still make use of equations (\ref{eq:sigv}) to find the rotation curve generated by any circularly symmetric disk. Instead of inserting equation (\ref{eq:invint}) directly to equation (\ref{eq:sigv}) to get the transformation $\Sigma\to\varv_\mathrm c^2$, let us consider the Mellin transforms $R^2\to t$ \citep[chap.~VI]{Er54} first;
\begin{equation}\label{eq:sigmel}\begin{split}
\mathfrak M[\Sigma](t)
&=\int_0^\infty\!\dm\rho\,\rho^{t-1}\Sigma(\!\sqrt\rho)
=2\!\int_0^\infty\!\dm R\,R^{2t-1}\Sigma(R);
\\\mathfrak M[\varv_\mathrm c^2](t)
&=2\!\int_0^\infty\!\dm R\,R^{2t-1}\varv_\mathrm c^2(R).
\end{split}\end{equation}
Replacing $\Sigma$ and $\varv_c^2$ with equation (\ref{eq:sigv}) leads these to
\begin{align*}\begin{split}
\mathfrak M[\Sigma](t)&=\frac1\pi\int_0^\infty\!\dm h\,h\Lambda(h)\!
\int_0^\infty\!\frac{R^{2t-1}\dm R}{(R^2+h^2)^{3/2}}
\\&=\frac{\Gamma(t)\Gamma(\frac32-t)}{\pi^{3/2}}\!
\int_0^\infty\!\dm h\,h^{2t-2}\Lambda(h);
\end{split}\\\mathfrak M[\varv_\mathrm c^2](t)
&=\frac{2G\Gamma(t+1)\Gamma(\frac12-t)}{\!\sqrt\pi}\!
\int_0^\infty\!\dm h\,h^{2t-1}\Lambda(h).
\end{align*}
The integrals in the right-hand sides are recognized as the Mellin transform of $\Lambda$ and related to $\mathfrak M[\Sigma](t)$ and $\mathfrak M[\varv_\mathrm c^2](t)$:
\begin{equation}\label{eq:melr}\begin{split}
\mathfrak M[\Lambda](t)
&=\int_0^\infty\!\dm\eta\,\eta^{t-1}\Lambda(\!\sqrt\eta)
=2\!\int_0^\infty\!\dm h\,h^{2t-1}\Lambda(h)
\\&=\frac{2\pi^{3/2}\mathfrak M[\Sigma](t+\frac12)}
{\Gamma(t+\frac12)\Gamma(1-t)}
=\frac{\!\sqrt\pi\,\mathfrak M[\varv_\mathrm c^2](t)}
{G\Gamma(t+1)\Gamma(\frac12-t)}.
\end{split}\end{equation}
Utilizing the Mellin inversion theorem,
\begin{equation}\label{eq:mit}
g(t)=\int_0^\infty\!\dm x\,x^{t-1}f(x)\
\Rightarrow\
f(x)=\frac1{2\pi\im}
\int\limits_{\gamma-\im\infty}^{\gamma+\im\infty}\frac{g(t)}{x^t}\dm t,
\end{equation}
the weight $\Lambda(h)$ is in principle recoverable as in
\begin{equation}\label{eq:lmel}\begin{split}
\Lambda(h)&=\frac1{2\pi\im}\!\int\limits_\mathcal C\frac{\dm t}{h^{2t}}
\frac{2\pi^{3/2}\mathfrak M[\Sigma](t+\frac12)}
{\Gamma(t+\frac12)\Gamma(1-t)}
\\&=\frac1{2\pi\im}\int\limits_\mathcal C\frac{\dm t}{h^{2t}}
\frac{\!\sqrt\pi\,\mathfrak M[\varv_\mathrm c^2](t)}
{G\Gamma(t+1)\Gamma(\frac12-t)}.
\end{split}\end{equation}
Here the specific integration path of equation (\ref{eq:mit})
is replaced with a generic complex contour $\mathcal C$ that includes
the point at infinity, which is
permissible provided the path deformation does not encounter
any poles or branch cuts of the integrand.
Equation (\ref{eq:melr}) also directly relates $\mathfrak M[\Sigma](t)$
and $\mathfrak M[\varv_\mathrm c^2](t)$ to each other,
resulting in the formulae;
\begin{equation}\label{eq:vsmel}\begin{split}
\varv_\mathrm c^2(R)&=\frac{2\pi G}{2\pi\im}
\!\int\limits_\mathcal C\frac{\dm s}{R^{2t}}
\frac{\Gamma(t+1)\Gamma(\frac12-t)}{\Gamma(t+\frac12)\Gamma(1-t)}
\mathfrak M[\Sigma](t+\tfrac12);
\\\Sigma(R)&=\frac1{2\pi\im}\!\int\limits_\mathcal C\frac{\dm s}{R^{2t}}
\frac{\Gamma(t)\Gamma(\frac32-t)}{\Gamma(t+\frac12)\Gamma(1-t)}
\frac{\mathfrak M[\varv_\mathrm c^2](t-\tfrac12)}{2\pi G}.
\end{split}\end{equation}
These are valid in general even when a proper weight $\Lambda$ does not exist. In principle, one can derive a similar formula for $\Phi$ via the double Mellin transform of equation (\ref{eq:genpot}) (see e.g.,  \cite{fdisk}, eq.~II.8). However, since $\Phi$ is a function of two variables $(R,z)$, the result involves a double integral expression, which is less practical. Although these formulae appear rather too formal and to be of little practical use, their true value actually lies with ubiquity of the Mellin--Barnes integral representations for many classes of special functions of interest. In particular, the Me{\ij}er-G functions recently have become more widely accepted in many applications, thanks in part to their ready availability in various computer algebra packages.

Suppose that $\Sigma(R)$ is representable as the Me{\ij}er-G function (Appendix~\ref{app:meijer}; see also \cite{Er53}, chap.~5.3 or \cite{DLMF}, \S~16.17) as in
\begin{equation}\label{eq:sigmeij}
\Sigma(R)=\Sigma_0\,G^{m,n}_{p,q}\!\left(\frac{R^2}{R_0^2}\vrule
\stack{a_1,\cdots,a_p}{b_1,\dots,b_q}\right),
\end{equation}
with constants $\Sigma_0$ and $R_0$. By definition,
%
\begin{equation*}
\mathfrak M[\Sigma](t)=\Sigma_0R_0^{2t}
\frac{\prod_{i=1}^n\Gamma(1-a_i-t)}{\prod_{i=n+1}^p\Gamma(t+a_i)}
\frac{\prod_{j=1}^m\Gamma(t+b_j)}{\prod_{j=m+1}^q\Gamma(1-b_j-t)},
\end{equation*}
and so follows equations (\ref{eq:lmel}) and (\ref{eq:vsmel}) that both $\Lambda(h)$ and $\varv_\mathrm c^2(R)$ are also expressible as the Me{\ij}er-G function,
\begin{equation}\label{eq:vmeij1}\begin{split}
\frac{\Lambda(h)}{2\pi^{3/2}\Sigma_0R_0}
&=G^{m,n}_{p+1,q+1}\!\left(\frac{h^2}{R_0^2}
\vrule\stack{a_1+\frac12,\cdots,a_p+\frac12,\frac12}
{b_1+\frac12,\dots,b_q+\frac12,0}\right);
\\\frac{\varv_\mathrm c^2(R)}{2\pi G\Sigma_0R_0}&=
G^{m+1,n+1}_{p+2,q+2}\!\left(\frac{R^2}{R_0^2}
\vrule\stack{\frac12,a_1+\frac12,\cdots,a_p+\frac12,\frac12}
{1,b_1+\frac12,\dots,b_q+\frac12,0}\right).
\end{split}\end{equation}
Similarly if the rotation curve is expressible as
\begin{equation}\label{eq:vcmeij}
\varv_\mathrm c^2(R)=\varv_0^2\,G^{m,n}_{p,q}\!\left(\frac{R^2}{R_0^2}
\vrule\stack{a_1,\cdots,a_p}{b_1,\dots,b_q}\right),
\end{equation}
then $\Lambda(h)$ and $\Sigma(R)$ are given by
\begin{equation*}\begin{split}
\Lambda(h)&=\frac{\!\sqrt\pi\,\varv_0^2}G
G^{m,n}_{p+1,q+1}\!\left(\frac{h^2}{R_0^2}
\vrule\stack{a_1,\cdots,a_p,1}{b_1,\dots,b_q,\frac12}\right);
\\\Sigma(R)&=\frac{\varv_0^2}{2\pi GR_0}
G^{m+1,n+1}_{p+2,q+2}\!\left(\frac{R^2}{R_0^2}
\vrule\stack{-\frac12,a_1-\frac12,\cdots,a_p-\frac12,\frac12}
{0,b_1-\frac12,\dots,b_q-\frac12,0}\right).
\end{split}\end{equation*}
Here both transformations $\Sigma\to\varv_\mathrm c^2$ and $\varv_\mathrm c^2\to\Sigma$ appear to complicate the result, for it sends $G^{m,n}_{p,q}$ to $G^{m+1,n+1}_{p+2,q+2}$ in general. However, if the gamma functions in the Mellin transforms of equation (\ref{eq:vsmel}) cancel another factors in the transformation kernel, the results can be simplified.

For completeness, we also consider the forward transformation of $\Lambda(h)$ given by the Me{\ij}er-G function;
\begin{equation}\label{eq:m_edz_w}
\Lambda(h)=\Lambda_0\,G^{m,n}_{p,q}\!\left(\frac{h^2}{h_0^2}
\vrule\stack{a_1,\cdots,a_p}{b_1,\dots,b_q}\right),
\end{equation}
which then results in
\begin{align}
\Sigma(R)&=\frac{\Lambda_0}{2\pi^{3/2}h_0}
G^{m+1,n+1}_{p+1,q+1}\!\left(\frac{R^2}{h_0^2}
\vrule\stack{-\frac12,a_1-\frac12,\cdots,a_p-\frac12}
{0,b_1-\frac12,\dots,b_q-\frac12}\right);
\label{eq:m_edz_s}\\\label{eq:m_edz_v}
\varv_\mathrm c^2(R)&=\frac{G\Lambda_0}{\!\sqrt\pi}
G^{m+1,n+1}_{p+1,q+1}\!\left(\frac{R^2}{h_0^2}
\vrule\stack{\frac12,a_1,\cdots,a_p}{1,b_1,\dots,b_q}\right).
\end{align}
This is a straightforward consequence of the Me{\ij}er-G convolution theorem since equation (\ref{eq:sigv}) is equivalent to
\begin{align*}\begin{split}
2\pi^\frac32\Sigma(R)
&=\frac1R\int_0^\infty\!\frac{\dm(h^2)}{h^2}
G^{1,1}_{1,1}\!\left(\frac{R^2}{h^2}
\vrule\stack{0}{\frac12}\right)\Lambda(h)
\\&=\int_0^\infty\!\frac{\dm(h^2)}{h^2}
G^{1,1}_{1,1}\!\left(\frac{R^2}{h^2}\vrule\stack{-\frac12}{0}\right)
\frac{\Lambda(h)}{h};\end{split}
\\\frac{\!\sqrt\pi}G\varv_\mathrm c^2(R)
&=\int_0^\infty\!\frac{\dm(h^2)}{h^2}
G^{1,1}_{1,1}\!\left(\frac{R^2}{h^2}
\vrule\stack{\frac12}{1}\right)\Lambda(h).
\end{align*}

\subsection{The potential on the symmetry axis}

Under the axial symmetry, the potential at a point on the symmetry axis is written down as a single integral:
\begin{equation}\label{eq:psa}
\Phi(0,z)=-2\pi G
\!\int_0^\infty\!\frac{R\Sigma(R)\,\dm R}{\!\sqrt{R^2+z^2}}.
\end{equation}
However this converges only if
\footnote{In fact, $\lim_{R\to0}R^2\Sigma(R)=0$ is also necessary, but this must hold for any integrable surface density.}
$\lim_{R\to\infty}R\Sigma(R)=0$ (i.e.\ $p_\infty>1$ where $\Sigma\sim R^{-p_\infty}$ as $R\to\infty$), reflecting the condition for the zero-point to be set at infinity. On the other hand, following a similar trick for equation (\ref{eq:potshi}), we may obtain an alternative integral for the potential with the zero-point at the origin:
\begin{equation*}
\Phi(0,z)=2\pi G\!\int_0^\infty\!\dm R\,\Sigma(R)
\frac{\!\sqrt{R^2+z^2}-R}{\!\sqrt{R^2+z^2}},
\end{equation*}
which converges if $\lim_{R\to\infty}R^{-1}\Sigma(R)=\lim_{R\to0}R\Sigma(R)=0$ (i.e.\ $p_\infty>-1$ and $p_0<1$ where $\Sigma\sim R^{-p_0}$ as $R\to0$).

Equation (\ref{eq:psa}) is also recognized as a GST of order $1/2$. Hence it is also possible to invert $\Phi(0,z)$ for $\Sigma(R)$ based on the techniques discussed in Sect.~\ref{sec:inv}. However given ambiguity of the zero-point, it is more appropriate to consider inverting the gravitational acceleration instead. The gravitational acceleration on the symmetry axis due to the disk as a function of the height is
\begin{align}
\varg(z)&=\left\lvert\frac{\dm\Phi(0,z)}{\dm z}\right\rvert
=2\pi Gz\!\int_0^\infty\!\frac{R\Sigma(R)\,\dm R}{(R^2+z^2)^{3/2}}
\label{eq:gf}\\\nonumber
&=2\pi G\left[\Sigma(0)
+z\int_0^\infty\!\frac{\dm R}{\!\sqrt{R^2+z^2}}
\frac{\dm\Sigma(R)}{\dm R}\right]
\quad(z>0)
\end{align}
which converges assuming $\lim_{R\to\infty}\Sigma(R)/R=0$. If the total disk mass, $M_\mathrm t=2\pi\int_0^\infty\!\dm R\,R\Sigma(R)$ is finite, the acceleration decays like $\varg(z)\simeq GM_\mathrm t/z^2$ as $z\to\infty$, but the gravitational force due to an infinite-mass disk can fall off slower.

Equation (\ref{eq:gf}) is formally equivalent to equation (\ref{eq:sigv}), and so the inversion formula for $\varg(z)\to\Sigma(R)$ follows that for $R\Sigma(R)\to\Lambda(h)$. If the analytic continuation of $\varg(z)$ for the complex right half-plane (i.e.\ $\Re[z]>0$ and $|\arg(z)|<\pi/2$) is available, the integral inversion formula is given by
\begin{equation*}
\pi G\Sigma(R)
=\frac1{2\pi\im}\int\limits_{-\im R}^{\im R}
\frac{\varg(z)\,\dm z}{\!\sqrt{z^2+R^2}},
\end{equation*}
which is analogous to equations (\ref{eq:invint}). Alternatively by considering the Mellin transform of $\varg(z)$:
\begin{equation*}\begin{split}
\mathfrak M[\varg](t)&=\int_0^\infty\!\dm(z^2)\,z^{2(t-1)}\varg(z)
=2\int_0^\infty\!\dm z\,z^{2t-1}\varg(z)
\\&=4\!\sqrt\pi\,G\Gamma(t+\tfrac12)\Gamma(1-t)
\!\int_0^\infty\!\dm R\,R^{2t-1}\Sigma(R),
\end{split}\end{equation*}
we also derive an inversion formula similar to equation (\ref{eq:lmel});
\begin{equation*}
2\!\sqrt\pi\,G\,\Sigma(R)=\frac1{2\pi\im}\int\limits_\mathcal C\!
\frac{\dm t}{R^{2t}}\frac{\mathfrak M[\varg](t)}
{\Gamma(t+\frac12)\Gamma(1-t)}.
\end{equation*}
Moreover, together with equation (\ref{eq:melr}), we can establish that $\!\sqrt\pi\,\mathfrak M[\varg](t)=\Gamma(t)\Gamma(\frac32-t)\mathfrak M[\varv_\mathrm c^2](t-\frac12)$, and so follows that
\begin{equation*}\begin{split}
\varv_\mathrm c^2(R)
&=\frac{\!\sqrt\pi}{2\pi\im}\int\limits_\mathcal C\frac{\dm t}{R^{2t}}
\frac{\mathfrak M[\varg](t+\frac12)}{\Gamma(t+\frac12)(1-t)};
\\\varg(z)&=\frac2{\!\sqrt\pi}
\frac1{2\pi\im}\int\limits_\mathcal C\!\dm t\,
\frac{\Gamma(t)\Gamma(\frac32-t)}{z^{2t}}\!
\int_0^\infty\!\dm R\,R^{2t-2}\varv_\mathrm c^2(R)
\\&=\frac2{\!\sqrt\pi}
\int_0^\infty\!\dm R\,\frac{\varv_\mathrm c^2(R)}{R^2}\,
G^{1,1}_{1,1}\!\left(\frac{z^2}{R^2}\vrule\stack{-\frac12}{0}\right)
\\&=\int_0^\infty\!\frac{R\varv_\mathrm c^2(R)\,\dm R}{(R^2+z^2)^{3/2}}
=\frac{\varv_\mathrm c^2(0)}z
+\int_0^\infty\!\frac{\dm R}{\!\sqrt{R^2+z^2}}
\frac{\dm\varv_\mathrm c^2(R)}{\dm R}.
\end{split}\end{equation*}
Therefore the disk with the surface density of equation (\ref{eq:sigmeij}) produces the vertical acceleration profile of the form
\begin{equation*}
\varg(z)=2\!\sqrt\pi\,G\Sigma_0\,
G^{m+1,n+1}_{p+1,q+1}\!\left(\frac{z^2}{R_0^2}\vrule
\stack{0,a_1,\dotsc,a_p}{\frac12,b_1,\dotsc,b_q}\right),
\end{equation*}
whereas the vertical acceleration due to the disk generating rotation curve of equation (\ref{eq:vcmeij}) is found to be
\begin{equation*}
\varg(z)=\frac{\varv_0^2}{\!\sqrt\pi\,R_0}
G^{m+1,n+1}_{p+1,q+1}\!\left(\frac{z^2}{R_0^2}\vrule
\stack{-\frac12,a_1-\frac12,\dotsc,a_p-\frac12}
{0,b_1-\frac12,\dotsc,b_q-\frac12}\right).
\end{equation*}
Conversely if the vertical acceleration profile is known to be
\begin{equation*}
\varg(z)=\varg_0\,G^{m,n}_{p,q}\left(\frac{z^2}{z_0^2}\vrule
\stack{a_1,\dotsc,a_p}{b_1,\dotsc,b_q}\right),
\end{equation*}
then the disk surface density and the rotation curve should be
\begin{equation*}\begin{split}
\Sigma(R)&=\frac{\varg_0}{2\!\sqrt\pi\,G}
G^{m,n}_{p+1,q+1}\left(\frac{R^2}{z_0^2}\vrule
\stack{a_1,\dotsc,a_p,\frac12}{b_1,\dotsc,b_q,0}\right);
\\\varv_\mathrm c^2(R)&=\!\sqrt\pi\,z_0\varg_0\,
G^{m,n}_{p+1,q+1}\left(\frac{R^2}{z_0^2}\vrule
\stack{a_1+\frac12,\dotsc,a_p+\frac12,\frac12}{b_1+\frac12,\dotsc,b_q+\frac12,0}\right).
\end{split}\end{equation*}

\subsection{The Qian Hypergeometric disks}
\label{sec:qhd}

Note that the generalized hypergeometric function is a special case of the Me{\ij}er-G functions (cf.\ eq~\ref{eq:mhgf}): that is,
\begin{multline}\label{eq:hypg_m}
x^{2t}{_pF_q}\!\left(\stack{a_1,\dotsc,a_p}{b_1,\dotsc,b_q};-x^2\right)
=\frac{\prod_{j=1}^q\Gamma(b_j)}{\prod_{i=1}^p\Gamma(a_i)}\\\times
G^{1,p}_{p,q+1}\!\left(x^2\vrule
\stack{1-a_1+t,\dotsc,1-a_p+t}{t,1-b_1+t,\dotsc,1-b_q+t}\right).
\end{multline}
Hence the disk with the surface density expressible to be
\begin{equation}\label{eq:hypg1}
\frac{\Sigma(R)}{\Sigma_0}
=\frac{\prod_{i=1}^p\Gamma(a_i)}{\prod_{j=1}^q\Gamma(b_j)}
{_pF_q}\!\left(\stack{a_1,\dotsc,a_p}{b_1,\dotsc,b_q};
-\frac{R^2}{R_0^2}\right)
\end{equation}
generates the mid-plane rotation curve behaving like
\begin{multline}\label{eq:hypg2}
\frac{\varv_\mathrm c^2(R)}{G\Sigma_0R_0}
=\frac{\pi^{3/2}\prod_{i=1}^p\Gamma(a_i+\frac12)}
{\prod_{j=1}^q\Gamma(b_j+\frac12)}\\\times\frac{R^2}{R_0^2}\,
{_{p+1}F_{q+1}}\!\left(\stack{\frac32,a_1+\frac12,\dotsc,a_p+\frac12}
{2,b_1+\frac12,\dotsc,b_q+\frac12};-\frac{R^2}{R_0^2}\right),
\end{multline}
while the mid-plane potential is recoverable through integrating $\varv_\mathrm c^2=R[\dm\Phi(R,0)/\dm R]=2x(\dm\Phi/\dm x)$ (where $x=R^2/R_0^2$),
\begin{multline*}
\frac{\Phi(R,0)}{G\Sigma_0R_0}
=-\frac{\pi^{3/2}\prod_{i=1}^p\Gamma(a_i-\frac12)}
{\prod_{j=1}^q\Gamma(b_j-\frac12)}\\\times
{_{p+1}F_{q+1}}\!\left(\stack{\frac12,a_1-\frac12,\dotsc,a_p-\frac12}
{1,b_1-\frac12,\dotsc,b_q-\frac12};-\frac{R^2}{R_0^2}\right).
\end{multline*}
Equations (\ref{eq:hypg1}) and (\ref{eq:hypg2}) reproduce equations (3.3) and (3.5) of \cite{Qi92} for the axisymmetric case ($m=0$), and so the formulae utilizing the Me{\ij}er-G function supersede those of \cite{Qi92} based on the generalized hypergeometric functions. In principle, the modifications to include the non-axisymmetric cases are possible following similar procedures as \cite{Qi92} establishing the integral transforms for the higher-order Fourier components although we shall not pursue this approach in this paper.

The vertical acceleration due to the disk of the same surface density and the weight function resulting in the same disk are both formally given by the Me{\ij}er-G function:
\begin{align}
\frac{\varg(z)}{G\Sigma_0}&=2\!\sqrt\pi\,
G^{2,p+1}_{p+1,q+2}\!\left(\frac{z^2}{R_0^2}\vrule
\stack{0,1-a_1,\dotsc,1-a_p}{0,\frac12,1-b_1,\dotsc,1-b_p}\right);
\nonumber\\\label{eq:hypgw}
\frac{\Lambda(h)}{\Sigma_0R_0}&=2\pi^\frac32\,
G^{0,p}_{p,q+1}\!\left(\frac{h^2}{R_0^2}
\vrule\stack{\frac32-a_1,\cdots,\frac32-a_p}
{\frac32-b_1,\dots,\frac32-b_q,0}\right).
\end{align}
Although equation (\ref{eq:hypg_m}) is not applicable here, both of these Me{\ij}er-G functions are reducible to the hypergeometric series; ${_{p+1}F_{q+1}}$ for $\varg(z)$ and ${_pF_q}$ for $\Lambda(h)$. For instance, $\varg(z)$ reduces to the hypergeometric series of the form (here $\zeta=z/R_0$)
\begin{multline*}
\frac{\varg(z)}{2\pi G\Sigma_0}=\sum_{k=0}^\infty
\frac{\Gamma(1+k/2)\Gamma(a_i+k/2)}{\Gamma(b_j+k/2)}\frac{(-2\zeta)^k}{k!}
\\=\frac{\prod_{i=1}^p\Gamma(a_i)}{\prod_{j=1}^q\Gamma(b_j)}
{_{p+1}F_{q+1}}\!\left(\stack{1,a_1,\dotsc,a_p}{\frac12,b_1,\dotsc,b_q};\zeta^2\right)
\\-\frac{\!\sqrt\pi\,\prod_{i=1}^p\Gamma(a_i+\frac12)}{\prod_{j=1}^q\Gamma(b_j+\frac12)}
\zeta{_pF_q}\!\left(\stack{a_1+\frac12,\dotsc,a_p+\frac12}
{b_1+\frac12,\dotsc,b_q+\frac12};\zeta^2\right),
\end{multline*}
With $\Lambda$ found, the potentials due to these disks follow equation (\ref{eq:genpot}), which may be evaluated as a real quadrature. In the subsequent sections, we shall consider when this integral can be resolved by analytic means. Following \cite{Qi92}, we restrict ourselves to the cases that both $\Sigma(R)$ and $\varv_\mathrm c^2(R)$ are expressible through at most a single ${_2F_1}$-function.

Before exploring this, we finish this section by considering the disk generating the rotation curve given in the form of
\begin{equation*}
\frac{\varv_\mathrm c^2(R)}{\varv_0^2}
=\frac{\prod_{i=1}^p\Gamma(a_i)}{\prod_{j=1}^q\Gamma(b_j)}
\frac{R^2}{R_0^2}\,{_pF_q}\!\left(\stack{a_1,\dotsc,a_p}{b_1,\dotsc,b_q};
-\frac{R^2}{R_0^2}\right).
\end{equation*}
The surface density of this razor-thin disk is then found to be
\begin{multline}\label{eq:hypgv}
\Sigma(R)=\frac{\varv_0^2}{GR_0}
\frac{\prod_{i=1}^p\Gamma(a_i-\frac12)}
{4\!\sqrt\pi\,\prod_{j=1}^q\Gamma(b_j-\frac12)}\\\times
{_{p+1}F_{q+1}}\!\left(\stack{\frac32,a_1-\frac12,\dotsc,a_p-\frac12}
{1,b_1-\frac12,\dotsc,b_q-\frac12};-\frac{R^2}{R_0^2}\right),
\end{multline}
while the vertical acceleration along the axis is given by
\begin{equation*}\begin{split}
\varg(z)&=\frac{\varv_0^2}{\!\sqrt\pi\,R_0}
G^{2,p+1}_{p+1,q+2}\!\left(\frac{z^2}{R_0^2}\vrule
\stack{-\frac12,\frac32-a_1,\dotsc,\frac32-a_p}
{0,\frac12,\frac32-b_1,\dotsc,\frac32-b_n}\right)
\\&=\sum_{k=0}^\infty
\frac{\Gamma(\frac32+k/2)\Gamma(a_i-\frac12+k/2)}
{k!\Gamma(b_j-\frac12+k/2)}
\left(-\frac{2z}{R_0}\right)^k
\\&=\frac{\!\sqrt\pi\,\prod_{i=1}^p\Gamma(a_i-\frac12)}
{2\prod_{j=1}^q\Gamma(b_j-\frac12)}
{_{p+1}F_{q+1}}\!\left(\stack{\frac32,a_i-\frac12}{\frac12,b_j-\frac12};
\frac{z^2}{R_0^2}\right)
\\&\qquad-\frac{2\prod_{i=1}^p\Gamma(a_i)}{\prod_{j=1}^q\Gamma(b_j)}
\zeta\,{_{p+1}F_{q+1}}\!\left(\stack{2,a_i}{\frac32,b_j};
\frac{z^2}{R_0^2}\right).
\end{split}\end{equation*}
Finally the weight $\Lambda(h)$ corresponding to the disk with the surface density of equation (\ref{eq:hypgv}) is
\begin{equation*}
\Lambda(h)=\frac{\!\sqrt\pi\,\varv_0^2}G\,
G^{0,p}_{p,q+1}\!\left(\frac{h^2}{R_0^2}
\vrule\stack{2-a_1,\cdots,2-a_p}{2-b_1,\dots,2-b_q,\frac12}\right).
\end{equation*}

\section{The Kuzmin--Mestel--Toomre disks}
\label{sec:mkt}

As the simplest example, we first consider the disk generated by the \citetalias{EdZ92} weight in the form of equation (\ref{eq:kmt_w}). In particular, let us consider the disk with the surface density of
\begin{equation}\label{eq:kmt}
\Sigma(R)=\Sigma_0\left(1+\frac{R^2}{R_0^2}\right)^{-\gamma}
=\frac{\Sigma_0R_0^{2\gamma}}{(R^2+R_0^2)^\gamma}.
\end{equation}
Here $\Sigma_0$ and $R_0$ are the scale constants and $a$ controls the shape. The surface density at center is finite, $\Sigma(0)=\Sigma_0$ whereas $\Sigma\sim R^{-2\gamma}$ as $R\to\infty$ with the $\gamma=0$ case corresponding to an infinite uniform plate. If $\gamma>1$, the mass contained in the disk increases with the radius like $M(R)/M_\mathrm t=1-(1+\varrho^2)^{-(\gamma-1)}$, where $\varrho=R_0/R$ and $M_\mathrm t=\pi R_0^2\Sigma_0/(\gamma-1)$ is the total mass of the disk. If $\gamma\le1$ on the other hand, the mass grows without any bound: in particular, $M(R)/(\pi R_0^2\Sigma)=\ln(1+\varrho^2)$ for $\gamma=1$ and $M(R)/(\pi R_0^2\Sigma)=[(1+\varrho^2)^{1-\gamma}-1]/(1-\gamma)$ for $\gamma<1$. If $\gamma-\frac12=n$ is a positive integer, the model is the same as the Model-$n$ of \cite{To63}, while the $\gamma=\frac12$ and $\frac32$ cases are respectively identified with the cored \citeauthor{Me63} disks and the \citeauthor{Ku56} disks. Hereafter we shall refer to this family as the KMTd (after \citeauthor{Ku56}--\citeauthor{Me63}--\citeauthor{To63} disk).

Given that the Mellin transform ($R^2\to t$ as defined in eq.~\ref{eq:sigmel}) of equation (\ref{eq:kmt}) is found to be
\begin{equation*}
\mathfrak M[\Sigma](t)
=R_0^{2t}\Sigma_0\frac{\Gamma(t)\Gamma(\gamma-t)}{\Gamma(\gamma)}
\quad(0<t<\gamma),
\end{equation*}
the surface density of equation (\ref{eq:kmt}) is also expressed as
\begin{equation*}
\frac{\Sigma(R)}{\Sigma_0}=\frac1{\Gamma(\gamma)}
G^{1,1}_{1,1}\!\left(\varrho^2\vrule\stack{1-\gamma}{0}\right)
={_1F_0}\!\left(\stack{\gamma}{};-\varrho^2\right).
\end{equation*}
where again $\varrho\equiv R/R_0$ (and henceforth). Therefore the rotation curve due to the KMTd is
\begin{equation}\begin{split}\label{eq:kmtvc}
\frac{\varv_\mathrm c^2(R)}{GR_0\Sigma_0}&=\frac{2\pi}{\Gamma(\gamma)}
G^{1,2}_{2,2}\!\left(\varrho^2\vrule
\stack{\frac12,\frac32-\gamma}{1,0}\right)
\\&=\frac{\pi^{3/2}\Gamma(\gamma+\frac12)}{\Gamma(\gamma)}\varrho^2
{_2F_1}\!\left(\stack{\frac32,\gamma+\frac12}2;-\varrho^2\right);
\end{split}\end{equation}
for $\gamma>-\frac12$, whereas the vertical acceleration is
\begin{align}
\frac{\varg(z)}{G\Sigma_0}&=\frac{2\!\sqrt\pi}{\Gamma(\gamma)}
\,G^{2,2}_{2,2}\!\left(\frac{z^2}{R_0^2}\vrule
\stack{0,1-\gamma}{0,\frac12}\right)
\nonumber\\
&=2\pi\left[{_2F_1}\!\left(\stack{1,\gamma}{\frac12};
\zeta^2\right)
-\frac{\!\sqrt\pi\,\Gamma(\gamma+\frac12)}{\Gamma(\gamma)}
\frac\zeta{(1-\zeta^2)^{\gamma+1/2}}\right]
\nonumber\\\label{eq:kmtgz}
&=\frac{2\pi}{2\gamma+1}
{_2F_1}\!\left(\stack{\gamma,1}{\gamma+\frac32};1-\zeta^2\right),
\end{align}
where $\zeta\equiv z/R_0$ and henceforth. While the last line follows Kummer's analytic continuation formula for the Gauss-${_2F_1}$ hypergeometric functions (see \cite{Er53}, eq.~2.10.(1)), it can be also derived directly from equation (\ref{eq:gf}). If $\gamma=0$, then $\varv_\mathrm c^2=0$ and $\varg=2\pi G\Sigma_0$, which are consistent with the translation symmetry implying no net gravitational force within the infinite uniform plate and the scale invariance leading to the constant force off the plate. For $-\frac12<\gamma<0$, we find $\varv_\mathrm c^2<0$, which reflects the fact that the gravitational force directs radially outward in the corresponding cases. If $\gamma\le-\frac12$, the gravity from matters at infinity dominates and the integral for the gravitational force diverges. Although a formal treatment is still possible, an infinite disk is an artificial construction in any case and so this is not a physical problem.

Equations (\ref{eq:kmtvc}) and (\ref{eq:kmtgz}) indicate that the rotation curve and the vertical acceleration as $r\to\infty$ behave like $\varv_\mathrm c^2\simeq GM_\mathrm t/R$ and $\varg\simeq GM_\mathrm t/z^2$ for finite-mass disks ($\gamma>1$), or
\begin{equation*}\begin{split}
\varv_\mathrm c^2(R)&\simeq2\pi GR_0\Sigma_0
\times\begin{cases}\varrho^{-1}[\ln(4\varrho)-1]&(\gamma=1)
\\A_\gamma\,\varrho^{1-2\gamma}&(1>\gamma>-\frac12)\end{cases},
\\\varg(z)&\simeq2\pi G\Sigma_0
\times\begin{cases}\zeta^{-2}[\ln(2\zeta)-1]&(\gamma=1)
\\B_\gamma\,\zeta^{-2\gamma}&(1>\gamma>-\frac12)\end{cases};
\\A_\gamma&=\frac{\Gamma(\frac12+\gamma)\Gamma(1-\gamma)}
{\Gamma(\gamma)\Gamma(\frac32-\gamma)},\quad
B_\gamma=\frac{\Gamma(\frac12+\gamma)\Gamma(1-\gamma)}{\!\sqrt\pi}.
\end{split}\end{equation*}
Here $\varv_\mathrm c(R)$ for $\gamma<\frac12$ diverges as $R\to\infty$ and the corresponding potential also diverges like $\Phi\sim r^{2\gamma-1}$ asymptotically (note $\varg\sim z^{-2\gamma}$). On the other hand, the rotation curve due to the $\gamma=\frac12$ case (the cored Mestel disk) is asymptotically flat and the potential diverges logarithmically.

The proper \citetalias{EdZ92} weight function of the KMTd exists for $0<\gamma<\frac32$ and is formally given by (cf.\ eq.~\ref{eq:kmt_w})
\begin{equation}\label{eq:edzw}\begin{split}
\Lambda(h)&=\frac{2\pi^{3/2}R_0\Sigma_0}{\Gamma(\gamma)}
G^{0,1}_{1,1}\!\left(\frac{h^2}{R_0^2}
\vrule\stack{\frac32-\gamma}{0}\right)
\\&=\frac{2\pi^{3/2}R_0^{2\gamma}\Sigma_0}
{\Gamma(\gamma)\Gamma(\frac32-\gamma)}
\frac{\Theta(h^2-R_0^2)}{(h^2-R_0^2)^{\gamma-1/2}}
\end{split}\end{equation}
where $\Theta(x)$ is the Heaviside unit step. Hence the full potential due to the KMTd can be written down as
\begin{equation}\label{eq:kmtp}
\Phi(R,z)
=-\frac{2\pi^{3/2}GR_0^{2\gamma}\Sigma_0}
{\Gamma(\gamma)\Gamma(\frac32-\gamma)}\int_{R_0}^\infty
\frac{(h^2-R_0^2)^{1/2-\gamma}\dm h}{\!\sqrt{R^2+(|z|+h)^2}},
\end{equation}
which converges for $\frac12<\gamma<\frac32$. The non-convergence for $\gamma\le\frac12$ is due to the impossibility of setting the zero-point for the potential at infinity. If the alternative zero-point $\Phi(0,0)=0$ is chosen instead, the adjustment similar to equation (\ref{eq:potshi}) leads to the zero-point-shifted potential in a quadrature;
\begin{multline}\label{eq:kmtps}
\Phi(R,z)
=-\frac{2\pi^{3/2}GR_0^{2\gamma}\Sigma_0}
{\Gamma(\gamma)\Gamma(\frac32-\gamma)}
\\\times\int_{R_0}^\infty\!\dm h\,(h^2-R_0^2)^{\frac12-\gamma}
\left[\frac1{\!\sqrt{R^2+(|z|+h)^2}}-\frac1h\right],
\end{multline}
which is convergent if $0<\gamma<\frac32$. Note that equation (\ref{eq:kmtps}) with $\gamma=\frac12$ recovers the potential of the cored Mestel disk in equation (\ref{eq:cmdp}) with $\varv_\infty^2=2\pi G\Sigma_0R_0$ and $a=c=R_0$.

With either integral in equation (\ref{eq:kmtp}) or (\ref{eq:kmtps}), one can also derive the expression for the mid-plane rotation curve, which recovers equation (\ref{eq:kmtvc}). On the other hand, the vertical acceleration follows the potential as
\begin{equation}\label{eq:kmtgz2}\begin{split}
\varg&=\frac{2\pi^{3/2}GR_0^{2\gamma}\Sigma_0}
{\Gamma(\gamma)\Gamma(\frac32-\gamma)}
\int_0^\infty\frac{(h+2R_0)^{1/2-\gamma}h^{1/2-\gamma}\dm h}{(z+R_0+h)^2}
\\&=\frac{2\pi G\Sigma_0}{2\gamma+1}
{_2F_1}\!\left(\stack{2\gamma,2}{\gamma+\frac32};\frac{1-\zeta}2\right),
\end{split}\end{equation}
which is equivalent to equation (\ref{eq:kmtgz}) thanks to the quadratic transformations for the ${_2F_1}$-functions (e.g., \cite{DLMF}, eq.~15.8.18). Although the integral in equation (\ref{eq:kmtgz2}) converges only if $0<\gamma<\frac32$, the last expression is valid for any $\gamma>-\frac12$.

If $\gamma\ge\frac32$ or $\gamma\le0$, equation (\ref{eq:sigv}) with $\Lambda$ given by equation (\ref{eq:edzw}) diverges. The proper weight for $\gamma=\frac32$ is instead $\Lambda=2\pi R_0^2\Sigma_0\deltaup(h-R_0)=4\pi R_0^3\Sigma_0\deltaup(h^2-R_0^2)$ where $\deltaup(x)$ is the Dirac delta. This is equivalent to identifying with $\lim_{\xi\downarrow0^+}[(x-y)^{\xi-1}/\Gamma(\xi)]=\deltaup(x-y)$ in equation (\ref{eq:edzw}). Equation (\ref{eq:genpot}) with the Dirac delta weight then recovers the potential of the Kuzmin disk (with $a=R_0$ and $M_\mathrm t=2\pi\Sigma_0R_0^2$) in equation (\ref{eq:kdp}). For $\gamma>\frac32$ however, a proper non-negative weight does not exist because, as noted earlier in Sect.~\ref{sec:melt}, any surface density expressible in the form of equation (\ref{eq:sigv}) with $\Lambda\ge0$ cannot decay faster than $\sim R^{-3}$ asymptotically.

\subsection{The recursion relation and the Weyl integrals}
\label{sec:kmtrw}

None the less, the potential due to the KMTd for $\gamma>\frac32$ is still found by following \cite{To63}. In particular, observe that equation (\ref{eq:kmt}) obeys the recursion relation:
\begin{equation}\label{eq:kmtdr}
\frac\partial{\partial(R_0^2)}\frac{\Sigma(R;\gamma)}{\Sigma_0R_0^{2\gamma}}
=-\gamma\frac{\Sigma(R;\gamma+1)}{\Sigma_0R_0^{2(\gamma+1)}}.
\end{equation}
The linearity of the Poisson equation indicates that the potential due to the disk also follows the analogous relation that
\begin{equation}\label{eq:mktpn}
\left[\frac\partial{\partial(R_0^2)}\right]^m
\frac{\Phi(\bm r;\gamma)}{\Sigma_0R_0^{2\gamma}}
=(-1)^m(\gamma)_m\frac{\Phi(\bm r;\gamma+m)}{\Sigma_0R_0^{2(\gamma+m)}},
\end{equation}
where $(a)_n=\prod_{j=0}^{n-1}(a+j)=\Gamma(a+n)/\Gamma(a)$ is the Pochhammer symbol (rising sequential product). That is, if $m=\lceil\gamma-\frac32\rceil$ is the integer ceiling of $\gamma-\frac32$, the potential due to the KMTd is obtained from the $m$-th derivative of equation (\ref{eq:kmtp}). In particular, the potential for $\gamma=\frac32+m$ with a non-negative integer $m$ -- corresponding to the \citeauthor{To63} Model-$(m+1)$ -- is given by
\begin{multline}\label{eq:tmrn}
\Phi(R,z)=-\frac{\pi GR_0^{3+2m}\Sigma_0}{(-1)^m(\frac12)_{m+1}}
\frac{\partial^m}{\partial(R_0^2)^m}
\frac1{R_0\!\sqrt{R^2+(|z|+R_0)^2}}
\\=-\frac{\pi GR_0\Sigma_0}{(\frac12)_{m+1}}
\frac{\partial^m}{\partial x^m}
\frac{x^m}{\sqrt{(1+x^{1/2}|z|)^2+xR^2}}\biggr\rvert_{x=R_0^{-2}}.
\end{multline}
Here the second line follows theorem A3 of \cite{An11}. With a formal identification of $(\partial/\partial x)^{-1}=\int\!\dm x$, equation (\ref{eq:tmrn}) may also be extended to $m=-1$ (i.e.\ $\gamma=\frac12$): namely,
\begin{multline}\label{eq:cmpi}
\Phi(R,z)=2\pi GR_0\Sigma_0\int\!
\frac{\dm R_0}{\!\sqrt{(R_0+|z|)^2+R^2}}
\\=2\pi GR_0\Sigma_0\ln\Bigl[R_0+|z|+\!\sqrt{R^2+(|z|+R_0)^2}\Bigr]+C,
\end{multline}
which is equivalent to equation (\ref{eq:kmtps}) with $\gamma=\frac12$. In other words, the cored Mestel disk is recognized as the \citeauthor{To63} Model-0 disk \citep{EdZ92}.

The general expression of the potential for the KMTd with $\gamma>\frac12$ can be written down compactly by introducing the Weyl integrals \citep[chap.~XIII]{Er54}:
\begin{equation}\label{eq:weyl}
{_b^-I}_x^\eta\,f(x)
=\begin{cases}\displaystyle
\frac1{\Gamma(\eta)}\int_x^b\!(y-x)^{\eta-1}f(y)\,\dm y
&(\eta>0)\\f(x)&(\eta=0)\end{cases},
\end{equation}
for $\eta\ge0$, and, if $\eta=-\xi<0$, then
\begin{equation*}
{_b^-I}_x^{-\xi}\,f(x)
=\biggl(-\frac\dm{\dm x}\biggr)^{\lfloor\xi\rfloor+1}
{_b^-I}_x^{1-\{\xi\}}f
=\biggl(-\frac\dm{\dm x}\biggr)^{\lceil\xi\rceil}
{_b^-I}_x^{\lceil\xi\rceil-\xi}f.
\end{equation*}
Here $\lfloor\xi\rfloor$, $\lceil\xi\rceil$ and $\{\xi\}=\xi-\lfloor\xi\rfloor$ are the integer floor, the integer ceiling, and the fractional part of $\xi>0$ respectively. If $m$ is a non-negative integer, ${_b^-I}_x^{-m}f(x)=(-1)^mf^{(m)}(x)$, and thus the Weyl integral may be considered as a generalization of differentiations. The Weyl integral for a complex variable is equivalent to the Riemann--Liouville fractional integral (see e.g., \cite{AvHB}), and the explicit transformations of Weyl's to the Riemann--Liouville integral restricted for a real positive variable also exist (e.g., change of the integration variable $y\to y^{-1}$). Finally note the composite rule ${_b^-I}_x^\xi\,{_b^-I}_x^\eta f={_b^-I}_x^{\xi+\eta}f$, which is valid for $\eta>-1$ and any $\xi$. Then combining equations (\ref{eq:kmtp}) and (\ref{eq:mktpn}) results in the general expression of the potential by means of the Weyl integral ($\gamma>\frac12$)
\begin{equation}\label{eq:weylpot}
\Phi=-\frac{\pi^{3/2}GR_0^{2\gamma}\Sigma_0}{\Gamma(\gamma)}
{_\infty^-I}_{R_0^2}^{\frac32-\gamma}
\left[\frac1{R_0\!\sqrt{R^2+(|z|+R_0)^2}}\right].
\end{equation}
If $\gamma<\frac32$, this reproduces equation (\ref{eq:kmtp}), while setting $\gamma=\frac32+m$ recovers equation (\ref{eq:tmrn}). Lastly we establish that this satisfies the recursion relation in equation (\ref{eq:mktpn}) following $(\dm/\dm x)^n({_b^-I}_x^\eta f)=(-1)^n\,{_b^-I}_x^{\eta-n}f$ and $(\gamma)_n=\Gamma(\gamma+n)/\Gamma(\gamma)$. Thanks to mathematical induction, equation (\ref{eq:weylpot}) is therefore the potential (with the zero-point at infinity) generated by the KMTd for $\gamma>\frac12$.

\subsection{The prolate spheroidal coordinates}
\label{sec:ecd}

Equation (\ref{eq:weylpot}) is still too formal and the differentiation, albeit good for analytic purposes, is not suitable for numerical evaluations, especially for a higher order. Nevertheless we can utilize equation (\ref{eq:mktpn}) to derive alternative integral quadratures (and other closed form expressions in terms of known special functions) for the potential with any $\gamma>\frac12$.

One such possibility involves the \emph{prolate} spheroidal coordinates (or the elliptic coordinates in the meridional plane). First let us define the distances $r_\pm$ to two points on the $R=0$ axis with $z=\mp R_0$ as in $r_\pm^2=R^2+(z\pm R_0)^2$. In the meridional plane, the loci of constant sums $(r_++r_-)$ and differences $|r_+-r_-|$ respectively trace the set of confocal ellipses and hyperbolae. Rotating around the $R=0$ symmetry axis, these sets together with the meridional planes constitute the mutually orthogonal coordinate surfaces for the prolate spheroidal coordinate \citep{Ly03,An13}. We choose the coordinate variables $(\lambda,\mu,\phi)$ to be the usual azimuth $\phi$ and
\begin{equation}\label{eq:pscv}
\lambda^\frac12=\frac{r_++r_-}{2R_0},
\qquad\mu^\frac12=\frac{r_+-r_-}{2R_0},
\end{equation}
which are consistent with equation (6.4) of \cite{EdZ92}, although the more common choice is $(\cosh^{-1}\lambda^{1/2},\cos^{-1}\mu^{1/2},\phi)$. The triangle inequality implies that $|\mu^{1/2}|\le1\le\lambda^{1/2}$ and so $0\le\mu\le1\le\lambda$. Technically two points symmetric about the $z=0$ plane would result in the same $(\lambda,\mu)$ but $\mu^{1/2}<0$ if $z<0$. Since the potential due to a razor-thin disk is reflection symmetric about the disk plane, all subsequent results involving the coordinate variables $(\lambda,\mu)$ are interpreted to be valid for $z\ge0$ and the result for $z<0$ is recovered through a simple reflection, e.g., $\Phi(R,-z)=\Phi(R,z)$ etc.

To find the inverse transformation $(\lambda,\mu)\to(R,z)$, note
\begin{equation}\label{eq:lmz}
(\lambda\mu)^\frac12=\frac{r_+^2-r_-^2}{4R_0^2}=\frac z{R_0}=\zeta.
\end{equation}
Next, $4R_0^2(\lambda,\mu)=(r_+^2+r_-^2+2r_+r_-,r_+^2+r_-^2-2r_+r_-)$ and so
\begin{equation*}
\lambda+\mu=\frac{r_+^2+r_-^2}{2R_0^2}=\frac{r^2+R_0^2}{R_0^2}
,\quad\lambda-\mu=\frac{r_+r_-}{R_0^2}
\end{equation*}
where $r^2=R^2+z^2$. It follows that
\begin{equation}\label{eq:lmR}
\varrho^2=\frac{R^2}{R_0^2}=\lambda+\mu-\lambda\mu-1=(\lambda-1)(1-\mu).
\end{equation}
%
Then the potentials due to the Kuzmin disk and the cored Mestel disk of the scale length $R_0$ are expressible to be
\begin{equation}\label{eq:kmplm}\begin{split}
\Phi_\mathrm K&=-\frac{2\pi GR_0\Sigma_0}{\lambda^{1/2}+\mu^{1/2}},\\
\Phi_\mathrm M&=2\pi GR_0\Sigma_0\ln(\lambda^\frac12+1)(1+\mu^\frac12).
\end{split}\end{equation}
Lastly, either from equation (\ref{eq:pscv}) with $\partial r_\pm/\partial R=R/r_\pm$ and $\partial r_\pm/\partial z=(z\pm R_0)/r_\pm$, or inverting the partial derivatives of equations (\ref{eq:lmz}) and (\ref{eq:lmR}) with respect to $(R,z)$, we find
\begin{equation}\label{eq:lmpd}
\begin{pmatrix}
\dfrac{\partial\lambda}{\partial(R^2)}&
\dfrac{\partial\lambda}{\partial(z^2)}
\smallskip\\\dfrac{\partial\mu}{\partial(R^2)}&
\dfrac{\partial\mu}{\partial(z^2)}
\end{pmatrix}
=\frac1{r_+r_-}
\begin{pmatrix}
\lambda&\lambda-1\\-\mu&1-\mu
\end{pmatrix}
\end{equation}
which may be used to recover the partial derivatives of any function $f(\lambda,\mu)$ with respect to $(R,z)$.

Now evaluating the definite integral
\begin{equation*}
\int_0^\infty\frac{t^{1/2}\dm t}{(t+\lambda)(t+\mu)}
=\frac\pi{\lambda^{1/2}+\mu^{1/2}}
\end{equation*}
indicates that -- NB: $R_0^2(t+\lambda)(t+\mu)=(t+1)tR_0^2+r^2t+z^2$;
\begin{equation}\label{eq:kzir}
\frac\pi{R_0\!\sqrt{R^2+(z+R_0)^2}}
=\int_0^\infty\frac{t^{1/2}\dm t}{(t+1)tR_0^2+r^2t+z^2}.
\end{equation}
Inserting this to equation (\ref{eq:weylpot}) for $z>0$ then yields
\begin{multline}\label{eq:crkmtp}
\frac{-\Phi(\bm r)}{G\Sigma_0}
=R_0^{2\gamma}\frac{\!\sqrt\pi}{\Gamma(\gamma)}
{_\infty^-I}_{R_0^2}^{\frac32-\gamma}
\int_0^\infty\frac{t^{1/2}\dm t}{(t+1)tR_0^2+r^2t+z^2}
\\=R_0\frac{\!\sqrt\pi\,\Gamma(\gamma-\frac12)}{\Gamma(\gamma)}
\!\int_0^\infty\!\dm t\,
\frac{t^{\gamma-1}(t+1)^{\gamma-3/2}}{[(t+\lambda)(t+\mu)]^{\gamma-1/2}}.
\end{multline}
Apart from differences in notation, this is identical to equation (6.2) of \cite{EdZ92}. Here we have utilized, for $a,b,x>0$,
\begin{equation}\label{eq:wyrel}
{_\infty^-I}_x^\xi(ax+b)^{-p}
=\frac{\Gamma(p-\xi)}{\Gamma(p)}
\frac{(ax+b)^{\xi-p}}{a^\xi},
\end{equation}
which can be proven for $p>\xi>0$ via direct calculations. Combining with $(\dm/\dm x)^m(ax+b)^{-p}=(-1)^m(p)_ma^m(ax+b)^{-p-m}$, equation (\ref{eq:wyrel}) is also proven to hold for any $\xi<1$ if $p\ge1$. Hence equation (\ref{eq:crkmtp}) is a valid integral for the potential with $\gamma>\frac12$. The last integral in equation (\ref{eq:crkmtp}) actually converges for $\gamma>0$. Since it also obeys the relation in equation (\ref{eq:mktpn}), equation (\ref{eq:crkmtp}) is in fact a valid potential (up to an immaterial constant offset) for $\gamma>0$, except for the $\gamma=\frac12$ case when the integral results in $\pi$ and the factor $\Gamma(\gamma-\frac12)$ diverges. The direct differentiations result in
\begin{equation}\label{eq:kmtf}\begin{split}
\frac{\partial\Phi}{\partial R}&=\frac{2G\Sigma_0R}{R_0}
\frac{\!\sqrt\pi\,\Gamma(\gamma+\frac12)}{\Gamma(\gamma)}
\!\int_0^\infty\!\dm t\,
\frac{t^\gamma(t+1)^{\gamma-3/2}}{[(t+\lambda)(t+\mu)]^{\gamma+1/2}};
\\\frac{\partial\Phi}{\partial z}&=\frac{2G\Sigma_0z}{R_0}
\frac{\!\sqrt\pi\,\Gamma(\gamma+\frac12)}{\Gamma(\gamma)}
\!\int_0^\infty\!\dm t\,
\frac{t^{\gamma-1}(t+1)^{\gamma-1/2}}{[(t+\lambda)(t+\mu)]^{\gamma+1/2}}.
\end{split}\end{equation}
which are indeed valid if $\gamma>0$. For $\gamma=\frac12$, this reduces to
\begin{equation*}\begin{split}
\frac{\partial\Phi}{\partial R}&=\frac{2\pi G\Sigma_0R}{R_0(\lambda-\mu)}
\left(\frac{\lambda^{1/2}}{\lambda^{1/2}+1}
-\frac{\mu^{1/2}}{1+\mu^{1/2}}\right)
\\&=\frac{2\pi GR_0\Sigma_0}{r_+}\frac{R}{R_0+z+r_+}
=\frac{2\pi GR_0\Sigma_0}R\left(1-\frac{R_0+z}{r_+}\right);
\\\frac{\partial\Phi}{\partial z}
&=\frac{2\pi G\Sigma_0z}{R_0(\lambda-\mu)}
\left(\frac1{\mu^{1/2}}-\frac1{\lambda^{1/2}}\right)
=\frac{2\pi GR_0\Sigma_0}{r_+},
\end{split}\end{equation*}
which are consistent with equation (\ref{eq:cmpi}) assuming $z>0$ (if $z<0$, then $r_+\to r_-$ and $\partial\Phi/\partial z$ switches the sign). If we identify with $\lim_{\gamma\downarrow0+}[t^{\gamma-1}/\Gamma(\gamma)]=\deltaup(t)$, equations (\ref{eq:crkmtp}) and (\ref{eq:kmtf}) also extend to include the $\gamma=0$ case; namely, the infinite plate of the uniform surface density $\Sigma_0$, for which $\Phi=2\pi GR_0\Sigma_0\!\sqrt{\lambda\mu}=2\pi G\Sigma_0|z|$ and $\partial\Phi/\partial z=2\pi G\Sigma_0z/|z|$, while $\partial\Phi/\partial R=0$: that is, at any location off the plane, the gravitational force is always vertical toward the mid-plane and independent of the height (cf.\ \cite{Ro92}, \S~22).

\subsubsection{the Carlson-R and Appell-$F_1$ functions}

The integrals in equations (\ref{eq:crkmtp}) and (\ref{eq:kmtf}) are also in the form of the Carlson-R functions (see Appendix~\ref{app:carl}):
\begin{equation}\label{eq:carlpot}\begin{split}
\Phi(\bm r)&=-\frac{2\pi GR_0\Sigma_0}{2\gamma-1}
R_{-\frac12}\bigl(\gamma-\tfrac12,\tfrac32-\gamma,\gamma-\tfrac12;\mu,1,\lambda\bigr),
\\\frac{\partial\Phi}{\partial R}&=\frac{4\gamma\pi G\Sigma_0R\,
R_{-\frac32}\bigl(\gamma+\tfrac12,\tfrac32-\gamma,\gamma+\tfrac12;
\mu,1,\lambda\bigr)}{(2\gamma+3)(2\gamma+1)R_0};
\\\frac{\partial\Phi}{\partial z}&=\frac{2\pi G\Sigma_0z\,
R_{-\frac32}\bigl(\gamma+\tfrac12,\tfrac12-\gamma,\gamma+\tfrac12;
\mu,1,\lambda\bigr)}{(2\gamma+1)R_0},
\end{split}\end{equation}
which are valid for $\gamma>-\frac12$ (but $\gamma\ne\frac12$ for $\Phi$). This is also equivalent to the bivariate hypergeometric series; that is,
\begin{multline}\label{eq:laurpot}
\Phi(\bm r)=-\frac{2\pi GR_0\Sigma_0}{(2\gamma-1)\!\sqrt\lambda}
F_1\Bigl(\tfrac12;\gamma-\tfrac12,\tfrac32-\gamma;\gamma+\tfrac12;
1-\frac\mu\lambda,1-\frac1\lambda\Bigr)
\\=-\frac{2\pi G\Sigma_0|z|}{(2\gamma-1)\lambda^\gamma}
F_1\Bigl(\gamma;\tfrac32-\gamma,\gamma-\tfrac12;\gamma+\tfrac12;
1-\mu,1-\frac\mu\lambda\Bigr),
\end{multline}
where $F_1(a;b_1,b_2;c;x_1,x_2)$ is the Appell-F$_1$ function,
\begin{equation}\label{eq:apf1}
F_1(a;b_1,b_2;c;x_1,x_2)=
\sum_{m,n=0}^\infty\frac{(a)_{m+n}}{(c)_{m+n}}
\frac{(b_1)_m(b_2)_n}{m!n!}x_1^mx_2^n,
\end{equation}
%
which is identical to the two-variable case of the Lauricella-${\rm F_D}$ series; i.e., $F_1=F^{(2)}_D$ with $F^{(n)}_D$ as defined in equation (\ref{eq:laud}). Similar expressions for the partial derivatives of the potential in terms of the $F_1$-series exist too.
Since the $F_1$-series of equation (\ref{eq:apf1}) converges for $|x_1|,|x_2|<1$ \cite[\S~16.13]{DLMF} and $0<\mu\le1\le\lambda$ for $z\ne0$, the bivariate power-series defined by equation (\ref{eq:laurpot}) with $\gamma>-\frac12$ also converges everywhere off the mid-plane.

For $z=0$, equations (\ref{eq:carlpot}) with $(\lambda,\mu)=(1-R^2/R_0^2,0)$ reduce to the ${_2F_1}$-hypergeometric functions;
\begin{equation*}\begin{split}
\Phi(R,0)&=-\frac{\pi^{3/2}\Gamma(\gamma-\frac12)GR_0\Sigma_0}
{\Gamma(\gamma)(1+R^2/R_0^2)^{1/2}}\,
{_2F_1}\!\left(\stack{\frac12,\frac32-\gamma}1;\frac{R^2}{R^2+R_0^2}\right);
\\\frac{\partial\Phi}{\partial R}\Biggr\rvert_{z=0}
&=\frac{\pi^{3/2}\Gamma(\gamma+\frac12)G\Sigma_0}
{\Gamma(\gamma)(1+R^2/R_0^2)^{3/2}}\frac R{R_0}
{_2F_1}\!\left(\stack{\frac32,\frac32-\gamma}2;\frac{R^2}{R^2+R_0^2}\right),
\end{split}\end{equation*}
which are consistent with equation (\ref{eq:kmtvc}) thanks to the Euler--Pfaff transformation. The expression for $\partial\Phi/\partial z$ in equation (\ref{eq:carlpot}) with $z=0$ is indeterminate, but there exists a transformation resulting in the proper limit:
\begin{equation*}\begin{split}
\frac{\partial\Phi}{\partial z}&=
\frac{2\pi G\Sigma_0}{(2\gamma+1)\,\lambda^\gamma}
R_{-\gamma}\bigl(\gamma+\tfrac12,\tfrac12-\gamma,\gamma+\tfrac12;
1,\mu,\mu/\lambda\bigr)
\\&\stackrel{\mu\to0}{=}2\pi G\Sigma_0(1+R^2/R_0^2)^{-\gamma},
\end{split}\end{equation*}
which is consistent with the surface density in equation (\ref{eq:kmt}).

If $R=0$, then $\lambda=1$ or $\mu=1$, and equation (\ref{eq:laurpot}) becomes a single-variable hypergeometric series. In fact, setting $(\lambda,\mu)=(1,z^2/R_0^2<1)$ and $(\lambda,\mu)=(z^2/R_0^2>1,1)$ in equations (\ref{eq:carlpot}) result in
\begin{equation*}\begin{split}
\Phi(0,z)&=-\frac{2\pi GR_0\Sigma_0}{2\gamma-1}
{_2F_1}\!\left(\stack{\frac12,\gamma-\frac12}{\gamma+\frac12};
1-\frac{z^2}{R_0^2}\right)
\\&=-\frac{2\pi GR_0\Sigma_0}{2\gamma-1}{_2F_1}\!\left(
\stack{1,2\gamma-1}{\gamma+\frac12};\frac{R_0-|z|}{2R_0}\right).
\end{split}\end{equation*}
Differentiating this with respect to $z$ then recovers equation (\ref{eq:kmtgz}) for $(\partial\Phi/\partial z)|_{R=0}$, whereas the axial symmetry necessitates $(\partial\Phi/\partial R)|_{R=0}=0$, which is confirmed by the corresponding result in equation (\ref{eq:carlpot}). The result is consistent with the direct integral of the $r^{-1}$-potential from the mass element, that is, $\Phi(0,z)=-2\pi G\int_0^\infty\!\dm R\,R\Sigma(R)/(R^2+z^2)^{1/2}$ and so on.

\subsection{The bivariate hypergeometric series in $(R,z)$}

Although equation (\ref{eq:laurpot}) provides a series expression for the potential converging everywhere, it involves the elliptic coordinates $(\lambda,\mu)$. In order to obtain the power series of the potential in terms of the cylindrical coordinates $(R,z)$, we need to consider the Mellin transform of equation (\ref{eq:crkmtp}). Utilizing $R_0^2(t+\lambda)(t+\mu)=(t+1)tR_0^2+t(R^2+z^2)+z^2$, we can find
\begin{equation*}
\frac{\mathfrak M_{R^2\to v}\mathfrak M_{z^2\to u}(-\Phi)}
{G\Sigma_0R_0^{2\gamma}}
=\frac{\!\sqrt\pi\,\Gamma(\gamma-\frac12)}{\Gamma(\gamma)}
\int_0^\infty\!\dm t\,t^{\gamma-1}(t+1)^{\gamma-\frac32}\mathcal I;
\end{equation*}
where
\begin{equation*}\begin{split}
\mathcal I&=\int_0^\infty\!\dm R^2\int_0^\infty\!\dm z^2
\frac{R^{2(v-1)}z^{2(u-1)}}{[(t+1)tR_0^2+t(R^2+z^2)+z^2]^{\gamma-1/2}}
\\&=R_0^{2(u+v-\gamma)+1}
\frac{\Gamma(v)\Gamma(u)\Gamma(\gamma-\frac12-v-u)}{\Gamma(\gamma-\frac12)}
t^{u-\gamma+\frac12}(t+1)^{v-\gamma+\frac12},
\end{split}\end{equation*}
and so follows
\begin{multline*}
\frac{\mathfrak M_{R^2\to v}\mathfrak M_{z^2\to u}(-\Phi)}
{GR_0\Sigma_0}=\frac{\!\sqrt\pi}{\Gamma(\gamma)}R_0^{2(u+v)}
\\\times\frac{\Gamma(v)\Gamma(u)\Gamma(u+\frac12)
\Gamma(\frac12-v-u)\Gamma(\gamma-\frac12-v-u)}{\Gamma(1-v)}.
\end{multline*}
The Mellin--Barnes integral representation of the potential is thus given by ($\varrho=R/R_0$ and $\zeta=z/R_0$)
\begin{multline}\label{eq:mgpot}
\Phi(R,z)=-\frac{\pi GR_0\Sigma_0}{\Gamma(\gamma)}\frac1{(2\pi\im)^2}
\\\times\iint\limits_{\mathcal C_v\times\mathcal C_u}\!\frac{2\dm v\,\dm u}{\varrho^{2v}(2\zeta)^{2u}}
\frac{\Gamma(v)\Gamma(2u)\Gamma(\frac12-v-u)
\Gamma(\gamma-\frac12-v-u)}{\Gamma(1-v)},
\end{multline}
%
This is in the form of the Me{\ij}er-G/Fox-H function of two variables\footnote{It appears that there is no established standard notation for the bivariate Me{\ij}er-G/Fox-H function. In this paper, to avoid unnecessary confusion, we leave the explicit double Mellin--Barnes integral for bivariate cases.} \citep{Ag65,Sh65,MG72}, but some may find it too formal. None the less, we can add up the residues contributed from the isolated poles at $(v,2u)=(-m,-n)$ where $m,n$ are non-negative integers, resulting in the bivariate hypergeometric series for the potential in the upper half-space ($z>0$): that is,
\begin{multline}\label{eq:f4pot}
\frac{-\Phi(R,z)}{\pi GR_0\Sigma_0}
=\sum_{m,n=0}^\infty\frac{(-\varrho^2)^m(-2\zeta)^n}{(m!)^2n!}
\frac{\Gamma(m+\frac{n+1}2)\Gamma(m+\frac{n-1}2+\gamma)}{\Gamma(\gamma)}
\\=\frac{\!\sqrt\pi\,\Gamma(\gamma-\frac12)}{\Gamma(\gamma)}
F_4\bigl(\gamma-\tfrac12,\tfrac12;1,\tfrac12;-\varrho^2,\zeta^2\bigr)
\\-2|\zeta|\,F_4\bigl(1,\gamma;1,\tfrac32;-\varrho^2,\zeta^2\bigr),
\end{multline}
where $F_4(a,b;c_1,c_2;x_1,x_2)$ is the Appell-F$_4$ functions:
\begin{equation*}
F_4(a,b;c_1,c_2;x_1,x_2)
=\sum_{m,n=0}^\infty\frac{(a)_{m+n}(b)_{m+n}}{(c_1)_m(c_2)_n}
\frac{x_1^mx_2^n}{m!n!}.
\end{equation*}
Since the $F_4$-series converges absolutely if $|x_1|^{1/2}+|x_2|^{1/2}<1$ \cite[\S~16.13]{DLMF}, equation (\ref{eq:f4pot}) converges for $R+|z|<R_0$ (and $\gamma\ne\frac12$). If $\gamma=0$, every term in the sum except for $(m,n)=(0,1)$ is zero and so $\Phi=2\pi G\Sigma_0z$. If $\gamma=\frac12$, the $(m,n)=(0,0)$ term is infinite but technically constant independent of $\rho$ and $\zeta$. Since an additive constant in the potential is physically immaterial, skipping $(m,n)=(0,0)$ in the sum with $\gamma=\frac12$ actually results in a power series for the potential of the cored Mestel disk.

Equation (\ref{eq:mgpot}) can be made to converge to its analytic continuation in the whole upper half-space with a proper choice of the integration path. In fact, unless $\gamma$ is an integer, utilizing the analytic continuations \cite[eq.~16.16.10]{DLMF}\footnote{There is a sign error in equation 5.11.(9) of \cite{Er53}.} of the $F_4$-series, we are able to obtain alternative series expressions:
\begin{multline*}
\frac{-\Phi(R,z)}{\pi GR_0\Sigma_0}=\frac1{\gamma-1}
\frac1{|\zeta|}F_4\Bigl(1,\tfrac12;1,2-\gamma;
-\frac{\varrho^2}{\zeta^2},\frac1{\zeta^2}\Bigr)
\\+\frac{\Gamma(1-\gamma)\Gamma(\gamma-\frac12)}{\!\sqrt\pi\,\zeta^{2\gamma-1}}
F_4\Bigl(\gamma-\tfrac12,\gamma;1,\gamma;
-\frac{\varrho^2}{\zeta^2},\frac1{\zeta^2}\Bigr),
\end{multline*}
which converges for $|z|-R>R_0$, or
\begin{multline*}
\frac{-\Phi(R,z)}{\pi GR_0\Sigma_0}=\frac1{\gamma-1}
\frac1\varrho F_4\Bigl(\tfrac12,\tfrac12;\tfrac12,2-\gamma;
-\frac{\zeta^2}{\varrho^2},-\frac1{\varrho^2}\Bigr)
\\+\frac{\Gamma(1-\gamma)\Gamma(\gamma-\frac12)}
{\Gamma(\gamma)\Gamma(\frac32-\gamma)\varrho^{2\gamma-1}}
F_4\Bigl(\gamma-\tfrac12,\gamma-\tfrac12;\tfrac12,\gamma;
-\frac{\zeta^2}{\varrho^2},-\frac1{\varrho^2}\Bigr)
\\-\frac{2|\zeta|}{\varrho^{2\gamma}}
F_4\Bigl(\gamma,\gamma;\tfrac32,\gamma;
-\frac{\zeta^2}{\varrho^2},-\frac1{\varrho^2}\Bigr),
\end{multline*}
which converges for $R-|z|>R_0$.
The same results may also be obtained directly from equation (\ref{eq:mgpot}) by summing up residues for poles for $\Gamma(\frac12-v-u)\Gamma(\gamma-\frac12-v-u)$. If $\gamma$ is an integer, the whole integrand of equation (\ref{eq:mgpot}) contains second-order poles, the residue of which requires more care to calculate (the resulting series typically involves the Euler--Mascheroni constant and the logarithmic terms).

\subsection{The Toomre-$n$ disks: a half-integer index $\gamma$}

Equation (\ref{eq:tmrn}) indicates that the potential due to the disk surface density profile of equation (\ref{eq:kmt}) with a non-negative integer $\gamma-\frac32=m$ can be written down as an elementary function of the coordinate variables. However it involves explicit derivatives of arbitrary order, which is impractical unless $m$ is a relatively small integer. The integral representation in equation (\ref{eq:crkmtp}) with an integer $m=\gamma-\frac32$ also reduces to a closed form involving only elementary functions (e.g., through the partial fraction decomposition, the integrand is reducible to a sum of analytically integrable terms like $t^{-1/2}(t+a)^{-n}$ where $a=\lambda$ or $\mu$ and $n$ is a positive integer), but deriving general formula via this procedure is challenging. Equations (\ref{eq:carlpot}) and (\ref{eq:laurpot}) are still valid but the presence of special functions defeats any practical advantages afforded by purely algebraic expressions. 

One route to derive explicit analytic formulae for the potential of the \citeauthor{To63} Model-$n$ uses the Burchnall--Chaundy expansion of the $F_4$-series \cite[eq.~16.16.7]{DLMF}; namely,
\begin{multline*}
F_4[a,b;c_1,c_2;x(1-y),y(1-x)]
=\sum_{k=0}^\infty\frac{(a)_k(b)_k(\tilde c)_k}{(c_1)_k(c_2)_k}
\\\times\frac{(xy)^k}{k!}
{_2F_1}\!\left(\stack{a+k,b+k}{c_1+k};x\right)
{_2F_1}\!\left(\stack{a+k,b+k}{c_2+k};y\right),
\end{multline*}
where $\tilde c=1+a+b-c_1-c_2$. If one of the upper indices of the $F_4$-function coincides with one of its lower indices, then
\begin{equation*}\begin{split}
F_4&[a,b;c,b;x(1-y),y(1-x)]
\\&=\sum_{k=0}^\infty\frac{(a)_k(\tilde c)_k(xy)^k}
{k!(c)_k[(1-x)(1-y)]^{a+k}}
{_2F_1}\!\left(\stack{a+k,c-b}{c+k};\frac x{x-1}\right)
\\&=\frac1{(1-x)^a(1-y)^a}F_1\!\left(a;c-b,\tilde c;c;
\frac x{x-1},\frac x{1-x}\frac y{1-y}\right)
\\&=R_{-a}\!\left[b-\tilde c,c-b,\tilde c;(1-x)(1-y),1-y,1-x-y\right]
\end{split}\end{equation*}
where $\tilde c=1+a-c$. This follows the definitions of ${_2F_1}$- and $F_1$-series together with $(a)_k(a+k)_m=(a)_{k+m}$. Given $\varrho^2=(\lambda-1)(1-\mu)$ and $\zeta^2=\lambda\mu$, the $F_4$-series in equation (\ref{eq:f4pot}) respectively transform to
\begin{align*}\begin{split}
F_4&\bigl(\gamma-\tfrac12,\tfrac12;1,\tfrac12;-\varrho^2,\zeta^2\bigr)
\\&=R_{-\gamma+\frac12}\bigl[\tfrac12,1-\gamma,\gamma-\tfrac12;
1-\mu,\lambda(1-\mu),\lambda-\mu\bigr]
\\&=\frac{\!\sqrt\lambda}{(\lambda-\mu)^{\gamma-1/2}}
R_{\gamma-\frac32}\biggl[1-\gamma,\tfrac12,\gamma-\tfrac12;
1,\lambda,\frac{\lambda(1-\mu)}{\lambda-\mu}\biggr];
\end{split}\\\begin{split}
\zeta&F_4\bigl(1,\gamma;1,\tfrac32;-\varrho^2,\zeta^2\bigr)
\\&=\!\sqrt{\lambda\mu}R_{-\gamma}\bigl[\tfrac32-\gamma,\gamma-\tfrac12,\tfrac12;
\lambda(1-\mu),\lambda-\mu,\lambda\bigr]
\\&=\frac{\!\sqrt\mu}{(\lambda-\mu)^{\gamma-1/2}}
R_{\gamma-\frac32}\biggl[\tfrac12,\tfrac32-\gamma,\gamma-\tfrac12;
1-\mu,1,\frac{\lambda(1-\mu)}{\lambda-\mu}\biggr].
\end{split}\end{align*}
%
According to equation (\ref{eq:Rmpoly}), the $R_m$-function with a non-negative integer $m=\gamma-\frac32$ is a polynomial. Following these transformations and further manipulations of multiple sums, equation (\ref{eq:f4pot}) with a non-negative integer $m=\gamma-\frac32$ is found to be equivalent to
\begin{equation}\label{eq:tnpot}
\frac{-\Phi(\bm r)}{\pi GR_0\Sigma_0}=\frac{2^{m+1}m!}{(2m+1)!!}
\frac{\!\sqrt\lambda p_m(\lambda,\mu)
-(-1)^m\!\sqrt\mu p_m(\mu,\lambda)}{(\lambda-\mu)^{m+1}}
\end{equation}
where $(2m+1)!!=1\cdot3\dotsm(2m-1)\cdot(2m+1) =2^{m+1}\cdot(\frac12)_{m+1}=2^m\cdot(\frac32)_m$ is the double factorial and $p_m(x,y)$ is defined to be
\begin{equation*}
p_m(x,y)=\sum_{k=0}^m\sum_{j=k}^m\frac{(m+k)!}{k!j!(m-j)!}
\frac{(\frac12)_{j-k}}{(j-k)!}
\frac{(x-1)^jy^k}{(y-x)^k}.
\end{equation*}
In fact, $(x-y)^mp_m(x,y)$ is a $m$-th polynomial of $(x,y)$: e.g.,
\begin{equation*}\begin{split}
2&(x-y)p_1(x,y)=x^2+x-5xy+3y;
\\8&(x-y)^2p_2(x,y)=3x^4+x^3(2-18y)
\\&\qquad+x^2(3-28y+63y^2)+10xy(3-7y)+15y^2,
\end{split}\end{equation*}
although these become rather messy as $m$ gets larger. Since $p_0=1$, equation (\ref{eq:tnpot}) with $m=0$ recovers $\Phi_\mathrm K$ in equation (\ref{eq:kmplm}). The potential due to $\Sigma=\Sigma_0(1+\varrho^2)^{-5/2}$ is identified with the $m=1$ case,
%
which simplifies to
\begin{equation*}
\Phi(\bm r)=-\frac{2\pi GR_0\Sigma_0}3
\frac{1+\lambda+\mu+3\!\sqrt{\lambda\mu}}
{\bigl(\!\sqrt\lambda+\!\sqrt\mu\,\bigr)^3}.
\end{equation*}
This is also consistent with equation (\ref{eq:mktpn}) applied for the potential of the Kuzmin disk following
\begin{equation}\label{eq:lmspd}
\frac{\partial\lambda}{\partial(R_0^2)}
=-\frac{\lambda}{R_0^2}\frac{\lambda-1}{\lambda-\mu},\quad
\frac{\partial\mu}{\partial(R_0^2)}
=-\frac{\mu}{R_0^2}\frac{1-\mu}{\lambda-\mu}.
\end{equation}

\subsection{The integer index cases with elliptic integrals}
\label{sec:ee}

According to the criteria in \S~19.16(iii) of \cite{DLMF}, $R$-functions in equation (\ref{eq:carlpot}) with a positive integer $\gamma$ are an elliptic integral. In particular, if $\gamma=1$,
\begin{gather}
\frac{-\Phi(\bm r)}{2\pi GR_0\Sigma_0}
=R_F(\mu,1,\lambda)=\frac{\mathrm F(\varphi,\kappa)}{\!\sqrt{\lambda-\mu}}
=\frac{\mathbf K(\kappa)-\mathrm F(\phi_\mu,\kappa)}{\!\sqrt{\lambda-\mu}};
\nonumber\\\label{eq:potg1}\biggl(\text{where}\
\kappa^2=\frac{\lambda-1}{\lambda-\mu},\
\cos^2\!\varphi=\frac\mu\lambda,\
\sin^2\!\phi_\mu=\mu\biggr).
\end{gather}
The last expression is consistent with the addition theorem of elliptic integrals and also follows equation (\ref{eq:f4pot}). Here $R_F$ is one of the Carlson symmetric elliptic integrals as in equation (\ref{eq:Rdef}), which is an alternative to Legendre's integrals of the first kind, $\mathbf K$ or $\mathrm F$ (see App.~\ref{app:lei}).

The acceleration field $\del\Phi$ (and the potential for any integer index other than one) additionally involves the elliptic integrals of the second kind; either the Legendre form $\mathrm E(\varphi,\kappa)$, or the Carlson basis function $R_D$. From the integral representation in equation (\ref{eq:kmtf}),
\begin{equation*}\begin{split}
\frac{\partial\Phi}{\partial R}&=\frac{2\pi G\Sigma_0}3\frac R{R_0}
\frac{\lambda R_D(\mu,1;\lambda)-\mu R_D(\lambda,1;\mu)}{\lambda-\mu};
\\\frac{\partial\Phi}{\partial z}&=\frac{2\pi G\Sigma_0}3\frac z{R_0}
\frac{(\lambda-1)R_D(\mu,1;\lambda)+(1-\mu)R_D(\lambda,1;\mu)}{\lambda-\mu},
\end{split}\end{equation*}
which are consistent with the direct differentiations of equation (\ref{eq:potg1}) using equations (\ref{eq:lmpd}). These are also alternatively expressible by means of $R_F(\mu,1,\lambda)$ and $R_D(\mu,1;\lambda)$, or Legendre's integrals $\mathrm F(\varphi,\kappa)$ and $\mathrm E(\varphi,\kappa)$ utilizing
\begin{equation*}\begin{split}
\frac{\lambda-1}3R_D(\mu,1;\lambda)
&=\frac{\mathrm F(\varphi,\kappa)-\mathrm E(\varphi,\kappa)}
{\!\sqrt{\lambda-\mu}},
\\
\frac{1-\mu}3R_D(\lambda,1;\mu)
&=\frac{\lambda-1}3R_D(\mu,1;\lambda)-R_F(\mu,1,\lambda)
+\frac1{\!\sqrt{\lambda\mu}}
\\&=\frac1{|\zeta|}-\frac{\mathrm E(\varphi,\kappa)}{\!\sqrt{\lambda-\mu}},
\end{split}\end{equation*}
with $(\varphi,\kappa)$ as defined in equation (\ref{eq:potg1}). In the $z=0$ limit, $\lim_{z\downarrow0^+}(\partial\Phi/\partial z)=2\pi G\Sigma_0/\lambda$ and $(\lambda,\mu)=(1+\varrho^2,0)$, which is consistent with the surface density of equation (\ref{eq:kmt}) with $\gamma=1$. The rotation curve follows $\partial\Phi/\partial R$ in the same limit,
\begin{equation*}
\frac{\varv_\mathrm c^2}{2\pi GR_0\Sigma_0}
=\frac{\varrho^2}3R_D(0,1;1+\varrho^2)
=\frac{\mathbf K(\kappa_0)-\mathbf E(\kappa_0)}{\!\sqrt{1+\varrho^2}},
\end{equation*}
where $\kappa_0^2=\varrho^2/(1+\varrho^2)=R^2/(R^2+R_0^2)$, which compares to the behavior of the mid-plane potential; $-\Phi(R,0)/(2\pi GR_0\Sigma_0)=R_F(0,1,1+\varrho^2)=(1+\varrho^2)^{-1/2}\mathbf K(\kappa_0)$.

The potentials and accelerations for $\Sigma\propto(1+\varrho^2)^{-m}$ with any positive integer $m$ are expressible using only algebraic functions of $(\mu,\lambda)$ with the elliptic integral basis set consisting of either $\set{R_F(\mu,1,\lambda),R_D(\mu,1;\lambda)}$ or $\set{\mathrm F(\varphi,\kappa),\mathrm E(\varphi,\kappa)}$ thanks to the recurrence relation in equation (\ref{eq:mktpn}) together with the partial derivatives of equation (\ref{eq:lmspd}). In practice, not only deriving such an expression is rather tedious but also the resulting expressions are not necessarily any more useful than unresolved integral representations such as equations (\ref{eq:crkmtp}) and (\ref{eq:kmtf}). Here we only provide the explicit expression for $\gamma=2$ (cf.\ \cite{EdZ92}, table~3)
\begin{equation*}\begin{split}
\frac{-\Phi(\bm r)}{\pi GR_0\Sigma_0}&=R_F
-\frac{\lambda(\lambda-1)R_D
+\mu(1-\mu)R_D(\lambda,1;\mu)}{3(\lambda-\mu)}
\\&=\frac1{\lambda-\mu}\left[\lambda R_F
-\frac{(\lambda-1)(\lambda+\mu)}3R_D
-\!\sqrt{\frac\mu\lambda}\right]
\\&=\frac{(\lambda+\mu)\mathrm E(\varphi,\kappa)-\mu\mathrm F(\varphi,\kappa)}
{(\lambda-\mu)^{3/2}}-\frac1{\lambda-\mu}\!\sqrt{\frac\mu\lambda},
\end{split}\end{equation*}
where $R_F=R_F(\mu,1,\lambda)$ and $R_D=R_D(\mu,1;\lambda)$. The first line follows either applying equation (\ref{eq:mktpn}) on equation (\ref{eq:potg1}) or resolving equation (\ref{eq:crkmtp}) via the partial fraction decomposition. On the mid-plane, this results in $-\Phi(R,0)/(\pi GR_0\Sigma_0)=R_F(0,1,1+\varrho^2)-(\varrho^2/3)R_D(0,1;1+\varrho^2)=(1+\varrho^2)^{-1/2}\mathbf E(\kappa_0)$ with the corresponding circular speed given by
\begin{equation*}\begin{split}
\varv_\mathrm c^2
&=\frac{\pi GR_0\Sigma_0\varrho^2}{1+\varrho^2}\left[
R_F(0,1,1+\varrho^2)+\frac{1-\varrho^2}3R_D(0,1;1+\varrho^2)\right]
\\&=\pi GR_0\Sigma_0\frac{\mathbf K(\kappa_0)+(\varrho^2-1)\mathbf E(\kappa_0)}
{(1+\varrho^2)^{3/2}}.
\end{split}\end{equation*}

\section{The Qian--Kalnajs--Mestel disks}

The \citetalias{EdZ92} weight of the KMTd is in the form of $G^{0,1}_{1,1}$ with one index fixed to be zero (eq.~\ref{eq:edzw}). Let us generalize this to $G^{0,1}_{1,1}$ with two free indices. In particular, consider the disk with the surface density of the form
\begin{equation}\label{eq:kmq}\begin{split}
\frac{\Sigma(R)}{\Sigma_0}
&={_2F_1}\!\left(\stack{a,\frac32}{b};-\frac{R^2}{R_0^2}\right)
\\&=\frac{2\Gamma(b)}{\sqrt\pi\,\Gamma(a)}
G^{1,2}_{2,2}\!\left(\frac{R^2}{R_0^2}
\vrule\stack{-\frac12,1-a}{0,1-b}\right),
\end{split}\end{equation}
which is one of the particular cases (with the redefined scale constants) of equation (\ref{eq:hypg1}) for which both $\Sigma(R)$ and $\varv_\mathrm c^2(R)$ are expressible through a single ${_2F_1}$-series. This is equivalent to the $m=0$ case of equation (4.7) of \cite{Qi92}. As is noted by \cite{Qi92}, this family encompasses the so-called \citet{Ka76}--\citeauthor{Me63} disk family \citep{EdZ92}, which corresponds to $(a,b)=(m/2,1+m/2)$ with $m$ being a positive integer. Since $\Sigma=\Sigma_0/(1+\varrho^2)^a$ with $\varrho=R/R_0$ if $b=\frac32$, the family includes all the KMTd with the cored Mestel disk resulting from $(a,b)=(\frac12,\frac32)$ and the $a=b$ case reducing to the Kuzmin disk. Also included in this family is the projected isochrone found in equation (6.21) of \cite{EdZ92}, which results from $(a,b)=(2,\frac52)$. Henceforth we refer to this family as the QKMd (after \citeauthor{Qi92}--\citeauthor{Ka76}--\citeauthor{Me63} disks).

Equation (\ref{eq:kmq}) indicates that $\Sigma(R)/\Sigma_0\simeq1-3a\varrho^2/(2b)$ for $R\ll R_0$, and so the surface density for $b>0$ is finite at the center with $\Sigma(0)=\Sigma_0$ (the $a=0$ case becoming the uniform plate $\Sigma=\Sigma_0$). On the other hand, the asymptotic behavior of $\Sigma(R)$ as $R\to\infty$ is found to be (assuming $b\ne a$ and $b\ne\frac32$)
\begin{equation*}
\frac\Sigma{\Sigma_0}\simeq\begin{cases}\displaystyle
\frac{2\Gamma(\frac32-a)\Gamma(b)}{\!\sqrt\pi\,\Gamma(b-a)}
\frac1{\varrho^{2a}}&(a<\tfrac32)
\smallskip\\\displaystyle
\frac{2\Gamma(b)}{\!\sqrt\pi\,\Gamma(b-\frac32)}
\frac{\ln(4\varrho^2)-\digamma(b-\frac32)-\gammaup-2}{\varrho^3}&(a=\tfrac32)
\smallskip\\\displaystyle
\frac{\Gamma(a-\frac32)\Gamma(b)}{\Gamma(a)\Gamma(b-\frac32)}
\frac1{\varrho^3}&(a>\tfrac32)\end{cases},
\end{equation*}
where $\gammaup\approx0.5772$ is the Euler--Mascheroni constant and $\digamma(x)=\dm\ln|\Gamma(x)|/\dm x$ is the digamma function. Note that $\Sigma(R)\to0$ as $R\to\infty$ if $a>0$. The enclosed mass, $M(R)=2\pi\int_0^R\dm\tilde R\tilde R\Sigma(\tilde R)$, is given by
\begin{equation*}
M(R)=\frac{2\!\sqrt\pi\,R_0^2\Sigma_0\Gamma(b)}{\Gamma(a)}
G^{1,3}_{3,3}\!\left(\frac{R^2}{R_0^2}\vrule
\stack{1,\frac12,2-a}{1,0,2-b}\right),
\end{equation*}
which behaves like $M(R)\simeq\pi\Sigma_0R^2$ as $R\to0$ and like
\begin{equation*}
\frac{M(R)}{R_0^2\Sigma_0}\simeq\begin{cases}
\dfrac{2\pi(b-1)}{a-1}&(a>1)
\\2\pi(b-1)\left[\ln\dfrac{\varrho^2}4-\digamma(b-1)-\gammaup\right]
&(a=1)
\smallskip\\\dfrac{2\!\sqrt\pi\,\Gamma(b)\Gamma(\frac32-a)}
{(1-a)\Gamma(b-a)}\varrho^{2(1-a)}
&(a<1)\end{cases}
\end{equation*}
as $R\to\infty$. If $a,b>1$, the total mass of the disk is a well-defined positive finite value $M_\mathrm t=2\pi R_0^2\Sigma_0(b-1)/(a-1)$. The non-positive mass for $b\le1$ is the consequence of the negative surface density at large radii, which is unphysical. In fact, $b\ge\min(a,\frac32)$ is also necessary for $\Sigma$ to be non-negative. For the disk model with the finite total mass, the mass profile can be expressed slightly simpler as
\begin{equation*}\begin{split}
1-\frac{M(R)}{M_\mathrm t}
&=\frac{\Gamma(b-1)}{\!\sqrt\pi\,\Gamma(a-1)}
G^{1,2}_{2,2}\!\left(\frac{R^2}{R_0^2}\vrule
\stack{\frac12,2-a}{0,2-b}\right)
\\&={_2F_1}\!\left(\stack{\frac12,a-1}{b-1};-\frac{R^2}{R_0^2}\right).
\end{split}\end{equation*}

The gravitational force due to the disk is found for models with $a>-\frac12$ and the resulting mid-plane rotation curve is 
\begin{equation}\label{eq:kmqv}
\varv_\mathrm c^2(R)=
\frac{2\pi\Gamma(a+\frac12)\Gamma(b)}{\Gamma(a)\Gamma(b+\frac12)}
\frac{G\Sigma_0R^2}{R_0}
{_2F_1}\!\left(\stack{\frac32,a+\frac12}{b+\frac12};-\frac{R^2}{R_0^2}\right),
\end{equation}
whose $R\to\infty$ behavior, assuming $b\ne1$, is like
\begin{equation*}
\frac{\varv_\mathrm c^2(R)}{GR_0\Sigma_0}\simeq\begin{cases}\displaystyle
\frac{\Gamma(a+\frac12)\Gamma(1-a)\Gamma(b)}{\Gamma(a)\Gamma(b-a)}
\frac{4\!\sqrt\pi}{\varrho^{2a-1}}&(a<1)
\smallskip\\\displaystyle
2\pi(b-1)\frac{\ln(4\varrho^2)-\digamma(b-1)-\gammaup-2}{\varrho}&(a=1)
\\\displaystyle
\frac{b-1}{a-1}\frac{2\pi}\varrho
=\frac1{R_0\Sigma_0}\frac{M_\mathrm t}{R}&(a>1)\end{cases}.
\end{equation*}
If $b=1$, then $\varv_\mathrm c^2\propto\varrho^2(1+\varrho^2)^{-(a+1/2)}$ and so $\varv_\mathrm c^2\sim\varrho^{1-2a}$ as $R\to\infty$ for any $a>0$ (but only physical if $a\le b=1$). For finite mass ($a>1$) disks, this is consistent with the Keplerian behavior, $\varv_\mathrm c^2\simeq GM_\mathrm t/R$ as $R\to\infty$, whereas the asymptotic behavior $\Sigma\sim R^{-2a}$ for $a<1$ results in a slower fall-off of the circular speed $\varv_\mathrm c\sim R^{a-1/2}$ with the $a=\frac12$ case leading to the asymptotically flat rotation curve, as suggested by $\Sigma\sim R^{-1}$ as $R\to\infty$.

On the other hand, the vertical force on the symmetry axis is given by ($\zeta=z/R_0$)
\begin{align}
\frac{\varg(z)}{G\Sigma_0}&=\frac{4\Gamma(b)}{\Gamma(a)}
G^{2,3}_{3,3}\!\left(\frac{z^2}{R_0^2}\vrule
\stack{0,-\frac12,1-a}{0,\frac12,1-b}\right)
\nonumber\\&
=\frac{2\pi\Gamma(b)}{\Gamma(a)}\sum_{k=0}^\infty
\frac{(k+1)\Gamma(a+k/2)}{\Gamma(b+k/2)}(-\zeta)^k
\label{eq:kmqgz}\\\nonumber
&=4\pi\left[F^+(\zeta^2)
-\frac{\Gamma(b)\Gamma(a+\frac12)}{\Gamma(a)\Gamma(b+\frac12)}
\zeta{_2F_1}\!\left(\stack{2,a+\frac12}{b+\frac12};\zeta^2\right)\right]
\end{align}
where the even-part function is
\begin{equation*}\begin{split}
F^+(\zeta^2)&=\frac12\,
{_3F_2}\!\left(\stack{1,\frac32,a}{\frac12,b};\zeta^2\right)
=\sum_{k=0}^\infty(k+\tfrac12)\frac{(a)_k}{(b)_k}\zeta^{2k}
\\&={_2F_1}\!\left(\stack{2,a}b;\zeta^2\right)
-\frac12\,{_2F_1}\!\left(\stack{1,a}b;\zeta^2\right)
\\&=\frac{b-1}{1-\zeta^2}-\frac{b-\frac32-(a-\frac12)\zeta^2}{1-\zeta^2}
{_2F_1}\!\left(\stack{1,a}b;\zeta^2\right).
\end{split}\end{equation*}
Also note that $\varg(0)=2\pi G\Sigma_0$, which relates to the finite central surface density, whereas the asymptotic behavior as $z\to\infty$ is 
\begin{equation*}
\frac{\varg(z)}{G\Sigma_0}\simeq\begin{cases}\displaystyle
\frac{4\pi^2(1-2a)\Gamma(b)}{\sin(2a\pi)\Gamma(a)\Gamma(b-a)}
\frac1{\zeta^{2a}}&(a<1)
\smallskip\\\displaystyle
2\pi(b-1)\frac{\ln\zeta^2-\digamma(b-1)-\gammaup-2}{\zeta^2}&(a=1)
\\\displaystyle
\frac{b-1}{a-1}\frac{2\pi}{\zeta^2}
=\frac{M_\mathrm t}{\Sigma_0}\frac1{z^2}&(a>1)\end{cases},
\end{equation*}
which are valid for $b\ne1$ -- NB: $\lim_{a\to1/2}(1-2a)\pi/\sin(2a\pi)=\lim_{a\to1/2}\Gamma(2a)\Gamma(2-2a)=1$. If $b=1$ on the other hand,
\begin{equation*}\begin{split}
\frac{\varg(z)}{G\Sigma_0}
&=\frac{2\!\sqrt\pi\,\Gamma(a+\frac12)}{\Gamma(a+2)}
{_2F_1}\!\left(\stack{a,\frac32}{a+2};1-\zeta^2\right)
\\&\!\!\!\stackrel{z\to\infty}\simeq\begin{cases}\displaystyle
\frac{2\pi(1-2a)}{\cos(a\pi)}\frac1{\zeta^{2a}}
&(a<\frac32)\smallskip\\\displaystyle
\frac{4[\ln(4\zeta^2)-3]}{\zeta^3}
&(a=\frac32)\smallskip\\\displaystyle
\frac{2\!\sqrt\pi\,\Gamma(a-\frac32)}{\Gamma(a)}\frac1{\zeta^3}
&(a>\frac32)\end{cases},\end{split}
\end{equation*}
although only those models with $-\frac12<a\le b=1$ result in
the non-negative surface density and the finite gravity.

\subsection{The potentials due to the Qian--Kalnajs--Mestel disks}

The \citetalias{EdZ92} weight for the QKMd follows equation (\ref{eq:hypgw}), which is an elementary function ($b>a>0$):
\begin{equation}\label{eq:kmqsw}\begin{split}
\Lambda(h)&=\frac{4\pi\Gamma(b)}{\Gamma(a)}\Sigma_0R_0\,
G^{0,1}_{1,1}\!\left(\frac{h^2}{R_0^2}
\vrule\stack{\frac32-a}{\frac32-b}\right)
\\&=\frac{4\pi\Sigma_0R_0^{2a}\Gamma(b)}{\Gamma(a)\Gamma(b-a)}
\frac{(h^2-R_0^2)^{b-a-1}}{h^{2b-3}}\Theta(h-R_0).
\end{split}\end{equation}
Once equation (\ref{eq:kmqsw}) is chosen as the weight, equations (\ref{eq:sigv}) result in the $_2F_1$-hypergeometric functions in equations (\ref{eq:kmq}) and (\ref{eq:kmqv}). Hence the potential generated by the QKMd can be written down as a quadrature like
\begin{equation}\label{eq:potrz}
\frac{\Phi(R,z)}{G\Sigma_0}
=-\frac{4\pi R_0^{2a}\Gamma(b)}{\Gamma(a)\Gamma(b-a)}
\int_{R_0}^\infty\frac{(h^2-R_0^2)^{b-a-1}\dm h}
{h^{2b-3}\!\sqrt{R^2+(|z|+h)^2}},
\end{equation}
which converges if $b>a>\frac12$. The change of the integration variable, $h\to x$, given by $2(h+|z|)=(x+|z|)-R^2/(x+|z|)$ leads this to a manifest integral form of the Carlson-R function;
\begin{align}
\frac{\Phi(R,z)}{G\Sigma_0}
&=-\frac{2^{2a+1}\pi R_0^{2a}\Gamma(b)}{\Gamma(a)\Gamma(b-a)}\!
\!\int_{R_0+r_+}^\infty\!\dm x
\frac{(x+|z|)^{2a-2}}{(x^2-r^2)^{2b-3}}\mathcal Q^{b-a-1};
\nonumber\\\label{eq:kmdp}
\mathcal Q&=\bigl[(x-R_0)^2-r_+^2\bigr]\bigl[(x+R_0)^2-r_-^2\bigr],
\end{align}
where $r^2=R^2+z^2$ and $r_\pm^2=R^2+(|z|\pm R_0)^2$. Provided that both $2a$ and $2b$ are integers, this is either resolved into an expression involving only elementary functions (if $b-a$ is also an integer) or an elliptic integral (if $b-a-1/2$ is an integer). The reduction to elliptic integrals however is simpler if the elliptic coordinates $(\lambda,\mu)$ defined in Sect.~\ref{sec:ecd} -- i.e.\ $(R/R_0)^2=(\lambda-1)(1-\mu)$ and $(z/R_0)^2=\lambda\mu$ -- and an alternative change of the integration variable, $h\to y$, given by $h/R_0=[\!\sqrt{\lambda(y+\lambda)}-\!\sqrt{\mu(y+\mu)}]/[\!\sqrt{\lambda(y+\mu)}-\!\sqrt{\mu(y+\lambda)}]$ are used. This then reduces equation (\ref{eq:potrz}) to
\begin{gather}\label{eq:poteltr}
\Phi(\bm r)=-\frac{2\pi GR_0\Sigma_0\Gamma(b)}{\Gamma(a)\Gamma(b-a)}\!
\int_0^\infty\!\frac{\mathcal F^{2a-2}\mathcal G^{3-2b}\dm y}
{\!\sqrt{(y+\mu)(y+1)(y+\lambda)}};
\\\nonumber\begin{split}
\mathcal F&=
\frac{\!\sqrt{\lambda(y+\mu)}-\!\sqrt{\mu(y+\lambda)}}{\lambda-\mu}
=\frac y{\!\sqrt{\lambda(y+\mu)}+\!\sqrt{\mu(y+\lambda)}},
\\\mathcal G&=
\frac{\!\sqrt{\lambda(y+\lambda)}-\!\sqrt{\mu(y+\mu)}}{\lambda-\mu}
=\frac{y+\lambda+\mu}{\!\sqrt{\lambda(y+\lambda)}+\!\sqrt{\mu(y+\mu)}}.
\end{split}\end{gather}
In general, this is not in the form of a single Carlson-R, although the $(a,b)=(1,\frac32)$ case recovers equation (\ref{eq:potg1}). Instead, if $2a$ and $2b$ are both integers, this is expressible as a finite sum of Carlson-R's (which reduces to the Carlson symmetric basis or an elementary function) expanding $\mathcal F^{2a-2}\mathcal G^{3-2b}$.

Equation (\ref{eq:potrz}) is also recognized as the Weyl integral. In fact, the generalization given by ($r^2=R^2+z^2$)
\begin{multline}\label{eq:potwy}
\frac{\Phi(R,z)}{G\Sigma_0}
=-2\pi R_0^{2a}\frac{\Gamma(b)}{\Gamma(a)}\,{_\infty^-I_{R_0^2}^{b-a}}
\left[\frac{R_0^{2(1-b)}}{\!\sqrt{R^2+(|z|+R_0)^2}}\right]
\\=-2R_0^{2a}\frac{\Gamma(b)}{\Gamma(a)}
\int_0^\infty\!\dm t\,t^\frac12\,{_\infty^-I_{R_0^2}^{b-a}}
\left[\frac{R_0^{3-2b}}{(t+1)tR_0^2+r^2t+z^2}\right]
\end{multline}
provides the general formula for the potential due to the QKMd with $a,b>\frac12$. This is proven by establishing the recursion relation and its generalization via the fractional integral applicable to the surface density in equation (\ref{eq:kmq}), viz.\
\begin{multline*}
\left[-\frac\partial{\partial(R_0^2)}\right]^m\,\frac1{R_0^{2a}}
{_2F_1}\!\left(\stack{a,\frac32}{b};-\frac{R^2}{R_0^2}\right)
\\=\frac{(a)_m}{R_0^{2(a+m)}}
{_2F_1}\!\left(\stack{a+m,\frac32}{b};-\frac{R^2}{R_0^2}\right);
\end{multline*}and\begin{multline*}
{_\infty^-I_{R_0^2}^\xi}\Biggl[\frac1{R_0^{2a}}
{_2F_1}\!\left(\stack{a,\frac32}{b};-\frac{R^2}{R_0^2}\right)\Biggr]
\\=\frac{\Gamma(a-\xi)}{\Gamma(a)}\frac1{R_0^{2(a-\xi)}}
{_2F_1}\!\left(\stack{a-\xi,\frac32}{b};-\frac{R^2}{R_0^2}\right).
\end{multline*}
Since the model with $a=b$ is the Kuzmin disk, we find
\begin{equation*}
{_2F_1}\!\left(\stack{a,\frac32}{b};-\frac{R^2}{R_0^2}\right)
=R_0^{2a}\frac{\Gamma(b)}{\Gamma(a)}\,{_\infty^-I_{R_0^2}^{b-a}}
\frac1{R_0^{2b}}\left(1+\frac{R^2}{R_0^2}\right)^{-\frac32},
\end{equation*}
and then equation (\ref{eq:potwy}) follows the linearity of the Poisson equation. Utilizing the formula
\begin{equation*}
{_\infty^-I_x^\xi}\left[\frac{x^\eta}{(ax+b)^p}\right]
=\frac{\Gamma(p-\xi-\eta)\,x^{\eta+\xi}}
{\Gamma(p-\eta)\,(ax+b)^p}\,
{_2F_1}\!\left(\stack{p,\xi}{p-\eta};
\frac b{ax+b}\right),
\end{equation*}
equation (\ref{eq:potwy}) also reduces to a quadrature given by
\begin{gather}\label{eq:potcr}\begin{split}
\frac{\Phi(\bm r)}{GR_0\Sigma_0}
=-&\frac{2\Gamma(b)\Gamma(a-\frac12)}{\Gamma(a)\Gamma(b-\frac12)}\!
\\&\times\int_0^\infty\!\frac{\dm t\,t^{1/2}}{(t+\lambda)(t+\mu)}
{_2F_1}\!\left(\stack{1,b-a}{b-\frac12};\mathcal X\right);
\end{split}\\\nonumber
\mathcal X=1-\frac{t(t+1)}{(t+\mu)(t+\lambda)}=\frac1{\lambda-\mu}
\left[\frac{(\lambda-1)\lambda}{t+\lambda}+\frac{(1-\mu)\mu}{t+\mu}\right],
\end{gather}
where $(\lambda,\mu)$ are the elliptic coordinates defined in Sect~\ref{sec:ecd}, and so $(t+1)tR_0^2+r^2t+z^2=R_0^2(t+\lambda)(t+\mu)$. If $(a,b)=(\gamma,\frac32)$ here, equation (\ref{eq:potcr}) reproduces equation (\ref{eq:crkmtp}).

Equation (\ref{eq:potrz}) with $b=\frac12$ does not converge but equation (\ref{eq:potcr}) reduces to a definite limit for $b=\frac12$ via $\lim_{c\to0}{_2F_1}(a,b;c;x)/\Gamma(c)=abx{_2F_1}(a+1,b+1;2;x)$:
\begin{equation}\label{eq:potbh}
\frac{\Phi(\bm r)}{GR_0\Sigma_0}
=\frac{2\!\sqrt\pi\,\Gamma(a+\frac12)}{\Gamma(a)}\!\int_0^\infty\!
\frac{\dm t\,t^{a-1}(t+1)^{a-3/2}}{[(t+\mu)(t+\lambda)]^{a-1/2}}
\,\mathcal X,
\end{equation}
which converges for $a>0$. This can be extended to the $a=0$ case by adopting $\lim_{a\downarrow0^+}t^{a-1}/\Gamma(a)=\deltaup(t)$ and then $\Phi=2\pi G\Sigma_0R_0(\lambda\mu)^{1/2}=2\pi G\Sigma_0|z|$. It is verifiable that equation (\ref{eq:potbh}) is consistent with the ``potential'' due to the surface density in equation (\ref{eq:kmq}) with $b=\frac12$ which reduces to
\begin{equation*}
\Sigma(R)=\Sigma_0R_0^{2a}\frac{(1-2a)R^2+R_0^2}{(R^2+R_0^2)^{a+1}},
\end{equation*}
although this is non-negative everywhere only if $a\le\frac12$ (and $\Sigma\to0$ as $R\to\infty$ for $a>0$). The $z=0$ potential resulting from equation (\ref{eq:potbh}) with $\mu=0$ and $\lambda=1+R^2/R_0^2$,
\begin{equation}\label{eq:pkmqr0}
\Phi(R,0)
=\frac{\pi^{3/2}\Gamma(a+\frac12)}{\Gamma(a)}\frac{G\Sigma_0R^2}{R_0}
{_2F_1}\!\left(\stack{\frac32,a+\frac12}2;-\frac{R^2}{R_0^2}\right),
\end{equation}
is consistent with the rotation curve in equation (\ref{eq:kmqv})
\begin{equation*}
\varv_\mathrm c^2(R)
=\frac{2\pi^{3/2}\Gamma(a+\frac12)}{\Gamma(a)}
\frac{G\Sigma_0R^2}{R_0}
{_2F_1}\!\left(\stack{\frac32,a+\frac12}1;-\frac{R^2}{R_0^2}\right),
\end{equation*}
but the zero-point for the potential in equation (\ref{eq:pkmqr0}) is at the origin and the potential for $a<\frac12$ diverges like $\sim R^{1-2a}$ as $R\to\infty$. By contrast, $\lim_{R\to\infty}\Phi(R,0)=2\pi GR_0\Sigma_0$ for $a=\frac12$, whereas $\lim_{R\to\infty}\Phi=0$ if $a>\frac12$, and $\Phi(R,0)$ is not monotonic for those cases (and so $\varv_\mathrm c^2<0$ for large radii and also $\Sigma<0$). On the other hand, equation (\ref{eq:potbh}) with $(\lambda,\mu)=(z^2/R_0^2>1,1)$ or $(\lambda,\mu)=(1,z^2/R_0^2<1)$ results in the potential along the symmetry axis expressible using a ${_2F_1}$-function;
\begin{equation*}
\Phi(0,z)=\frac{2\pi G\Sigma_0|z|}{2a+1}\,
{_2F_1}\!\left(\stack{a,1}{a+\frac32};1-\frac{z^2}{R_0^2}\right).
\end{equation*}
Although it is not immediately obvious, this is consistent with the earlier result in equation (\ref{eq:kmqgz}) with $b=\frac12$.

Equation (\ref{eq:potbh}) is further reducible to closed-form expressions utilizing the Carlson-R functions,
\begin{equation*}\begin{split}
\frac{\Phi(\bm r)}{GR_0\Sigma_0}&=2\pi\,\biggl[
R_{-\frac12}\bigl(a-\tfrac12,\tfrac32-a,a-\tfrac12;\mu,1,\lambda\bigr)
\\&\qquad-\frac{2a}{2a+1}
R_{-\frac12}\bigl(a+\tfrac12,\tfrac12-a,a+\tfrac12;\mu,1,\lambda\bigr)\biggr]
\\\phantom{\frac{\Phi(\bm r)}{GR_0\Sigma_0}}
&=\frac{2\pi}{2a+1}\Biggl[\frac{\lambda(\lambda-1)}{\lambda-\mu}\,
R_{-\frac32}\bigl(a-\tfrac12,\tfrac32-a,a+\tfrac12;\mu,1,\lambda\bigr)
\\&\qquad+\frac{\mu(1-\mu)}{\lambda-\mu}\,
R_{-\frac32}\bigl(a+\tfrac12,\tfrac32-a,a-\tfrac12;\mu,1,\lambda\bigr)\Biggr].
\end{split}\end{equation*}
The result for $a=0$ is consistent with the potential due to an infinite plate ($\Phi\propto|z|$) whereas the result with $a=\frac12(=b)$ is reducible to $\Phi=2\pi G\Sigma_0R_0[1-(\!\sqrt\lambda+\!\sqrt\mu)^{-1}]$, which is the zero-point-shifted potential due to the Kuzmin disk.

If $a-b$ is a non-negative integer, equation (\ref{eq:potwy}) is purely differential and so reducible to a completely analytic expression. In particular, if $a=b+1$, we have
\begin{equation*}\begin{split}
\Sigma(R)&=\frac{\Sigma_0R_0^3}{(R^2+R_0^2)^{3/2}}
\left(1-\frac3{2b}\frac{R^2}{R^2+R_0^2}\right);
\\\Phi(\bm r)&=-\frac{2\pi G\Sigma_0R_0^2}{\!\sqrt{R^2+(|z|+R_0)^2}}
\left(1-\frac1b+\frac{R_0}{2b}\frac{|z|+R_0}{R^2+(|z|+R_0)^2}\right),
\\\varv_\mathrm c^2(R)&=
\frac{2\pi G\Sigma_0R_0^2R^2}{(R_0+R^2)^{3/2}}
\left(1+\frac1{2b}-\frac3{2b}\frac{R^2}{R^2+R_0^2}\right);
\\\varg(z)
&=\frac{2\pi G\Sigma_0R_0^2}{(z+R_0)^2}\left(1-\frac1b\frac z{z+R_0}\right)
\quad(z>0).
\end{split}\end{equation*}
Here the surface density is non-negative everywhere if $b\ge\frac32$.

\subsection{The Kalnajs--Mestel disks}
\label{sec:kmd}

Equation (\ref{eq:potrz}) with a positive integer value of $b-a=n+1$, is reducible to a closed form consisting entirely of elementary functions if $2a-1$ is also a positive integer. In particular, if $n=0$ and $a\in\set{1,\frac32,2,\cdots}$, equation (\ref{eq:kmdp}) is resolved via the standard techniques such as partial fraction decompositions: e.g., if $(a,b)=(1,2)$,
\begin{equation}\label{eq:kd1}
\frac{\Phi(\bm r)}{GR_0\Sigma_0}
=-\frac{8\pi}\sigma\tanh^{-1}\biggl(\frac\sigma{1+\tau_+}\biggr),
\end{equation}
whereas, if $(a,b)=(\frac32,\frac52)$,
\begin{equation}\label{eq:kd3}
\frac{\Phi(\bm r)}{GR_0\Sigma_0}=
-\frac{6\pi}{\sigma^2}\left[\tau_+-1
-\frac{2|\zeta|}\sigma\tanh^{-1}\biggl(\frac\sigma{1+\tau_+}\biggr)
\right].
\end{equation}
Here $\sigma^2=(r/R_0)^2=\varrho^2+\zeta^2$ and $\tau_\pm^2=(r_\pm/R_0)^2=\varrho^2+(|\zeta|\pm1)^2$ with $(\varrho,\zeta)=(R/R_0,z/R_0)$. Also note $2\lambda^{1/2}=\tau_++\tau_-$ and $2\mu^{1/2}=\tau_+-\tau_-$. If $a=\frac12$, equation (\ref{eq:potrz}) does not converge, but the zero-point-shifted potential (cf. eq.~\ref{eq:potshi}) can still be obtained by the alternative kernel $[R^2+(|z|+t)^2]^{1/2}-t^{-1}$. In particular, for $(a,b)=(\frac12,\frac32)$, the same variable change as equation (\ref{eq:kmdp}) results in
\begin{multline*}
\frac{\Phi(\bm r)}{2\pi GR_0\Sigma_0}=\int_{R_0}^\infty\!\dm h
\left(\frac1h-\frac1{\!\sqrt{R^2+(|z|+h)^2}}\right)
\\=\int_{R_0+r_+}^\infty\!\frac{2(|z|x+r^2)\,\dm x}{(x+|z|)(x^2-r^2)}
=\sinh^{-1}\biggl(\frac{1+|\zeta|}\varrho\biggr)+\ln\frac\varrho2
\end{multline*}
which recovers the potential of a cored Mestel disk (eq.~\ref{eq:cmdp}) with the zero-point at the origin, $\Phi(0,0)=0$.

The corresponding $\Sigma(R)$ and $\varv_\mathrm c^2(R)$ are expressed as
\begin{equation*}\begin{split}
\frac{\Sigma(R)}{\Sigma_0}
&={_2F_1}\!\left(\stack{a,\frac32}{a+1};-\varrho^2\right)
=\frac a{\varrho^{2a}}B_{\kappa_0^2}\!\left(a,\tfrac32-a\right);
\\
\frac{\varv_\mathrm c^2(R)}{GR_0\Sigma_0}
&=\frac{2\pi a\,\varrho^2}{a+\frac12}
{_2F_1}\!\left(\stack{\frac32,a+\frac12}{a+\frac32};-\varrho^2\right)
=\frac{2\pi a}{\varrho^{2a-1}}B_{\kappa_0^2}\!\left(a+\tfrac12,1-a\right),
\end{split}\end{equation*}
where $\kappa_0^2=\varrho^2/(1+\varrho^2)$ and
\begin{equation*}
B_z(a,b)={\rm B}(z;a,b)=\int_0^z\!u^{a-1}(1-u)^{b-1}\dm u
\end{equation*}
is the incomplete beta function,
which is reducible to an elementary function using the formulae:
\begin{gather*}
B_{\kappa_0^2}\left(k+1,\tfrac12-k\right)
=\frac{2(-1)^kk!}{(\frac12)_k}\left[1-\frac1{\!\sqrt{1+\varrho^2}}
\sum_{j=0}^k\binom{\frac12}j\,\varrho^{2j}\right];
\\\begin{split}
B_{\kappa_0^2}&\left(k+\tfrac32,-k\right)
=\frac{2(-1)^k(\frac32)_k}{k!}
\\&\times\left[\sinh^{-1}\varrho-\frac\varrho{\!\sqrt{1+\varrho^2}}
\,\Biggl(1-\sum_{j=1}^k\frac{(j-1)!}2
\frac{(-\varrho^2)^j}{(\frac32)_j}\Biggr)\right],
\end{split}\end{gather*}
for any non-negative integer $k$. That is to say, if $a$ is a half integer, $\Sigma(R)$ involves $\sinh^{-1}\varrho$ -- except $a=\frac12$ for which $\Sigma\propto(1+\varrho^2)^{-1/2}$ -- but $\varv_\mathrm c^2(R)$ is entirely algebraic, whereas $\Sigma(R)$ is algebraic and $\varv_\mathrm c^2(R)$ involves $\sinh^{-1}\varrho$ for a (positive) integer value of $a$. The associated vertical acceleration on the symmetry axis is found to be ($z>0$)
\begin{equation*}
\frac{\varg(z)}{G\Sigma_0}
=\frac{4\pi a}{\zeta^{2a}}\,
{\rm B}\biggl(\frac\zeta{1+\zeta};2a,2-2a\biggr),
\end{equation*}
which can be expressed as an elementary function,
\begin{multline*}
\frac{\varg(z)}{4\pi aG\Sigma_0}=
m\left[\frac{(-1)^{m+1}}{\zeta^{m+1}}\ln(1+\zeta)
+\sum_{k=1}^m\frac{(-1)^k}{m+1-k}\frac1{\zeta^k}\right]
\\+\frac1{1+\zeta},
\end{multline*}
if $m=2a-1$ is a non-negative integer.

\subsubsection{The Kalnajs isochrone disk vs.\ the projected isochrone}

The mid-plane potential for the $(a,b)=(\frac32,\frac52)$ case
\begin{equation*}
-\Phi(R,0)=6\pi GR_0\Sigma_0\frac{\!\sqrt{1+\varrho^2}-1}{\varrho^2}
=\frac{GM_\mathrm t}{R_0+(R^2+R_0^2)^{1/2}}
\end{equation*}
has the same functional form as the spherical isochrone potential \cite{He59} of the same mass $M_\mathrm t=6\pi R_0^2\Sigma_0$ and scale length $h=R_0$, i.e.\ $\Phi(r)=-GM_\mathrm t/[h+\!\sqrt{h^2+r^2}]$. In other words, the mid-plane rotation curve due to the disk with 
\begin{equation*}\begin{split}
\Sigma(R)&=\frac{M_\mathrm tR_0}{2\pi R^3}
\left[\sinh^{-1}\biggl(\frac R{R_0}\biggr)
-\frac R{(R_0^2+R^2)^{1/2}}\right];
\\\varv_\mathrm c^2(R)
&=\frac{GM_\mathrm tR^2}
{(R_0^2+R^2)^{1/2}\,\bigl[R_0+(R_0^2+R^2)^{1/2}\bigr]^2}
\end{split}\end{equation*}
is identical to the circular speed curve of the spherical isochrone potential \citep{Ka76}. The vertical force along the symmetry axis due to this disk is given by
\begin{equation*}
\varg(z)=\frac{GM_\mathrm t}{z^2}\left[\frac{z+2R_0}{z+R_0}
-\frac{2R_0}z\ln\biggl(1+\frac z{R_0}\biggr)\right]
\end{equation*}
The expression for the full potential is provided in equation (\ref{eq:kd3}) and also found in \cite{EdZ92}.

However the surface density profile in equation (33) of \cite{He59} corresponds to the projected density of the spherical profile. Curiously its functional form (cf.\ \cite{EdZ92}, eq.~6.21)
\begin{equation*}
\frac{\Sigma(R)}{\Sigma_0}
=\frac3{2\varrho^3}\left(\tan^{-1}\varrho-\frac\varrho{1+\varrho^2}\right)
={_2F_1}\!\left(\stack{\frac32,2}{\frac52};-\frac{R^2}{R_0^2}\right)
\end{equation*}
is in the form of equation (\ref{eq:kmq}) with $(a,b)=(2,\frac52)$. Yet, the rotation curve due to the disk of the same density profile is
\begin{equation*}\begin{split}
\varv_\mathrm c^2(R)&=\frac{9\pi^2}{16}GR_0\Sigma_0\,
\varrho^2{_2F_1}\!\left(\stack{\frac32,\frac52}3;-\varrho^2\right)
\\&=\frac{GM_\mathrm t}{R_0}\left[R_F(0,1,1+\varrho^2)
-\frac{2+\varrho^2}3R_D(0,1;1+\varrho^2)\right]
\\&=\frac{GM_\mathrm t}{R_0}\frac{(2+\varrho^2)\mathbf E(\kappa_0)-2\mathbf K(\kappa_0)}
{\varrho^2\!\sqrt{1+\varrho^2}};\quad
\kappa_0^2=\frac{\varrho^2}{1+\varrho^2},
\end{split}\end{equation*}
where $M_\mathrm t=3\pi R_0^2\Sigma_0$ is the disk mass. The maximum of this curve is higher than the circular velocity due to the spherical isochrone of the same mass, which is expected. The expression of the potential due to this disk is provided in equation (6.23) of \cite{EdZ92}, which involves elliptic integrals\footnote{Equation (\ref{eq:potrz}) is an elliptic integral if $b-a$ is a half-integer and $2a$ is an integer. If $a$ is an integer and $b$ is a half-integer, $\Sigma(R)$ is elementary but $\varv_\mathrm c^2(R)$ requires elliptic integrals, whereas, if $a$ is a half-integer and $b$ is an integer, $\Sigma(R)$ is an elliptic integral but $\varv_\mathrm c^2(R)$ is elementary.}.

\subsection{The projected column density of the spherical profile}

As a brief digression, let us ask how a projected column density of the spherical profile is related to the disk surface density if the rotation curves due to both models are the same. If $\rho(r)$ is the spherical profile with the radius $r$, then its projected column density is given by
\begin{equation*}
\Sigma^\mathrm p(R)
=2\!\int_R^\infty\!\frac{r\rho(r)\,\dm r}{\!\sqrt{r^2-R^2}}.
\end{equation*}
Since the rotation curve of the spherical profile is found to be $\varv_{\rm rot}^2(r)=GM(r)/r=4\pi Gr^{-1}\!\int_0^r\dm\tilde r\,\tilde r^2\rho(\tilde r)$, the projected profile can be directly obtained from the rotation curve via:
\begin{equation*}
\Sigma^\mathrm p(R)
=\frac1{2\pi G}\!\int_R^\infty\!\frac{\dm r}{r\!\sqrt{r^2-R^2}}
\frac{\dm(r\varv_{\rm rot}^2)}{\dm r}.
\end{equation*}
From the Mellin transform $R^2\to t$ of this, we find
\begin{equation}\label{eq:melsph}\begin{split}
\mathfrak M[\Sigma^\mathrm p](t)
&=\frac{(1-t)\Gamma(t)}{\Gamma(t+\frac12)}
\frac{\mathfrak M[\varv_{\rm rot}^2]\bigl(t-\tfrac12\bigr)}{2\!\sqrt\pi\,G};
\\\frac{\mathfrak M[\varv_{\rm rot}^2](t)}{2\!\sqrt\pi\,G}
&=\frac{\Gamma(t+1)\,\mathfrak M[\Sigma^\mathrm p](t+\tfrac12)}
{(\frac12-t)\Gamma(t+\frac12)}.
\end{split}\end{equation}
Given the Me{\ij}er-G convolution theorem for the second equation, we can derive the direct integral formula for $\Sigma^\mathrm p\to\varv_{\rm rot}^2$,
\begin{multline*}
\varv_{\rm rot}^2(r)=\frac{2\pi G}r\!\int_0^r\!\dm R\,R\Sigma^\mathrm p(R)
\\+4G\!\int_r^\infty\!\dm R
\left[\frac Rr\sin^{-1}\Bigl(\frac rR\Bigr)
-\frac{R}{\!\sqrt{R^2-r^2}}\right]
\Sigma^\mathrm p(R),
\end{multline*}
which is verified directly. As for the question posed, we can compare equation (\ref{eq:melsph}) to (\ref{eq:vsmel}). That is to say, if $\varv_{\rm rot}^2(r)=\varv_\mathrm c^2(R=r)$, then the Mellin transform of $\Sigma(R)$ and $\Sigma^\mathrm p(R)$ should be related such that $\!\sqrt\pi\,\Gamma(2-t)\mathfrak M[\Sigma](t)=\Gamma(\frac32-t)\mathfrak M[\Sigma^\mathrm p](t)$.

For the disk surface density of equation (\ref{eq:kmq}), its mid-plane rotation curve given in equation (\ref{eq:kmqv}) is identical to the one due to the spherical profile:
\begin{equation*}\begin{split}
\rho(r)&=
\frac{3\Gamma(a+\frac12)\Gamma(b)}{2\Gamma(a)\Gamma(b+\frac12)}
\frac{\Sigma_0}{R_0}
{_2F_1}\!\left(\stack{\frac52,a+\frac12}{b+\frac12};
-\frac{r^2}{R_0^2}\right)
\\&=\frac{2\Gamma(b)}{\!\sqrt\pi\,\Gamma(a)}\frac{\Sigma_0}{R_0}
G^{1,2}_{2,2}\!\left(\frac{r^2}{R_0^2}\vrule
\stack{-\frac32,\frac12-a}{0,\frac12-b}\right),
\end{split}\end{equation*}
which projects to the column density of
\begin{equation*}
\frac{\Sigma^\mathrm p(R)}{\Sigma_0}
=\frac{2\Gamma(b)}{\Gamma(a)}
G^{1,2}_{2,2}\!\left(\frac{R^2}{R_0^2}\vrule\stack{-1,1-a}{0,1-b}\right)
=2\,{_2F_1}\!\left(\stack{2,a}{b};-\frac{R^2}{R_0^2}\right).
\end{equation*}
If $a=\frac32$, the functional form of the projected column density also belongs to the same family as equation (\ref{eq:kmq}) with $a=2$ (the scale factor of $2$ accounting for the mass difference; i.e.\ the spherical profile is twice as massive), which is indeed the case for the \citeauthor{Ka76} isochrone disk.

\subsection{The disks with asymptotically flat rotation curves}

We find equation (\ref{eq:kmqv}) with $a=\frac12$ tends to a finite non-zero limit as $R\to\infty$; that is,
\begin{equation*}
\lim_{R\to\infty}\frac{\varv_\mathrm c^2(R)}{GR_0\Sigma_0}
=\frac{4\!\sqrt\pi\,\Gamma(b)}{\Gamma(b-\frac12)}=\varv_\infty^2.
\end{equation*}
Since the flat rotation curve implies the logarithmic divergence of the potential, the integral for $\Phi(\bm r)$ in equation (\ref{eq:potrz}) does not converge. Instead, the zero-point-shifted potential is found similarly as equations (\ref{eq:potshi}); that is,
\begin{equation}\label{eq:kmqfrp}\begin{split}
\frac{\Phi(\bm r)}{\varv_\infty^2}
&=\int_{R_0}^\infty\!\dm h\left(1-\frac{R_0^2}{h^2}\right)^{b-\frac32}
\left(\frac1h-\frac1{\!\sqrt{R^2+(|z|+h)^2}}\right)
\\&=2\!\int_{R_0+r_+}^\infty\frac{|z|x+r^2}{x+|z|}
\frac{\mathcal Q^{b-3/2}\dm x}{(x^2-r^2)^{2b-2}}
\end{split}\end{equation}
which converges if $b>a=\frac12$. The last line utilizes the same change of the integration variable as equations (\ref{eq:kmdp}) with $\mathcal Q$ defined there, and is in a manifest form of the Carlson-R, although the general result is of a minimal practical interest.

If $b-\frac32$ is an integer,
the integrand is a rational function of $x$ and equation (\ref{eq:kmqfrp}) is expressible via elementary functions up to logarithms or inverse hyperbolic functions: e.g.,
\begin{equation*}\begin{split}
\frac\Phi{\varv_\infty^2}
&=f_0=\ln\biggl(\frac{1+|\zeta|+\tau_+}2\biggr)\qquad(b=\tfrac32);\\
\frac\Phi{\varv_\infty^2}&=f_0+\frac{f_1+\tau_+-1}{\sigma^2}
-\frac12\qquad(b=\tfrac52),
\end{split}\end{equation*}
where $\sigma$ and $\tau_+$ are as defined for equation (\ref{eq:kd1}) and
\begin{equation*}
f_1
=-\frac{2|\zeta|}\sigma\tanh^{-1}\biggl(\frac\sigma{1+\tau_+}\biggr).
\end{equation*}
The explicit expression for the potential quickly becomes rather complicated with increasing $b$; for instance, if $b=\frac72$
\begin{multline*}
\frac{\Phi}{\varv_\infty^2}=f_0
+\frac{4\sigma^4+3\varrho^2-2\zeta^2}{2\sigma^6}f_1
+\frac{(4\varrho^2-11\zeta^2)(\tau_+-1)}{6\sigma^6}
\\+\frac{5(2\sigma^2+|\zeta|)\tau_+}{6\sigma^4}
-\frac2{\sigma^2}-\frac34.
\end{multline*}

The general expression for the corresponding surface density may be found from (where $m=b-\frac32\in\set{1,2,3,\dotsc}$)
\begin{multline*}
\Sigma(R)=\frac{\varv_\infty^2m!}{2\pi\,GR_0(\frac32)_m}
{_2F_1}\!\left(\stack{\frac32,\frac12}{m+\frac32};-\varrho^2\right)
\\=\frac{\varv_\infty^2(1+\varrho^2)^{m-\frac12}}{2\pi GR_0}
\left[\frac1{\varrho^{2m}}
+\frac{(-1)^m}{(m-1)!}\,\biggl(\frac\partial{\partial\varrho^2}\biggr)^{m-1}
\!\frac{\sinh^{-1}\!\varrho}{\varrho^3\!\sqrt{1+\varrho^2}}\right]
\end{multline*}
which further reduces to
\begin{multline*}
\Sigma(R)=\frac{\varv_\infty^2}{2\pi GR_0}\Biggl[\sum_{k=1}^m\!
{_4F_3}\!\left(\stack{1,1,k+\frac12,k-m}{\frac32,k,k+1};1\right)
\\\times
\binom mk\frac{(\frac32)_{k-1}}{(k-1)!}\frac{\!\sqrt{1+\varrho^2}}{\varrho^{2k}}
\\-\frac{(\frac32)_{m-1}}{(m-1)!}
{_2F_1}\!\left(\stack{-m,1-m}{\frac12-m};-\varrho^2\right)
\frac{\sinh^{-1}\!\varrho}{\varrho^{2m+1}}\Biggr].
\end{multline*}
The expression for the rotation curve is relatively simpler:
\begin{equation*}\begin{split}
\varv_\mathrm c^2(R)
&=\frac{\varv_\infty^2}{2(m+1)}\frac{\varrho^2}{1+\varrho^2}
{_2F_1}\!\left(\stack{m+\frac12,1}{m+2};\frac{\varrho^2}{1+\varrho^2}\right)
\\&=m!\varv_\infty^2
\left[\sum_{k=0}^m\frac1{(m-k)!(\frac12)_k}\frac1{\varrho^{2k}}
-\frac{(1+\varrho^2)^{m-\frac12}}{(\frac12)_m\,\varrho^{2m}}\right],
\end{split}\end{equation*}
while the vertical force for $m\in\set{1,2,3,\dotsc}$ is in the form of
\begin{equation*}
\varg(z)=\frac{\varv_\infty^2}{R_0}\left[
\frac{2m(\zeta^2-1)^{m-1}}{\zeta^{2m+1}}\ln(1+\zeta)
+\frac{p_\varg(\zeta)}{\zeta^{2m}}\right],
\end{equation*}
where $p_\varg(\zeta)$ is a $(2m-1)$-th polynomial of $\zeta$ and $\varg(z)$ falls off asymptotically like $\varg(z)\simeq\varv_\infty^2/z$ as $z\to\infty$ independent of $b$. The resulting specific results for $b=\frac52$ ($m=1$) are
\begin{equation*}
\frac{\Sigma(R)}{\varv_\infty^2}
=\frac{\!\sqrt{1+\varrho^2}-\hat S}{2\pi GR_0\,\varrho^2};\quad
\frac{\varv_\mathrm c^2(R)}{\varv_\infty^2}=1-\frac2{1+\!\sqrt{1+\varrho^2}};
\end{equation*}
and $p_\varg(\zeta)=\zeta-2$, where $\hat S=\varrho^{-1}\sinh^{-1}\!\varrho$. For $b=\frac72$ ($m=2$), we have $p_\varg(\zeta)=\zeta^3-(8/3)\zeta^2-2\zeta+4$ and
\begin{equation*}\begin{split}
\frac{\Sigma(R)}{\varv_\infty^2}&
=\frac{(3+2\varrho^2)\sqrt{1+\varrho^2}
-(3+4\varrho^2)\hat S}{4\pi GR_0\,\varrho^4};
\\\frac{\varv_\mathrm c^2(R)}{\varv_\infty^2}&
=1-\frac43\frac{3+4\varrho^2}{2+3\varrho^2+2(1+\varrho^2)^{3/2}}.
\end{split}\end{equation*}
The $b=\frac32$ ($m=0$) case is simply the cored Mestel disk.

\subsubsection{the potential with elliptic integrals of the third kind}

Let us observe that equation (\ref{eq:kmqfrp}) is also reducible to
\begin{equation*}
\frac{\Phi(\bm r)}{\varv_\infty^2}
=2\!\int_{R_0+r_+}^\infty\!\frac{\dm x}{\!\sqrt{\mathcal Q}}
\left(|z|+\frac{R^2}{x+|z|}\right)
\left[1-\frac{4R_0^2(x+|z|)^2}{(x^2-r^2)^2}\right]^{b-1};
\end{equation*}
Given that $\mathcal Q$ is a quartic polynomial of $x$, if $b$ is a positive integer, this is recognized as an elliptic integral of the third kind, which requires an additional basis ($R_J$ or $\Pi$) besides those introduced earlier in Sect.~\ref{sec:ee} to be written down.

If $b=1$, the change of variable $u=2L_+M_+r_+/(x-R_0-r_+)$ in equation (\ref{eq:kmqfrp}) then results in
\begin{equation*}\begin{split}
\frac{\Phi(\bm r)}{\varv_\infty^2}
&=\int_0^\infty\frac{\dm u}{\!\sqrt{\mathcal R}}
\left[\tau_+-1-\frac{2\tau_+(\tau_+-|\zeta|-1)}{u+2\tau_+}\right]
\\&=2(\tau_+-1)\,R_F(M_+\tau_+,L_+M_+,L_+\tau_+)
\\&\phantom{=2(\tau_+}
-\tfrac43\tau_+L_-M_-\,R_J(M_+\tau_+,L_+M_+,L_+\tau_+;2\tau_+),
\end{split}\end{equation*}
where $(L_\pm,M_\pm,\tau_\pm)=(\lambda^{1/2}\pm1,1\pm\mu^{1/2},\lambda^{1/2}\pm\mu^{1/2})$ and $\mathcal R=(u+M_+\tau_+)(u+L_+M_+)(u+L_+\tau_+)$. Here also note that $(M_+\tau_+,L_+M_+,L_+\tau_+)=(\mu+\Xi,1+\Xi,\lambda+\Xi)$ with $\Xi=\lambda^{1/2}+\mu^{1/2}+\lambda^{1/2}\mu^{1/2}$, and so the duplication formula of elliptic integrals (see e.g., \cite{DLMF}, \S~19.26(iii)) indicates $2R_F(M_+\tau_+,L_+M_+,L_+\tau_+)=R_F(\mu,1,\lambda)$. In fact, also utilizing the duplication formula for $R_J$, one obtains
\begin{equation*}\begin{split}
\frac{\Phi(\bm r)}{\varv_\infty^2}
&=(\tau_+-1)\,R_F
-\tfrac23\tau_+L_-M_-\,R_J(\mu,1,\lambda;\nu_0)+\ln\nu_0
\\&=(1-\tau_-)\,R_F
+\tfrac23\tau_-L_-M_+\,R_J(\mu,1,\lambda;\nu_1)+\ln\nu_1,
\end{split}\end{equation*}
where $R_F=R_F(\mu,1,\lambda)$ with
\begin{equation*}\begin{split}
\nu_0&=\tau_+-|\zeta|=\lambda^{1/2}+\mu^{1/2}-\lambda^{1/2}\mu^{1/2}
=L_-M_-+1;
\\
\nu_1&=|\zeta|+\tau_-=\lambda^{1/2}\mu^{1/2}+\lambda^{1/2}-\mu^{1/2}
=L_-M_++1.
\end{split}\end{equation*}

Notice that these expressions are naturally expressed in the prolate spheroidal coordinates $(\lambda,\mu)$ defined in Sect.~\ref{sec:ecd}. However, ``simpler'' expressions in terms of the same elliptic coordinates also exist. In particular, equation (\ref{eq:kmqfrp}) following the same variable change as equation (\ref{eq:poteltr}) results in
\begin{equation}\label{eq:poteg}
\frac{\Phi(\bm r)}{\varv_\infty^2}
=\frac12\!\int_0^\infty\!\dm y\,\frac{\mathcal I}y
\left(\frac{\!\sqrt{\lambda(y+\lambda)}+\!\sqrt{\mu(y+\mu)}}
{y+\lambda+\mu}\right)^{2(b-1)}
\end{equation}
where
\begin{displaymath}
\mathcal I=\frac{\!\sqrt\lambda}{\!\sqrt{y+\lambda}}
+\frac{\!\sqrt\mu}{\!\sqrt{y+\mu}}-\frac1{\!\sqrt{y+1}}
-\frac{\!\sqrt{\lambda\mu}}{\!\sqrt{(y+\mu)(y+1)(y+\lambda)}}.
\end{displaymath}
Here $\mathcal I/y$ is indeterminate ($0/0$) as $y\to0$, and an obvious reduction to the Carlson-R also yields an indeterminate ($\infty-\infty$) expression. To find a properly convergent expression, one needs to employ the parameter change for the elliptic integral of the third kind (e.g., \cite{DLMF}, \S~19.7(iii) and 19.21(iii)) to cancel off the logarithmic singularities. Specifically, if $b=1$,
\begin{align}
\frac{\Phi(\bm r)}{\varv_\infty^2}
&=\lim_{\epsilon\downarrow0^+}
\left[\ln\frac{2\!\sqrt{\lambda\mu}}{\!\sqrt\epsilon}
-\frac{\!\sqrt{\lambda\mu}}3
R_J(\epsilon+\mu,\epsilon+1,\epsilon+\lambda;\epsilon)\right]
\nonumber\\&=|\zeta|\left[R_F(\mu,1,\lambda)
-\frac{\varrho^2}3R_J(\mu,1,\lambda;\vartheta)\right]
+\frac{\ln\vartheta}2
\nonumber\\&=\frac{\!\sqrt\lambda}{\!\sqrt\mu}
\frac{\mathrm F(\varphi,\kappa)-(1-\mu)\,\Pi(\varphi,\varpi^2,\kappa)}
{\!\sqrt{\lambda-\mu}}+\ln\!\sqrt{\lambda+\mu-\lambda\mu}
\nonumber\\&=\frac{\!\sqrt\mu\,\Pi(\varphi,\omega^2,\kappa)}
{\!\sqrt\lambda\sqrt{\lambda-\mu}}
+\ln\!\sqrt{\lambda\mu+\lambda-\mu}
\label{eq:blpd}\end{align}
where $\vartheta=\lambda+\mu-\lambda\mu=1+\varrho^2$ and
\begin{displaymath}
\kappa^2=\frac{\lambda-1}{\lambda-\mu},\quad
\cos^2\!\varphi=\frac\mu\lambda,\quad
\varpi^2=\frac{(\lambda-1)\mu}{\lambda-\mu},\quad
\omega^2=\frac{\lambda-1}\lambda.
\end{displaymath}
Note $0\le\varpi^2,\omega^2\le\kappa^2$, and the arguments $(\varphi,\kappa)$ are the same as defined for equation (\ref{eq:potg1}). By contrast, the corresponding potential for $R=0$ or $z=0$ is much simpler; $\Phi(R,0)=\varv_\infty^2\ln\!\sqrt{1+\varrho^2}$ and $\Phi(0,z)=\varv_\infty^2|\zeta|\,R_C(\zeta^2;1)$, where $R_C(x;y)=R_F(x,y,y)$ is an elementary function such that
\begin{equation}\label{eq:rcz1}
R_C(\zeta^2;1)
=\begin{cases}\dfrac{\cos^{-1}\!|\zeta|}{\sqrt{1-\zeta^2}}&(|\zeta|<1)\\
1&(|\zeta|=1)\\
\dfrac{\cosh^{-1}\!|\zeta|}{\sqrt{\zeta^2-1}}&(|\zeta|>1)\end{cases}.
\end{equation}
Hence the resulting rotation curve and the vertical force are 
\begin{equation}\label{eq:fzbp}
\frac{\varv_\mathrm c^2(R)}{\varv_\infty^2}=\frac{\varrho^2}{1+\varrho^2};\quad
\varg(z)=\frac{\varv_\infty^2}{R_0}\frac{R_C(\zeta^2;1)-\zeta}{1-\zeta^2}.
\end{equation}
Here if $z=R_0$ ($\zeta=1$), then $\varg(R_0)=(2\varv_\infty^2)/(3R_0)$. Interestingly, $\Phi(R,0)$ and $\varv_\mathrm c^2(R)$ of this model, $(a,b)=(\frac12,1)$, are the same as the so-called logarithmic potential of Binney, although the disk surface density generating this is given by
\begin{equation*}\begin{split}
\Sigma(R)&=\frac{\varv_\infty^2}{4GR_0}
{_2F_1}\!\left(\stack{\frac32,\frac12}{1};-\varrho^2\right)
\\&=\frac{\varv_\infty^2}{2\pi GR_0}\left[
R_F(0,1,1+\varrho^2)-\frac{\varrho^2}3R_D(0,1;1+\varrho^2)\right]
\\&=\frac{\varv_\infty^2\mathbf E(\kappa_0)}{2\pi G(R^2+R_0^2)^{1/2}};\quad
\left(\kappa_0^2=\frac{\varrho^2}{1+\varrho^2}
=\frac{R^2}{R^2+R_0^2}\right).
\end{split}\end{equation*}

For other integer values of $b$, the surface density profile can be discovered using the differential recursion, namely
\begin{equation*}
\Sigma(R)=\frac{\varv_\infty^2}{2\pi GR_0}
\frac{(1+\varrho^2)^{b-2}}{\bigl(-\frac12\bigr)_{b-1}}\,
\biggl(-\frac\partial{\partial\varrho^2}\biggr)^{b-1}\Bigl[\!\sqrt{1+\varrho^2}\,
\mathbf E(\kappa_0)\Bigr],
\end{equation*}
%
whereas it is also possible to find the general expressions for
\begin{gather*}\begin{split}
\frac{\varv_\mathrm c^2(R)}{\varv_\infty^2}
=&\frac1{2b-1}\frac{\varrho^2}{1+\varrho^2}\left[1
-\sum_{k=0}^{b-2}\frac{(\frac12-b)_k}{(2-b)_k}
\left(\frac{1+\varrho^2}{\varrho^2}\right)^k\right]
\\&+\frac{(\frac32)_{b-2}}{(b-2)!}\left(\frac{1+\varrho^2}{\varrho^2}\right)^{b-2}
\left(1-\frac{\tan^{-1}\!\varrho}\varrho\right);
\end{split}\\
\varg(z)=\frac{\varv_\infty^2}{R_0}\left[
\frac{(2b-3)(\zeta^2-1)^{b-2}}{\zeta^{2b-2}}R_C(\zeta^2;1)
+\frac{q_\varg(\zeta)}{\zeta^{2b-2}}\right].
\end{gather*}
where $q_\varg(\zeta)$ is another $(2b-3)$-th polynomial of $\zeta$ except $b=1$, for which $q_\varg(\zeta)=\zeta/(\zeta^2-1)$. For $b=2$, these result in
\begin{equation*}\begin{split}
\Sigma(R)&=\frac{\varv_\infty^2}{2\pi GR_0\varrho^2}
\left[\!\sqrt{1+\varrho^2}\,\mathbf E(\kappa_0)
-\frac{\mathbf K(\kappa_0)}{\!\sqrt{1+\varrho^2}}\right];
\\\frac{\varv_\mathrm c^2(R)}{\varv_\infty^2}
&=1-\frac{\tan^{-1}\!\varrho}\varrho;\quad
\frac{\varg(z)}{\varv_\infty^2}
=\frac{R_C(\zeta^2;1)+\zeta-\frac\pi2}{R_0\zeta^2},
\end{split}\end{equation*}
while corresponding potential is found from equation (\ref{eq:poteg}):
\begin{equation*}\begin{split}
\frac{\Phi(\bm r)}{\varv_\infty^2}
&=\frac{\Phi^{b=1}}{\varv_\infty^2}
-\frac{\sqrt{\lambda\mu}}3R_J(\mu,1,\lambda;\lambda+\mu)
+\frac{\tan^{-1}\!\sigma}\sigma-1
\\&=\frac{\sqrt\lambda}{\sqrt\mu}
\frac{\Pi(\varphi,\tilde\varpi^2,\kappa)
-(1-\mu)\,\Pi(\varphi,\varpi^2,\kappa)}{\sqrt{\lambda-\mu}}-1
\\&\qquad+\frac{\ln(\lambda+\mu-\lambda\mu)}2
+\frac{\tan^{-1}\!\sigma}\sigma
\\&=\frac{\mu\,\Pi(\varphi,\omega^2,\kappa)
-\lambda\,\Pi(\varphi,\tilde\omega^2,\kappa)}
{\sqrt{\lambda\mu(\lambda-\mu)}}-1
\\&\qquad+\frac{\ln(\lambda\mu+\lambda-\mu)}2
+\frac1\sigma\left(\frac\pi2+\tan^{-1}\!\frac{\lambda-1}\sigma\right)
\end{split}\end{equation*}
%
where $\Phi^{b=1}$ is the potential for the $b=1$ case given in equation (\ref{eq:blpd}) , while $\sigma^2=\varrho^2+\zeta^2=\lambda+\mu-1$ and
\begin{equation}\label{eq:alpp}
\tilde\varpi^2=-\frac\mu{\lambda-\mu},\quad
\tilde\omega^2=-\frac{\lambda-1}\mu.
\end{equation}
The specialized results for $R=0$ or $z=0$ are given by
\begin{equation*}\begin{split}
\frac{\Phi(R,0)}{\varv_\infty^2}
&=\ln\!\sqrt{1+\varrho^2}+\frac{\tan^{-1}\!\varrho}{\varrho}-1;
\\\frac{\Phi(0,z)}{\varv_\infty^2}
&=\frac{\frac\pi2+(\zeta^2-1)\,R_C(\zeta^2;1)}{|\zeta|}-1,
\end{split}\end{equation*}
which are also consistent with the earlier expression for $\varv_\mathrm c^2(R)=R\,\dm\Phi(R,0)/\dm R$ and $\varg(z)=\dm\Phi(0,z)/\dm z$.

\section{The extended Evans--de~Zeeuw disk families:
the beta distribution weights}

As shown in Appendix~\ref{app:mex}, all Me{\ij}er-G functions of $G^{m,n}_{2,2}$ (including $G^{1,2}_{2,2}$ in eq.~\ref{eq:hypg_m}) are reducible to ${_2F_1}$. Hence, from equations (\ref{eq:m_edz_w})--(\ref{eq:m_edz_v}), if we choose the \citetalias{EdZ92} weight to be in the form of $G^{m,n}_{1,1}$ (which are actually just geometric series; see App.~\ref{app:mex}), both $\Sigma(R)$ and $\varv_\mathrm c^2(R)$ are expressible as ${_2F_1}$. As shown in equation (\ref{eq:kmqsw}), the \citetalias{EdZ92} weight for the QKMd is given by $G^{0,1}_{1,1}$. In the next two chapters, we shall consider the remaining two cases, $G^{1,0}_{1,1}$ and  $G^{1,1}_{1,1}$ for the \citetalias{EdZ92} weight.


First suppose that the \citetalias{EdZ92} weight is given by
\begin{equation}\label{eq:eedzw}\begin{split}
\Lambda(h)&=\frac{2M_\mathrm t}{R_0}\frac{\Gamma(b)}{\Gamma(c)}
\,G^{1,0}_{1,1}\!\left(\frac{h^2}{R_0^2}\vrule
\stack{b-\frac12}{c-\frac12}\right)
\\&=\frac{2M_\mathrm t}{R_0^{2b-2}}\frac{\Gamma(b)}{\Gamma(c)}
\frac{h^{2c-1}(R_0^2-h^2)^{b-c-1}}{\Gamma(b-c)}\Theta(R_0-h)
\end{split}\end{equation}
with $b>c>0$. Then equations (\ref{eq:m_edz_w})--(\ref{eq:m_edz_v}) indicate that
\begin{align}\begin{split}
\Sigma(R)&=\frac{M_\mathrm t}{R_0^2}
\frac{\Gamma(b)}{\pi^{3/2}\Gamma(c)}
G^{2,1}_{2,2}\!\left(\frac{R^2}{R_0^2}
\vrule\stack{-\frac12,b-1}{0,c-1}\right)
\\&=\frac{\Gamma(b)\Gamma(c+\frac12)}{2\pi\Gamma(b+\frac12)\Gamma(c)}
\frac{M_\mathrm tR_0}{R^3}
{_2F_1}\!\left(\stack{\frac32,c+\frac12}{b+\frac12};
-\frac1{\varrho^2}\right);
\end{split}\label{eq:eedzs}\\\label{eq:eedzv}\begin{split}
\varv_\mathrm c^2(R)&=\frac{GM_\mathrm t}{R_0}
\frac{2\Gamma(b)}{\!\sqrt\pi\,\Gamma(c)}
G^{2,1}_{2,2}\!\left(\frac{R^2}{R_0^2}
\vrule\stack{\frac12,b-\frac12}{1,c-\frac12}\right)
\\&=\frac{GM_\mathrm t}R
{_2F_1}\!\left(\stack{\frac32,c}{b};-\frac1{\varrho^2}\right),
\end{split}\end{align}
where $\varrho=R/R_0$ with the $b=c$ limit being the Kuzmin disk. These behave like $\Sigma\sim R^{-3}$ and $\varv_\mathrm c^2\sim R^{-1}$ as $R\to\infty$. All disks in this family with $c>0$ have a finite total mass $M_\mathrm t$ with the enclosed mass profile given by 
\begin{equation*}\begin{split}
\frac{M(R)}{M_\mathrm t}&=\frac{\Gamma(b)}{\!\sqrt\pi\,\Gamma(c)}
G^{2,2}_{3,3}\!\left(\frac{R^2}{R_0^2}\vrule
\stack{1,\frac12;b}{1,c;0}\right)
\\&=1-\frac{\Gamma(b)}{\!\sqrt\pi\,\Gamma(c)}
G^{2,1}_{2,2}\!\left(\frac{R^2}{R_0^2}\vrule
\stack{\frac12,b}{0,c}\right)
\\&=1-\frac{\Gamma(b)\Gamma(c+\frac12)}{\Gamma(b+\frac12)\Gamma(c)}
\frac1\varrho{_2F_1}\!\left(\stack{\frac12,c+\frac12}{b+\frac12}
;-\frac1{\varrho^2}\right)
\end{split}\end{equation*}
With the weight in equation (\ref{eq:eedzw}), the potential of the family is written down as a real quadrature ($\varrho=R/R_0$ and $\zeta=z/R_0$);
\begin{equation}\label{eq:eedzp1}
\Phi(R,z)
=-\frac{2GM_\mathrm t}{R_0}
\frac{\Gamma(b)}{\Gamma(b-c)\Gamma(c)}\!\int_0^1\!
\frac{t^{2c-1}(1-t^2)^{b-c-1}\dm t}{\!\sqrt{\varrho^2+(|\zeta|+t)^2}}.
\end{equation}

\subsection{The dual KMT disks}

The surface density of the model with $b=1$ results in 
\begin{equation}\label{eq:dkmt}
\frac{\Sigma}{M_\mathrm t}
=\frac{\Gamma(c+\frac12)}{R_0^2\pi^{3/2}\Gamma(c)}
\frac{\varrho^{2c-2}}{(1+\varrho^2)^{c+1/2}}
\end{equation}
%
with the rotation curve and the vertical force given by
\begin{equation*}\begin{split}
\varv_\mathrm c^2(R)&=\frac{GM_\mathrm t}R
{_2F_1}\!\left(\stack{\frac32,c}{1};-\frac1{\varrho^2}\right);
\\\varg(z)
&=\frac{\Gamma(c+\frac12)}{\sqrt\pi\,\Gamma(c+2)}\frac{GM_\mathrm t}{z^2}
{_2F_1}\!\left(\stack{3,2c}{c+2};\frac{\zeta-1}{2\zeta}\right).
\end{split}\end{equation*}
%
Here the model with $0<c<b=1$ possesses a central density cusp, $\sim R^{-2(1-c)}$ as $R\to0$. If $c>b=1$ on the other hand, the weight in equation (\ref{eq:eedzw}) is not proper and equation (\ref{eq:eedzp1}) does not converge; that is to say, the surface density of equation (\ref{eq:dkmt}) with $c>1$ is not expressible as the Stieltjes transform. The corresponding disks are mostly of a formal interest given not only that $\Sigma\to0$ as $R\to0$ but also that $\varv_\mathrm c^2(R)<0$ for a sufficiently small $R$ for such models.

Nevertheless equation (\ref{eq:dkmt}) obeys the differential recursion similar to equation (\ref{eq:kmtdr}), namely $(\partial/\partial x)[R_0^{2c}\Sigma(R;c,b=1)]=-cR_0^{2c+2}\Sigma(R;c+1,b=1)$ where $x=R_0^{-2}$. Consequently the potential for any model with $b=1$ and $c>0$ may be obtained using the fractional calculus analogous to the KMTd in Sect.~\ref{sec:kmtrw}. In general, changing the integration variable in equation (\ref{eq:kzir}) via $t\to (z^2/R_0^2)t^{-1}$ leads to
\begin{equation*}
\frac1{\!\sqrt{R^2+(z+R_0)^2}}
=\frac z{\pi}\int_0^\infty\frac{t^{-1/2}\dm t}{z^2+r^2t+R_0^2t(1+t)},
\end{equation*}
and it follows that equation (\ref{eq:eedzp1}) may be recast to be
\begin{equation}\label{eq:eedzp2}
\Phi(\bm r)=-\frac{GM_\mathrm t\Gamma(b)|z|}{\pi\Gamma(c)R_0^{2(b-1)}}
\!\int_0^\infty\!\frac{\dm t}{\!\sqrt t}\,{_0^+I}_{R_0^2}^{b-c}
\biggl[\frac{R_0^{2(c-1)}}{z^2+r^2t+R_0^2t(1+t)}\biggr]
\end{equation}
using the Riemann--Liouville integral \cite{Er54}
\begin{equation*}
{_a^+I}_x^\eta\,f(x)
=\begin{cases}\displaystyle
\frac1{\Gamma(\eta)}\int_a^x\!(x-y)^{\eta-1}f(y)\,\dm y
&(\eta>0)\\f(x)&(\eta=0)\\\displaystyle
\biggl(\frac\dm{\dm x}\biggr)^{\lceil\xi\rceil}
{_a^+I}_x^{\lceil\xi\rceil-\xi}f
&(\eta=-\xi<0)\end{cases},
\end{equation*}
which extends the formula to include $b\le c$ cases. Since we find ${_0^+I}_x^{-\xi}[x^\xi/(Ax+B)]=\Gamma(\xi+1)B^\xi(Ax+B)^{-\xi-1}$, the general expression for the potential due to $\Sigma$ in equation (\ref{eq:dkmt}) from equation (\ref{eq:eedzp2}) with $b=1$ becomes
\begin{align}\label{eq:dkmtp}\begin{split}
\Phi(\bm r)&=-\frac{GM_\mathrm t|z|}\pi
\int_0^\infty\!\frac{\dm t}{\!\sqrt t}
\frac{(r^2t+z^2)^{c-1}}{[t(t+1)R_0^2+r^2t+z^2]^c}
\\&=-\frac{GM_\mathrm t}{\pi R_0}
\int_0^\infty\!\dm t\,t^{c-\frac12}
\frac{(\lambda+\mu-1+t)^{c-1}}{[(\lambda+t)(\mu+t)]^c}
\end{split}\\\nonumber&=-\frac{GM_\mathrm t\Gamma(c+\frac12)}{\!\sqrt\pi\,\Gamma(c+1)R_0}
R_{-\frac12}(c,1-c,c;\mu,\lambda+\mu-1,\lambda).
\end{align}
%
Note that the $c=0$ case results in $\Phi=-(GM_\mathrm t/R_0)(\lambda+\mu-1)^{-1/2}=-GM_\mathrm tr^{-1}$, that is, the potential of a point mass $M_\mathrm t$. Although equation (\ref{eq:dkmt}) with $c=0$ is technically unintegrable, one may instead identify it with $\lim_{c\to0}\varrho^{2c-2}/\Gamma(c)\to\deltaup(\varrho^2)=R_0^2\deltaup(R^2)$. Then the central point mass is indeed the formal limit corresponding to the $c=0$ case. If $c$ is a positive integer, the potential is expressible through elementary functions, but specific forms are easier to find through the differential recurrence relation instead; that is, for $c\in\set{1,2,3,\cdots}$,
\begin{equation*}\begin{split}
\Phi(\bm r)
&=\frac{(-1)^cGM_\mathrm t}{(c-1)!R_0^{2c}}
\left(\frac\partial{\partial(R_0^{-2})}\right)^{c-1}
\frac{R_0^2}{\!\sqrt{R^2+(|z|+R_0)^2}}
\\&=-\frac{GM_\mathrm t}{(c-1)!}
\left(\frac\partial{\partial R_0^2}\right)^{c-1}
\frac{R_0^{2(c-1)}}{\!\sqrt{R^2+(|z|+R_0)^2}}.
\end{split}\end{equation*}

For a half-integer value of $c$, the Carlson-R of equation (\ref{eq:dkmtp}) is actually an elliptic integral. If $c=\frac12$ in particular, 
\begin{align}
\Sigma(R)&=\frac{M_\mathrm tR_0}{\pi^2R(R^2+R_0^2)}\quad\Rightarrow\
\nonumber\\
\Phi(\bm r)&=-\frac{2GM_\mathrm t}{\pi R_0}R_F(\mu,\lambda+\mu-1,\lambda)
=-\frac{2GM_\mathrm t}{\pi R_0}
\frac{\mathrm F(\varphi,\tilde\kappa)}{\sqrt{\lambda-\mu}};
\nonumber\\&\left(\text{where }\ \cos^2\!\varphi=\frac\mu\lambda,\quad
\tilde\kappa^2=\frac{1-\mu}{\lambda-\mu}\right),
\label{eq:dkmte}\end{align}
which is also obtained by changing the integration variable of equation (\ref{eq:eedzp1}) similarly to equation (\ref{eq:poteltr}). Here $\varphi$ is the same as equation (\ref{eq:potg1}) but $\tilde\kappa^2=1-\kappa^2$. The specialized results for $R=0$ or $z=0$ are then found to be
\begin{align}\label{eq:rc1z}
\frac{\Phi(R,0)}{GM_\mathrm t/R_0}&=-\frac2\pi R_F(0,\varrho^2,1+\varrho^2)
=-\frac2\pi\frac{\mathbf K(\tilde\kappa_0)}{\sqrt{1+\varrho^2}}
\\\nonumber\frac{\Phi(0,z)}{GM_\mathrm t/R_0}&=-\frac2\pi R_C(1;\zeta^2)
=-\frac2\pi\begin{cases}
\dfrac{\mbox{sech}^{-1}|\zeta|}{\sqrt{1-\zeta^2}}&(|\zeta|<1)\\
1&(|\zeta|=1)\\
\dfrac{\sec^{-1}\!|\zeta|}{\sqrt{\zeta^2-1}}&(|\zeta|>1)\end{cases}
\end{align}
where $\tilde\kappa_0^2=(1+\varrho^2)^{-1}$ and $R_C(1;\zeta^2)=|\zeta|^{-1}R_C(\zeta^{-2};1)$, while
\begin{equation*}
\frac{R_0\varv_\mathrm c^2(R)}{GM_\mathrm t}=\frac2\pi
\frac{\mathbf E(\tilde\kappa_0)}{\sqrt{1+\varrho^2}};\quad
\frac{R_0^2\varg(z)}{GM_\mathrm t}=\frac2\pi
\frac{\zeta R_C(1;\zeta^2)-\zeta^{-1}}{\zeta^2-1}
\end{equation*}
with $R_0^2\varg(R_0)/(GM_\mathrm t)=4/(3\pi)$. The explicit expressions for $c\in\set{\frac32,\frac52,\frac72,\dotsc}$ may then be obtained by similarly applying the differential recursion relation on the $c=\frac12$ results; that is, $(\partial/\partial x)^m(R_0^{2c}f_c)=(-1)^m(c)_mR_0^{2(c+m)}f_{c+m}$ with $x=R_0^{-2}$ for any quantity $f_c$ linear to $\Sigma$.

\subsection{The dual QKM disks}

For arbitrary $b>c>0$, the pair of $\Sigma$ in equation (\ref{eq:eedzs}) and $\varv_\mathrm c^2$ in equation (\ref{eq:eedzv}) behaves as $R\to0$ like
\begin{equation*}
\frac{\Sigma}{M_\mathrm t/(\pi R_0^2)}\simeq\begin{cases}
\dfrac{b-1}{2(c-1)}&(c>1)\smallskip\\\displaystyle
\frac{b-1}2\,\biggl(\ln\frac4{\varrho^2}-H_{b-2}-2\biggr)
&(c=1)\smallskip\\\dfrac{\Gamma(b)\Gamma(c+\frac12)\Gamma(1-c)}
{\!\sqrt\pi\,\Gamma(b-c)\Gamma(c)\,\varrho^{2(1-c)}}&(c<1)\end{cases}
\end{equation*}
and
\begin{equation*}
\frac{\varv_\mathrm c^2}{GM_\mathrm t/R_0}\simeq\begin{cases}
\dfrac{\Gamma(b)\Gamma(c-\frac32)}{\Gamma(b-\frac32)\Gamma(c)}\,\varrho^2
&(c>\frac32)\\\displaystyle
\frac{2\Gamma(b)\,\varrho^2}{\!\sqrt\pi\,\Gamma(b-\frac32)}\,
\biggl(\ln\frac4{\varrho^2}-H_{b-\frac52}-2\biggr)
&(c=\frac32)\\
\dfrac{2\Gamma(b)\Gamma(\frac32-c)}{\!\sqrt\pi\,\Gamma(b-c)}\,\varrho^{2c-1}
&(c<\frac32)\end{cases},
\end{equation*}
where $H_z=\digamma(z+1)+\gammaup$ is the generalized Harmonic number. In other words, either $\Sigma(R)$ is cusped for $c\le1$ or $\Sigma(0)$ is constant for $c>1$, whereas $\varv_\mathrm c^2(R)\to0$ as $R\to0$ if $c>\frac12$. The $b>c=\frac12$ models possess the $R^{-1}$-cusp in $\Sigma$ and $\varv_\mathrm c^2(0)=2\Gamma(b)/[\!\sqrt\pi\,\Gamma(b-\frac12)]$ is finite. On the other hand, the vertical force resulting from the family is given by
\begin{equation*}
\varg(z)=\frac{2GM_\mathrm t}{\pi R_0^2}
\frac{\Gamma(b)}{\Gamma(c)}
G^{3,2}_{3,3}\!\left(\frac{z^2}{R_0^2}
\vrule\stack{0,-\frac12,b-1}{\frac12,0,c-1}\right)
\end{equation*}
which behaves like $\varg(z)\simeq GM_\mathrm tz^{-2}$ as $z\to\infty$, and for $z\to0$,
\begin{equation*}
\frac{\varg(z)}{GM_\mathrm t/R_0^2}\simeq\begin{cases}
\dfrac{b-1}{c-1}&(c>1)\\
(b-1)(\ln\zeta^{-2}-H_{b-2}-2)&(c=1)\\
\dfrac{2^{2c}\Gamma(c+\frac12)\Gamma(2-2c)\Gamma(b)}
{\!\sqrt\pi\,\Gamma(b-c)\,\zeta^{2(1-c)}}&(c<1)\end{cases}.
\end{equation*}
In principle, the integral of equation (\ref{eq:eedzp1}) is reducible to the Carlson-R: that is, the change of the integration variable, $t\to\tilde x$ given by $t=2(\sigma\tilde x+|\zeta|)/(\tilde x^2-1)$ leads to
\begin{align}
\frac{\Phi(\bm r)}{GM_\mathrm t/R_0}
&=-\frac{2^{2c+1}\Gamma(b)}{\Gamma(c)\Gamma(b-c)}
\!\int_{\sigma+\tau_+}^\infty\!\dm\tilde x\,
\frac{(\sigma\tilde x+|\zeta|)^{2c-1}}{(\tilde x^2-1)^{2b-2}}
\mathcal S^{b-c-1};
\nonumber\\\mathcal S&=\left[(\tilde x-\sigma)^2-\tau_+^2\right]
\left[(\tilde x+\sigma)^2-\tau_-^2\right],
\label{eq:pdqkmr}\end{align}
where $\sigma^2=\varrho^2+\zeta^2$ and $\tau_\pm^2=\varrho^2+(|\zeta|\pm1)^2$. This is reducible to elementary functions if $b-c$ and $2c$ are both integers, while it is an elliptic integral for an integer value of
$2c$ and a half-integer value of $b-c$.

The models with $b-c=1$ and $c\in\set{\frac12,1,\frac32,2,\cdots}$ may be seen as the formal dual of the Kalnajs--Mestel disk (cf.\ Sect.~\ref{sec:kmd}). Resolving the integral is then basically an exercise on the partial fraction decomposition, and in particular the potential of the $(b,c)=(\frac32,\frac12)$ case results in
\begin{equation}\label{eq:mdif}
\frac{-\Phi(\bm r)}{GM_\mathrm t/R_0}
=2\coth^{-1}(\tau_++\sigma),
\end{equation}
with the $R=0$ or $z=0$ specializations being 
\begin{equation*}
\frac{-\Phi(R,0)}{GM_\mathrm t/R_0}
=\sinh^{-1}\biggl(\frac{R_0}R\biggr);\quad
\frac{-\Phi(0,z)}{GM_\mathrm t/R_0}
=\ln\biggl(1+\frac{R_0}{|z|}\biggr).
\end{equation*}
We observe that this is identical to the potential of the Mestel-difference disk in equation (\ref{eq:pmd}) with $(c_1,c_2)=(R_0,0)$. In fact, equations (\ref{eq:eedzs}--\ref{eq:eedzv}) with $(b,c)=(\frac32,\frac12)$ result in
\begin{equation*}
\Sigma(R)=\frac{M_\mathrm t}{2\pi R_0}\left[\frac1R-\frac1{(R^2+R_0^2)^{1/2}}\right];
\quad\varv_\mathrm c^2(R)=\frac{GM_\mathrm t}{(R^2+R_0^2)^{1/2}},
\end{equation*}
whereas the vertical force profile is found to be
\begin{equation*}
\varg(z)
=\frac{GM_\mathrm t}{z(z+R_0)}=\frac{GM_\mathrm t}{R_0}\left(\frac1z-\frac1{z+R_0}\right).
\end{equation*}
That is to say, this is the difference between the scale-free and cored Mestel disks of the same asymptotic circular speed.

Similarly we also find
\begin{equation*}\begin{split}
\frac{-\Phi(\bm r)}{GM_\mathrm t/R_0}
&=2\left(\tau_+-\sigma-|\zeta\Phi^{\rm MD}|\right);
\\
\frac{-\Phi(\bm r)}{GM_\mathrm t/R_0}
&=\frac32\left[\tau_+-3|\zeta|(\tau_+-\sigma)
+(2\zeta^2-\varrho^2)|\Phi^{\rm MD}|\right];
\\
\frac{-\Phi(\bm r)}{GM_\mathrm t/R_0}
&=\frac23\left[(2-5|\zeta|)\tau_+
-(4\varrho^2-11\zeta^2)(\tau_+-\sigma)\right]
\\&\qquad
+2(3\varrho^2-2\zeta^2)|\zeta\Phi^{\rm MD}|,
\end{split}\end{equation*}
for the models with $(b,c)=(2,1)$, $(\frac52,\frac32)$ and $(3,2)$. Here $|\Phi^{\rm MD}|=2\coth^{-1}(\tau_++\sigma)$ is the rescaled equation (\ref{eq:mdif}). These are generated by the surface density profiles given by
\begin{align*}
\frac{\Sigma(R)}{M_\mathrm t/(\pi R_0^2)}&=
\sinh^{-1}\frac1\varrho-\frac1{\!\sqrt{1+\varrho^2}};
&(b,c)&=(2,1),
\\
\frac{\Sigma(R)}{M_\mathrm t/(\pi R_0^2)}&=
3\left(\frac{1+2\varrho^2}{2\!\sqrt{1+\varrho^2}}-\varrho\right);
&(b,c)&=(\tfrac52,\tfrac32),
\\
\frac{\Sigma(R)}{M_\mathrm t/(\pi R_0^2)}&=
\frac{1+3\varrho^2}{\!\sqrt{1+\varrho^2}}-3\varrho^2\sinh^{-1}\frac1\varrho;
&(b,c)&=(3,2).
\end{align*}
In general, equations (\ref{eq:eedzs}--\ref{eq:eedzv}) with $b-c=1$ result in
\begin{equation*}\begin{split}
\Sigma(R)&=\frac{M_\mathrm t}{\pi R^2}\frac{c}{2c+1}\frac1{\!\sqrt{1+\varrho^2}}
{_2F_1}\!\left(\stack{c,1}{c+\frac32};-\frac1{\varrho^2}\right);
\\\varv_\mathrm c^2(R)&=\frac{GM_\mathrm t}{(R^2+R_0^2)^{1/2}}
{_2F_1}\!\left(\stack{c-\frac12,1}{c+1};-\frac1{\varrho^2}\right),
\end{split}\end{equation*}
and $\varv_\mathrm c^2(R)$ of the model with $(b,c)=(n+1,n)$ follows the same function of $R$ as $R^2\Sigma(R)$ with $(b,c)=(n+\frac12,n-\frac12)$.

On the other hand, the simplest potential expressible with elliptic integrals is provided in equation (\ref{eq:dkmte}) with $(b,c)=(1,\frac12)$, for which $\Sigma(R)$ is elementary. In general, if $b$ is an integer but $c$ is a half-integer, then $\Sigma(R)$ is an elementary function (which may involve $\cot^{-1}\varrho$) but $\varv_\mathrm c^2(R)$ would require elliptic integrals to write down. By contrast, if $c$ is an integer but $b$ is a half-integer, $\varv_\mathrm c^2(R)$ is an elementary function but $\Sigma(R)$ is an elliptic integral; for example, if $(b,c)=(\frac32,1)$,
\begin{equation*}\begin{split}
\Sigma(R)&
=\frac{M_\mathrm t}{2\pi R_0^2}\frac{\mathbf K(\tilde\kappa_0)-\mathbf E(\tilde\kappa_0)}
{\!\sqrt{1+\varrho^2}};
\\\varv_\mathrm c^2(R)&=\frac{GM_\mathrm tR^2}{R^2+R_0^2}
=\frac{GM_\mathrm t}{R_0}\!\frac\varrho{1+\varrho^2},
\\
\varg(z)&=\frac{GM_\mathrm t}{R_0^2}\frac{R_C(1;\zeta^2)-1}{1-\zeta^2}\
\left(\stackrel{z=R_0}\longrightarrow\frac{GM_\mathrm t}{3R_0^2}\right).
\end{split}\end{equation*}
Although equation (\ref{eq:pdqkmr}) is in the form of the Carlson-R (and so it should be straightforward to express this with Carlson's symmetric bases if it is indeed an elliptic integral),
the final expressions are easier to read if we introduce the elliptic coordinates $(\lambda,\mu)$ again and consider the change of variable, $t=[\!\sqrt{\lambda(y+\mu)}-\!\sqrt{\mu(y+\lambda)}]/[\!\sqrt{\lambda(y+\lambda)}-\!\sqrt{\mu(y+\mu)}]$ for equation (\ref{eq:eedzp1}), which results in
\begin{equation*}
\frac{-\Phi(R,z)}{GM_\mathrm t/R_0}
=\frac{\Gamma(b)}{\Gamma(b-c)\Gamma(c)}\!\int_0^\infty\!
\frac{\mathcal F^{2c-1}\mathcal G^{2-2b}\dm y}
{\!\sqrt{(y+\mu)(y+\sigma^2)(y+\lambda)}}
\end{equation*}
where $\mathcal F$ and $\mathcal G$ are the same as in equation (\ref{eq:poteltr}). Also note that $0\le\mu\le\sigma^2=\lambda+\mu-1\le\lambda$. If $(b,c)=(1,\frac12)$, this reproduces to the potential in equation (\ref{eq:dkmte}). For the $(b,c)=(\frac32,1)$ case, it is straightforward to show that
\begin{equation*}\begin{split}
\frac{-\Phi(\bm r)}{GM_\mathrm t/R_0}
&=\cot^{-1}\sigma-\frac{|\zeta|}3R_J(\mu,\sigma^2,\lambda;\lambda+\mu)
\\&=\frac\pi2-|\zeta|\left[R_F(\mu,\sigma^2,\lambda)+
\frac{\varrho^2}3R_J(\mu,\sigma^2,\lambda;\lambda\mu)\right]
\\&=\cot^{-1}\sigma+\frac{\!\sqrt\lambda}{\!\sqrt\mu}
\frac{\Pi(\varphi,\tilde\varpi^2,\tilde\kappa)-\mathrm F(\varphi,\tilde\kappa)}
{\!\sqrt{\lambda-\mu}}
\\&=\frac\pi2-\frac{\!\sqrt\mu}{\!\sqrt\lambda}
\frac{(\lambda-1)\Pi(\varphi,\hat\varpi^2,\tilde\kappa)
+\mathrm F(\varphi,\tilde\kappa)}{\!\sqrt{\lambda-\mu}}
\\
&=\frac\pi2+\tan^{-1}\!\frac{1-\mu}\sigma-\frac{\!\sqrt\lambda}{\!\sqrt\mu}
\frac{\Pi(\varphi,\hat\omega^2,\tilde\kappa)}{\!\sqrt{\lambda-\mu}}
\end{split}\end{equation*}
where $(\varphi,\tilde\kappa)$ is as defined in equation (\ref{eq:dkmte}) and
\begin{displaymath}
\tilde\varpi^2=-\frac\mu{\lambda-\mu},\quad
\hat\varpi^2=\frac{\lambda(1-\mu)}{\lambda-\mu},\quad
\hat\omega^2=-\frac{1-\mu}\mu.
\end{displaymath}
Here $\tilde\varpi^2$ is the same as equation (\ref{eq:alpp}) but the transformed parameters $\hat\varpi^2$ and $\hat\omega^2$ are not. If $R=0$ or $z=0$, 
\begin{equation*}
\frac{-\Phi(R,0)}{GM_\mathrm t/R_0}=\cot^{-1}\varrho;\quad
\frac{-\Phi(0,z)}{GM_\mathrm t/R_0}=\frac\pi2-|\zeta|\,R_C(1;\zeta^2),
\end{equation*}
which are consistent with both the rotation curve and the vertical force in the upper half-space.

\section{The semi-infinite beta-prime weights}

Suppose that the \citetalias{EdZ92} weight is given by
\begin{equation*}
\Lambda(h)
=\frac{\Lambda_0}{\Gamma(\beta+\gamma-1)}\,
G^{1,1}_{1,1}\!\left(\frac{h^2}{R_0^2}
\vrule\stack{\frac32-\gamma}{\beta-\frac12}\right)
=\frac{\Lambda_0R_0^{2\gamma-1}h^{2\beta-1}}
{(h^2+R_0^2)^{\beta+\gamma-1}}.
\end{equation*}
Equation (\ref{eq:m_edz_s}) indicates the resulting surface density to be
\begin{equation}\label{eq:g2den}\begin{split}
\frac{\Sigma(R)}{\Lambda_0/R_0}&
=\frac1{2\pi^{3/2}\Gamma(\beta+\gamma-1)}\,
G^{2,2}_{2,2}\!\left(\frac{R^2}{R_0^2}
\vrule\stack{-\frac12,1-\gamma}{0,\beta-1}\right)
\\&=\frac{\Gamma(\beta+\frac12)\Gamma(\gamma)}
{4\pi\Gamma(\beta+\gamma+\frac12)}\,
{_2F_1}\!\left(\stack{\frac32,\gamma}{\beta+\gamma+\frac12}
;{1-\varrho^2}\right),
\end{split}\end{equation}
where $\varrho=R/R_0$, with the limiting behavior as $R\to0$:
\begin{equation*}
\frac{\Sigma(R)}{\Lambda_0/R_0}\simeq\begin{cases}
\dfrac{\Gamma(\beta-1)\Gamma(\gamma)}
{4\pi\Gamma(\beta+\gamma-1)}&(\beta>1)
\smallskip\\\dfrac1{4\pi}\,\biggl(
\ln\dfrac4{\varrho^2}-H_{\gamma-1}-2\biggr)
&(\beta=1)\smallskip\\\dfrac{\Gamma(\beta+\frac12)\Gamma(1-\beta)}
{2\pi^{3/2}}\dfrac1{\varrho^{2(1-\beta)}}&(\beta<1)\end{cases},
\end{equation*}
and that as $R\to\infty$:
\begin{equation*}
\frac{\Sigma(R)}{\Lambda_0/R_0}
\simeq\begin{cases}
\dfrac{\Gamma(\beta+\frac12)\Gamma(\gamma-\frac32)}
{4\pi\Gamma(\beta+\gamma-1)}\dfrac1{\varrho^3}
&(\gamma>\frac32)\smallskip\\\dfrac1{4\pi}\,\biggl[
\ln(4\varrho^2)-H_{\beta-\frac32}-2\biggr]\,\dfrac1{\varrho^3}
&(\gamma=\frac32)\\\dfrac{\Gamma(\gamma)\Gamma(\frac32-\gamma)}
{2\pi^{3/2}}\dfrac1{\varrho^{2\gamma}}&(\gamma<\frac32)\end{cases}.
\end{equation*}
Thus it is necessary $\beta>0$ for $\Sigma(R)$ to be integrable at the center -- note $\Sigma(0)$ is finite if $\beta>1$ -- and $\gamma>0$ for it to be non-negative. Next the mass profile 
follows as
\begin{equation*}
\frac{M(R)}{\Lambda_0R_0}
=\frac1{2\!\sqrt\pi\,\Gamma(\beta+\gamma-1)}\,
G^{2,3}_{3,3}\!\left(\varrho^2
\vrule\stack{1,\frac12,2-\gamma}{1,\beta,0}\right),
\end{equation*}
which grows near $R\sim0$ like
\begin{equation*}
\frac{M(R)}{\Lambda_0R_0}\simeq\begin{cases}
\dfrac{\Gamma(\beta-1)\Gamma(\gamma)}{4\Gamma(\beta+\gamma-1)}\,\varrho^2&(\beta>1)\smallskip\\\dfrac14\,\biggl(
\ln\dfrac4{\varrho^2}-H_{\gamma-1}-1\biggr)\,\varrho^2&(\beta=1)
\\\dfrac{\Gamma(\frac12+\beta)\Gamma(1-\beta)}{2\!\sqrt\pi\,\beta}\,
\varrho^{2\beta}&(\beta<1)\end{cases}.
\end{equation*}
On the other hand, as $R\to\infty$,
\begin{equation}\label{eq:g2ml}
\frac{M(R)}{\Lambda_0R_0}\simeq\begin{cases}
\dfrac{\Gamma(\beta)\Gamma(\gamma-1)}
{2\Gamma(\beta+\gamma-1)}
&(\gamma>1)\smallskip\\
\ln\dfrac{\varrho}2-\dfrac{H_{\beta-1}}2
&(\gamma=1)\\
\dfrac{\Gamma(\gamma)\Gamma(\frac32-\gamma)}
{2(1-\gamma)\!\sqrt\pi}\,\varrho^{2(1-\gamma)}&(\gamma<1)\end{cases}.
\end{equation}
Since $\Sigma(R)$ for $\gamma>1$ decays faster than $\sim R^{-2}$ as $R\to\infty$, the corresponding disk has a finite total mass. The mass profiles of these finite mass cases are also expressible to be
\begin{align*}
\frac{M(R)}{M_\mathrm t}
&=1-\frac1{\!\sqrt\pi\,\Gamma(\beta)\Gamma(\gamma-1)}
G^{2,2}_{2,2}\!\left(\varrho^2\vrule
\stack{\frac12,2-\gamma}{0,\beta}\right)
\\&=1-\frac{\Gamma(\beta+\frac12)\Gamma(\beta+\gamma-1)}
{\Gamma(\beta)\Gamma(\beta+\gamma-\frac12)}
{_2F_1}\!\left(\stack{\frac12,\gamma-1}{\beta+\gamma-\frac12}
;{1-\varrho^2}\right),
\end{align*}
where $M_\mathrm t$ is the total mass of the disk whose numerical value is inferable from the $\gamma>1$ case of equation (\ref{eq:g2ml}).

The circular speed of these disks follows equation (\ref{eq:m_edz_v});
\begin{equation}\label{eq:g2rot}\begin{split}
\frac{\varv_\mathrm c^2(R)}{G\Lambda_0}
&=\frac1{\!\sqrt\pi}\,G^{2,2}_{2,2}\!\left(\frac{R^2}{R_0^2}
\vrule\stack{\frac12,\frac32-\gamma}{1,\beta-\frac12}\right)
\\&=\frac{\Gamma(\beta)\Gamma(\gamma+\frac12)}
{2\Gamma(\beta+\gamma+\frac12)}
\frac1\varrho\,{_2F_1}\!\left(\stack{\frac32,\beta}
{\beta+\gamma+\frac12};1-\frac1{\varrho^2}\right).
\end{split}\end{equation}
which behaves asymptotically as $R\to\infty$ like
\begin{equation*}
\frac{\varv_\mathrm c^2(R)}{G\Lambda_0}=\begin{cases}
\dfrac{\Gamma(\beta)\Gamma(\gamma-1)}
{2\Gamma(\beta+\gamma-1)}\dfrac1\varrho
&(\gamma>1)\smallskip\\\dfrac12\,\biggl[
\ln(4\varrho^2)-H_{\beta-1}-2\biggr]\,\dfrac1\varrho
&(\gamma=1)\\
\dfrac{\Gamma(\gamma+\frac12)\Gamma(1-\gamma)}
{\!\sqrt\pi}\dfrac1{\varrho^{2\gamma-1}}
&(\gamma<1)\end{cases}.
\end{equation*}
Obviously, the rotation curve of a finite mass model approaches the Keplerian model, $\varv_\mathrm c^2\simeq GM_\mathrm t/R$ as $R\to\infty$. By contrast, the $\gamma=\frac12$ model exhibits an asymptotically flat rotation curve with $\varv_\mathrm c^2\to\varv_\infty^2=G\Lambda_0$. As for the behavior as $R\to0$, we find $\varv_\mathrm c\to0$ if $\beta>1$. Specifically,
\begin{equation*}
\frac{\varv_\mathrm c^2(R)}{G\Lambda_0}=\begin{cases}
\dfrac{\Gamma(\beta-\frac32)\Gamma(\gamma+\frac12)}
{2\Gamma(\beta+\gamma-1)}\,\varrho^2
&(\beta>\frac32)\smallskip\\\dfrac12\,\biggl(
\ln\dfrac4{\varrho^2}-H_{\gamma-\frac12}-2\biggr)
\,\varrho^2&(\beta=\frac32)\\
\dfrac{\Gamma(\beta)\Gamma(\frac32-\beta)}{\!\sqrt\pi}
\,\varrho^{2\beta-1}&(\beta<\frac32)\end{cases}.
\end{equation*}

From equation (\ref{eq:genpot}), the potential due to this disk
family is written as an integral quadrature
($\varrho=R/R_0$ and $\zeta=z/R_0$):
\begin{equation}\label{eq:g2pot}
\frac{\Phi(\bm r)}{G\Lambda_0}
=-\int_0^\infty\frac{t^{2\beta-1}}{(1+t^2)^{\beta+\gamma-1}}
\frac{\dm t}{\!\sqrt{\varrho^2+(|\zeta|+t)^2}},
\end{equation}
which converges if $\beta>0$ and $\gamma>\frac12$.
Alternatively, provided $\beta>\frac12$ and $\gamma>0$,
the zero-point-shifted potential for the model with a non-decaying
rotation curve may be set up as
\begin{equation}\label{eq:g2potd}
\frac{\Phi(\bm r)}{G\Lambda_0}
=\int_0^\infty\!\dm t\,\frac{t^{2\beta-2}}
{(t^2+1)^{\beta+\gamma-1}}
\frac{\sqrt{\varrho^2+(|\zeta|+t)^2}-t}{\!\sqrt{\varrho^2+(|\zeta|+t)^2}}.
\end{equation}

\subsection{Scale-free disks}

If $\beta+\gamma=1$, equation (\ref{eq:g2den}) results in
\begin{equation*}
\frac{\Sigma(R)}{\Lambda_0/R_0}
=\frac{\Gamma(\gamma)\Gamma(\frac32-\gamma)}{2\pi^{3/2}\varrho^{2\gamma}}
;\quad\frac{M(R)}{\Lambda_0R_0}
=\frac{\Gamma(\gamma)\Gamma(\frac32-\gamma)}{2\!\sqrt\pi}
\frac{\varrho^{2(1-\gamma)}}{1-\gamma},
\end{equation*}
which follow strict power-laws (also $\Lambda\propto\varrho^{1-2\gamma}$ and $\varv_\mathrm c^2\propto\varrho^{1-2\gamma}$). In other words, the potential of the scale-free disk, $\Sigma(R)\propto R^{-2\gamma}$ (with $0<\gamma<1$) may be examined through the current model with $\beta=1-\gamma$. Let us consider the scale-free disk with the enclosed mass growing with the radius $R$ like
\begin{equation*}
\frac{M(R)}{M_0}=\biggl(\frac R{R_0}\biggr)^{2-2\gamma}=\varrho^{2(1-\gamma)}
\end{equation*}
where $R_0$ is a fiducial length and $M_0=M(R_0)$ is the scale factor, which relates to the constant $\Lambda_0$ via $2\!\sqrt\pi\,(1-\gamma)M_0=\Lambda_0R_0\Gamma(\gamma)\Gamma(\frac32-\gamma)$. The corresponding surface density is then extended to include the $\gamma=1$ model (i.e.\ the point mass): 
\begin{align*}
\Sigma(R)&=\frac1{2\pi R}\frac{\dm M(R)}{\dm R}
=\frac{M_0}{\pi R_0^2}\frac{1-\gamma}{\varrho^{2\gamma}}
&(\gamma&<1);
\\\Sigma(R)&=\frac{M_0}{\pi R_0^2}\deltaup(\varrho^2)
=\frac{M_0}\pi\deltaup(R^2)=\frac{M_0}\pi\frac{\deltaup(R)}{2R}
&(\gamma&=1).
\end{align*}

Equation (\ref{eq:g2rot}) indicates that the rotation curve is like
\begin{equation*}
\varv_\mathrm c^2(R)
=C_\gamma\frac{GM_0}{R_0}\frac1{\varrho^{2\gamma-1}};\quad
C_\gamma=\frac{2\Gamma(\gamma+\frac12)\Gamma(2-\gamma)}
{\Gamma(\gamma)\Gamma(\frac32-\gamma)},
\end{equation*}
while the direct integration (eq.~\ref{eq:gf}) results in
\begin{equation}\label{eq:ssvert}
\varg(z)=D_\gamma\frac{GM_0}{R_0^2}\frac1{\zeta^\gamma};\quad
D_\gamma
=\frac{2\Gamma(\gamma+\frac12)\Gamma(2-\gamma)}{\!\sqrt\pi}.
\end{equation}
Both results are valid for $-\frac12<\gamma\le1$. If $\frac12<\gamma<1$, then $1<C_\gamma\le C_{\frac34}(\approx1.09422)$ and $D_{\frac34}(\approx0.927037)\le D_\gamma<1$, while $C_\gamma=D_\gamma=1$ for $\gamma=\frac12$ or $\gamma=1$, and $0<C_\gamma<1$ and $1<D_\gamma<2$ for $0<\gamma<\frac12$.
The $\gamma=1$ case corresponds to the point mass $M_0$ at the center, and the results are simply the Kepler rotation curve, $\varv_\mathrm c^2=GM_0/R$ and the inverse square force $\varg=GM_0/z^2$. The $\gamma=\frac12$ case ($\Sigma\propto R^{-1}$), for which $\varv_\mathrm c^2=GM(R)/R=GM_0/R_0=\varv_0^2$ and $\varg=\varv_0^2/z$, is the scale-free Mestel disk (see Sect.~\ref{sec:mestel}), which is essentially the disk analogue of a singular isothermal sphere. The $\gamma=0$ case (for which $C_\gamma=0$ and $D_\gamma=2$) is an infinite uniform disk: there should be no net gravitational force within the disk thanks to the translation symmetry and the vertical force is independent of the height \citep[\S~22]{Ro92}. If $-\frac12<\gamma<0$, we find $C_\gamma<0$, which indicates that the total gravity due to such disks would act outward in the centrifugal direction. If $\gamma\le-\frac12$, the force due to matters at infinity dominates, and so all the integrals accounting for the gravitational force diverge. Although the formal mathematical treatment is still possible, an infinite disk is an artificial construction and so this is not in fact a physical problem and shall not be discussed further.

The potential generated by the scale-free disk with $\Sigma\propto R^{-2\gamma}$, according to equation (\ref{eq:g2pot}), is given by
\begin{equation}\label{eq:sspoti}
\frac{-\Phi(\bm r)}{G\Lambda_0}
=\int_0^\infty\frac{t^{1-\gamma}\dm t}{\!\sqrt{\varrho^2+(|\zeta|+t)^2}}
=\int_0^\infty\!\dm x\frac{x^{1-\gamma}(x+\sigma)^{1-\gamma}}
{(x+x_0)^{2-\gamma}},
\end{equation}
where $2x_0=\sigma+|\zeta|$ and $\sigma^2=\varrho^2+\zeta^2$, which converges if $\frac12<\gamma<1$. The second expression follows changing the variable; $t=x(x+\sigma)/(x+x_0)$ and $t+|\zeta|=x+x_0-\varrho^2/[4(x+x_0)]$. The integral in equation (\ref{eq:sspoti}) then results in
\begin{multline}\label{eq:sspoth}
\frac{\Phi(\bm r)}{GM_0/R_0}=-\frac{D_\gamma}{2\gamma-1}
R_{1-2\gamma}\!\left(2-2\gamma,2\gamma-1;\frac{|\zeta|+\sigma}2,\sigma\right)
\\=-\frac{D_\gamma}{2\gamma-1}\frac1{\sigma^{2\gamma-1}}
{_2F_1}\!\left(\stack{2\gamma-1,2-2\gamma}1;\frac{\sigma-|\zeta|}{2\sigma}\right),
\end{multline}
where $D_\gamma$ is defined in equation (\ref{eq:ssvert}). Here the ${_2F_1}$-function is equivalent to the Legendre-P function, $P_{1-2\gamma}(x)=P_{2\gamma-2}(x)={_2F_1}(2\gamma-1,2-2\gamma;1;y)$ with $2y=1-x$, and so
\begin{equation}\label{eq:sspot}
\Phi(\bm r)=-\frac{\Gamma(\gamma-\frac12)\Gamma(2-\gamma)}{\!\sqrt\pi}
\frac{GM_0}{R_0}
\frac{P_{1-2\gamma}(|\cos\theta|)}{(r/R_0)^{2\gamma-1}},
\end{equation}
where $\cos\theta=z/r=\zeta/\sigma$ (i.e.\ $\theta$ is the usual zenith angle in the spherical polar coordinate). Furthermore,
\begin{equation*}
\Phi(R,z=0)=-\frac{\varv_\mathrm c^2(R)}{2\gamma-1};\quad
\Phi(R=0,z)=-\frac{z\varg(z)}{2\gamma-1},
\end{equation*}
given $P_{1-2\gamma}(1)=1$ and $\Gamma(\gamma)\Gamma(\frac32-\gamma)P_{1-2\gamma}(0)=\!\sqrt\pi$. Let us observe that $f=-Ar^\eta P_\eta(\cos\theta)$ with a fixed $\eta$ is the axisymmetric solution of the Laplace equation $\nabla^2\Phi=0$, obtainable via the separation of variables in the spherical polar coordinate $(r,\theta,\phi)$. Unlike the usual spherical harmonics basis, we concern the behavior of the solution limited in the upper half-space and so $\eta$ need not necessarily be an integer, for the singular behavior at $\theta=\frac\pi2$ is irrelevant. In other words, equation (\ref{eq:sspot}) satisfies the Laplace equation for $z\ne0$, and so corresponds to the potential due to an infinitesimally thin disk.

Equations (\ref{eq:sspoth}-\ref{eq:sspot}) are actually well-defined for any $\gamma\le1$ unless $\frac12-\gamma$ is a non-negative integer. In fact, they correspond to the true potential of the scale-free disk, $\Sigma\propto R^{-2\gamma}$ with $-\frac12<\gamma\le1$ and $\gamma\ne\frac12$, for they satisfy the Poisson equation. The zero-point of these expressions for $-\frac12<\gamma<\frac12$ is at the origin such that $\Phi(R=0,z=0)=0$. It is also easy to observe that equation (\ref{eq:sspot}) for $\gamma=1$ (cf.\ $P_0(x)=P_{-1}(x)=1$) reduces to the Kepler potential $\Phi=-GM_0/r$, while the $\gamma=0$ case (cf.\ $P_1(x)=R_{-2}(x)=x$) results in $\Phi=2\pi G\Sigma_0|z|$ where $\Sigma_0=M/(\pi R_0^2)$ is the constant surface density of the infinite uniform plate. The translation symmetry implies that the potential due to the uniform infinite plane must be a function of the height alone. However since $\nabla^2\Phi(z)=\Phi''(z)$, the only harmonic function $\Phi(z)$ of the height alone is the linear function of $z$. If $\gamma=\frac12$, equation (\ref{eq:sspot}) is technically an infinite constant, which is invalid. However the $\gamma=\frac12$ case is the scale-free Mestel disk, and its potential is already provided in equation (\ref{eq:mestel}), which follows equation (\ref{eq:sspoti}) via the subtraction of the infinite offset (see eq.~\ref{eq:potshi}).
%
%

\subsection{Algebraic cases with the oblate spheroidal coordinates}

The same change of the variable, $t=x(x+\sigma)/(x+x_0)$ with $2x_0=\sigma+|\zeta|$ transforms equation (\ref{eq:g2pot}) to
\begin{equation}\label{eq:g2pottr}
\frac{-\Phi(\bm r)}{G\Lambda_0}
=\int_0^\infty\!\dm x
\frac{[x(x+\sigma)]^{2\beta-1}(x+x_0)^{2\gamma-2}}
{[x^2(x+\sigma)^2+(x+x_0)^2]^{\beta+\gamma-1}}.
\end{equation}
If both $2\beta$ and $\beta+\gamma$ are integers, the integrand is a rational function of $x$ and the integral is resolved into an elementary algebraic function involving up to logarithms or inverse trigonometric/hyperbolic functions. The reduction is achieved via the integration by parts and the partial fraction decomposition utilizing the factorization of the quartic as in
\begin{multline*}
x^2(x+\sigma)^2+(x+x_0)^2
\\=[(x+x_0-\xi^{1/2}n_+)^2+n_+^2][(x+x_0+\xi^{1/2}n_-)^2+n_-^2].
\end{multline*}
Here $2n_\pm=1\pm\chi^{1/2}$, and $(\xi,\chi)=(\alpha-1,1-\delta)$ is the \emph{oblate} spheroidal coordinate component\footnote{This definition, consistent with the notation of \cite{EdZ92}, differs from \cite[sect.~III.A]{fdisk} in the square roots for $\alpha$ and $\delta$ as well as the overall scaling by $R_0$. This choice makes clear the manifest correspondence to the prolate spheroidal coordinate defined in Sect.~\ref{sec:ecd}.} defined such that
\begin{equation}\label{eq:oedad}
(\alpha^{1/2},\delta^{1/2})
=\left(\frac{\upsilon_++\upsilon_-}2,\frac{\upsilon_+-\upsilon_-}2\right),
\end{equation}
where $\upsilon_\pm^2=(\varrho\pm1)^2+\zeta^2$.
Also note that $(\alpha\delta,\xi\chi)=(\varrho^2,\zeta^2)$ and $\sigma^2=\varrho^2+\zeta^2=\alpha+\delta-1=\xi-\chi+1$, and $\xi+\chi=\alpha-\delta=\upsilon_+\upsilon_-$.

For instance, consider the model with $(\beta,\gamma)=(\frac12,\frac32)$:
\begin{equation}\label{eq:g2b1g3}\begin{split}
\frac{\Sigma(R)}{\Lambda_0/R_0}
&=\frac1{6\pi}\,{_2F_1}\!\left(\stack{\frac32,\frac32}{\frac52}
;{1-\varrho^2}\right)
=\frac{1-\varrho R_C(\varrho^2;1)}{2\pi\varrho(1-\varrho^2)};
\\\frac{\varv^2(R)}{G\Lambda_0}
&=\frac2{3\varrho}\,{_2F_1}\!\left(\stack{\frac32,\frac12}{\frac52};
1-\frac1{\varrho^2}\right)
=\frac{1-\varrho^2R_C(1;\varrho^2)}{1-\varrho^2},
\end{split}\end{equation}
where $R_C(\varrho^2;1)=(1-\varrho^2)^{-1/2}\cos^{-1}\varrho=(\varrho^2-1)^{-1/2}\cosh^{-1}\varrho$ and $R_C(1;\varrho^2)=(1-\varrho^2)^{-1/2}\mbox{sech}^{-1}\varrho=(\varrho^2-1)^{-1/2}\sec^{-1}\varrho$ (see also eqs.~\ref{eq:rcz1} and \ref{eq:rc1z}). The mass profile grows like $M(R)/M_\mathrm t=1-(2/\pi)R_C(\varrho^2;1)$ with the finite total mass being $M_\mathrm t=\Lambda_0R_0\pi/2$. The equation (\ref{eq:g2pottr}) is then reducible to
\begin{multline*}
\frac{-\Phi(\bm r)}{G\Lambda_0}
=\int_{x_0}^\infty\!\frac{x\,\dm x}
{[(x-\xi^{1/2}n_+)^2+n_+^2][(x+\xi^{1/2}n_-)^2+n_-^2]}
\\=\frac1{\xi+\chi}\!\int_{x_0}^\infty\!\left[
\frac{\chi^{1/2}x-2\xi^{1/2}n_-^2}{(x+\xi^{1/2}n_-)^2+n_-^2}
-\frac{\chi^{1/2}x-2\xi^{1/2}n_+^2}{(x-\xi^{1/2}n_+)^2+n_+^2}\right]\dm x
\end{multline*}
which is analytically integrable given that $\int(x^2+1)^{-1}\dm x=\tan^{-1}x$ and $\int x(x^2+1)^{-1}\dm x=\ln\!\sqrt{x^2+1}$. After some simplifications, the potential due to this disk is written down as
\begin{equation}\label{eq:g2potb1g3}\begin{split}
\frac{-\Phi(\bm r)}{G\Lambda_0}
&=\frac1{\xi+\chi}\left[
\xi^{1/2}\cos^{-1}\biggl(\frac\delta{\sigma+|\zeta|}\biggr)
+\chi^{1/2}\ln\Omega\right];
\\\Omega&=\frac{\sigma+|\zeta|}{\alpha+\xi^{1/2}\sigma-\chi^{1/2}}
=\frac{\alpha-\xi^{1/2}\sigma+\chi^{1/2}}{\sigma+\zeta}.
\end{split}\end{equation}
Here note $\delta/(\sigma+|\zeta|)=(\sigma-|\zeta|)/\alpha$. Specializing for $z=0$, we have $\alpha=\max(1,\varrho^2)$ and $\delta=\min(1,\varrho^2)$, and so $(\xi,\chi)=(\varrho^2-1,0)$ for $R\ge R_0$ or $(\xi,\chi)=(0,1-\varrho^2)$ for $R\le R_0$. Hence
\begin{equation*}
\frac{-\Phi(R,0)}{GM_t/R_0}
=\frac2\pi\,R_C(1;\varrho^2)
=\frac2\pi\frac1{\!\sqrt{1-\varrho^2}}
\ln\biggl(\frac{1+\!\sqrt{1-\varrho^2}}\varrho\biggr),
\end{equation*}
which is consistent with the circular speed in equation (\ref{eq:g2b1g3}). As for the vertical force along $R=0$, we have $(\alpha,\delta)=(1+\zeta^2,0)$ and $(\xi,\chi)=(\zeta^2,1)$ and so follows that
\begin{equation*}\begin{split}
\frac{\Phi(0,z)}{GM_\mathrm t/R_0}&=-
\frac1{1+\zeta^2}\left(\zeta-\frac2\pi\ln\zeta\right);
\\\frac{\varg(z)}{GM_\mathrm t/R_0^2}&=
\frac1{(1+\zeta^2)^2}\left[\zeta^2-1
+\frac2\pi\biggl(\zeta-2\zeta\ln\zeta+\frac1\zeta\biggr)\right].
\end{split}\end{equation*}

The surface density for $(\beta,\gamma)=(1,1)$ reduces to
\begin{equation*}
\frac{\Sigma(R)}{\Lambda_0/R_0}=\frac1{6\pi}\,
{_2F_1}\!\left(\stack{\frac32,1}{\frac52};{1-\varrho^2}\right)
=\frac{R_C(1;\varrho^2)-1}{2\pi(1-\varrho^2)},
\end{equation*}
and the mass profile grows like
\begin{multline*}
\frac{M(R)}{\Lambda_0R_0}=R_C(1;\varrho^2)+\ln\frac\varrho2
\\=\frac1{\!\sqrt{1-\varrho^2}}\left[
\ln\biggl(\frac{1+\!\sqrt{1-\varrho^2}}2\biggr)
-\frac{\varrho^2\ln(\varrho/2)}{1+\!\sqrt{1-\varrho^2}}\right],
\end{multline*}
which diverges logarithmically as $R\to\infty$. The potential is found from similar calculations utilizing the partial fraction decompositions, which reduces equation (\ref{eq:g2pottr}) to
%
\begin{equation*}
\frac{-\Phi(\bm r)}{G\Lambda_0}=\frac1{\xi+\chi}
\left[\chi^{1/2}\cos^{-1}\biggl(\frac\delta{\sigma+\zeta}\biggr)
-\xi^{1/2}\ln\Omega\right],
\end{equation*}
where $\Omega$ is same as equation (\ref{eq:g2potb1g3}). Specializing for $z=0$,
\begin{equation*}\begin{split}
\frac{-\Phi(R,0)}{G\Lambda_0}&=R_C(\varrho^2;1)
=\frac{\ln(\varrho+\!\sqrt{\varrho^2-1})}{\!\sqrt{\varrho^2-1}};
\\\frac{\varv_\mathrm c^2(R)}{G\Lambda_0}
&=\frac{\varrho[1-\varrho R_C(\varrho^2;1)]}{1-\varrho^2},
\end{split}\end{equation*}
which is consistent with equation (\ref{eq:g2rot}). As for $R=0$ cases,
\begin{equation*}\begin{split}
\frac{-\Phi(0,z)}{G\Lambda_0}
&=\frac1{1+\zeta^2}\left(\frac\pi2+\zeta\ln\zeta\right);
\\\frac{\varg(z)}{G\Lambda_0/R_0}
&=\frac{(\zeta^2-1)\ln\zeta-1+\pi\zeta-\zeta^2}{(1+\zeta^2)^2}.
\end{split}\end{equation*}

If $(\beta,\gamma)=(\frac32,\frac12)$, the surface density becomes
\begin{equation}\label{eq:g2b3g1}
\frac{\Sigma(R)}{\Lambda_0/R_0}
=\frac1{3\pi}\,{_2F_1}\!\left(\stack{\frac32,\frac12}{\frac52}
;{1-\varrho^2}\right)
=\frac{R_C(\varrho^2;1)-\varrho}{2\pi(1-\varrho^2)},
\end{equation}
which implies a linear asymptotic growth of the mass:
\begin{equation*}
\frac{M(R)}{\Lambda_0R_0}
=\varrho+R_C(\varrho^2;1)-\frac\pi2.
\end{equation*}
Although equations (\ref{eq:sspoti}) and (\ref{eq:g2pottr}) do not converge, the zero point shifted potential is found from equation (\ref{eq:g2potd}), which is transformed by the variable change $t=x(x+\sigma)/(x+x_0)$ to
%
\begin{equation}\label{eq:g2int}
\frac{\Phi(\bm r)}{G\Lambda_0}
=\int_0^\infty\left[|\zeta|+\frac{\varrho^2}{2(x+x_0)}\right]
\frac{x(x+\sigma)\,\dm x}{x^2(x+\sigma)^2+(x+x_0)^2}.
\end{equation}
In principle, this integral is also resolved into an elementary function. However the sum of the surface densities in equations (\ref{eq:g2b1g3}) and (\ref{eq:g2b3g1}) with the same scale constants is simply $(2\pi\varrho)^{-1}(\Lambda_0/R_0)$. In other words, the disk model with $(\beta,\gamma)=(\frac32,\frac12)$ is equivalent to the Mestel disk minus the $(\beta,\gamma)=(\frac12,\frac32)$ case with the matching scales. Since the potential of the Mestel disk, $\Sigma(R)=(2\pi\varrho)^{-1}(\Lambda_0/R_0)$ is $\Phi(\bm r)=G\Lambda_0\ln(\sigma+\zeta)$ up to an additive constant, the potential due to the disk model with $(\beta,\gamma)=(\frac32,\frac12)$, assuming $\Phi(0,0)=0$, is given by
\begin{equation*}
\frac{\Phi(\bm r)}{G\Lambda_0}
=\ln\biggl(\frac{\sigma+|\zeta|}2\biggr)+\frac1{\xi+\chi}\left[
\xi^{1/2}\cos^{-1}\biggl(\frac\delta{\sigma+|\zeta|}\biggr)
+\chi^{1/2}\ln\Omega\right],
\end{equation*}
which is also recovered from equation (\ref{eq:g2int}). The zero-point of the potential is verified for the specialized results:
\begin{equation*}
\frac{\Phi(R,0)}{G\Lambda_0}=R_C(1;\varrho^2)+\ln\frac\varrho2
;\quad\frac{\Phi(0,z)}{G\Lambda_0}
=\frac{\zeta(\pi+2\zeta\ln\zeta)}{2(1+\zeta^2)}.
\end{equation*}
Here the behavior of the potential in the disk has the identical functional dependence as the mass profile of the $(\beta,\gamma)=(1,1)$ case -- and so $\varv_\mathrm c^2(R)=\varrho^2[1-R_C(1;\varrho^2)]/(\varrho^2-1)$ similarly follows that of $R^2\Sigma(R)$ for the same case -- and it is easier to observe $\lim_{R\to0}\Phi(R,0)=0$ based on the equivalent logarithmic expression.

\subsection{Elliptic integral cases}

If both $2\beta$ and $2\gamma$ are integers but $\beta+\gamma$ is a half-integer, the integrals representing the potential involve a square root of a quartic polynomial, and so, in general, are elliptic integrals. The simplest such case is when $(\beta,\gamma)=(\frac12,1)$, for which
\begin{equation*}
\frac{\Sigma(R)}{\Lambda_0/R_0}
=\frac1{4\pi}{_2F_1}\!\left(\stack{\frac32,1}2
;{1-\varrho^2}\right)=\frac1{2\pi}\frac1{\varrho(\varrho+1)}
\end{equation*}
with $M(R)=\Lambda_0R_0\ln(1+\varrho)$. The circular speed behaves like
\begin{equation}\label{eq:vcb1g2}\begin{split}
\frac{\varv_\mathrm c^2(R)}{G\Lambda_0}
&=\frac\pi{4\varrho}\,{_2F_1}\!\left(\stack{\frac32,\frac12}2
;1-\frac1{\varrho^2}\right)
=\frac{R_D(0,\varrho^{-2};1)}{3\varrho}
\\&=\begin{cases}
\dfrac{\varrho[\mathbf K(\kappa_+)-\mathbf E(\kappa_+)]}{\varrho^2-1}
&(\varrho>1;\,\kappa_+^2=1-\varrho^{-2})
\\\dfrac{\mathbf E(\kappa_-)-\varrho^2\mathbf K(\kappa_-)}{1-\varrho^2}
&(\varrho<1;\,\kappa_-^2=1-\varrho^2)
\end{cases}
\\&=\frac1{1-\varrho}\left[\mathbf E(\kappa_1)
-\frac{2\varrho\mathbf K(\kappa_1)}{1+\varrho}\right]
;\quad\kappa_1=\frac{|1-\varrho|}{1+\varrho}.
\end{split}\end{equation}
The expression of the potential using the canonical elliptic integral basis set is found by applying the reduction formula \cite[eq.~19.29.24]{DLMF} on equation (\ref{eq:g2pot}) with $(\beta,\gamma)=(\frac12,1)$:
\begin{equation}\label{eq:potb1g2}\begin{split}
\frac{-\Phi(\bm r)}{G\Lambda_0}
&=\int_0^\infty\frac{\dm t}{\!\sqrt{1+t^2}\!\sqrt{\varrho^2+(|\zeta|+t)^2}}
\\&=2R_F(\sigma+|\zeta|,\sigma+|\zeta|+\delta,\sigma+|\zeta|+\alpha)
\\&=\frac2{\alpha^{1/2}}\mathrm F(\varphi,\kappa),
\end{split}\end{equation}
where $(\alpha,\delta)$ are the oblate spheroidal coordinates as in equation (\ref{eq:oedad}) and the arguments for the Legendre form are
\begin{equation}\label{eq:b1g2kp}
\kappa^2=\frac{\alpha-\delta}\alpha;\quad
\sin^2\varphi=\frac\alpha{\sigma+|\zeta|+\alpha}.
\end{equation}
The same result can also be derived by changing the integration variable to $x\to u=2x(x+\sigma)+2\!\sqrt{x^2(x+\sigma)^2+(x+x_0)^2}$ in equation (\ref{eq:g2pottr}) for $(\beta,\gamma)=(\frac12,1)$, given that $\dm u/\dm x=\!\sqrt{u(u+\delta)(u+\alpha)}/\!\sqrt{x^2(x+\sigma)^2+(x+x_0)^2}$.

The Gauss--Landen transformations of the elliptic integrals also admit additional equivalent expressions for the potential: 
\begin{multline}\label{eq:potb1g2b}
\frac{-\Phi(\bm r)}{G\Lambda_0}
=R_F\left[\left(\frac\zeta{1+\sigma}\right)^2,
\left(\frac{\delta+\sigma}{1+\sigma}\right)^2,
\left(\frac{\alpha+\sigma}{1+\sigma}\right)^2\right]
=\frac{\mathrm F(\psi,\kappa)}{\alpha^{1/2}}
\\=R_F\Biggl[\frac{\sigma-\varrho}2,\frac{\sigma+\varrho}2,
\left(\frac{1+\sigma}2\right)^2\Biggr]=\frac2{\upsilon_+}
\mathrm F\biggl(\sin^{-1}\!\frac{\upsilon_+}{1+\sigma},
\frac{\upsilon_-}{\upsilon_+}\biggr),
\end{multline}
where $\upsilon_\pm^2$ is as in equation (\ref{eq:oedad}) and $\kappa^2=(\alpha-\delta)/\alpha$ but
\begin{equation}\label{eq:b1g2ba}
\sin^2\psi=\frac{\alpha(1+\sigma)^2}{(\alpha+\sigma)^2},\quad
\cos^2\psi=\frac{\zeta^2}{(\alpha+\sigma)^2}.
\end{equation}
%
The second line of equation (\ref{eq:potb1g2b}) follows the descending Landen transformation of equation (\ref{eq:potb1g2}), whereas the descending Gauss transformation of the first line of equation (\ref{eq:potb1g2b}) leads to the same expression of the second line. Furthermore, changing the integration variable in equation (\ref{eq:potb1g2}) via $t\to w=2[t(t+\zeta)+\!\sqrt{t^2+1}\!\sqrt{\varrho^2+(\zeta+t)^2}]=[\!\sqrt{t^2+1}+\!\sqrt{\varrho^2+(\zeta+t)^2}]^2-(1+\sigma^2)$ (cf. $\dm w/\dm t=\!\sqrt{w+1+\sigma^2}\!\sqrt{w^2-4\varrho^2}/[\!\sqrt{t^2+1}\!\sqrt{\varrho^2+(\zeta+t)^2}]$) also results in the second line. Finally equation (\ref{eq:potb1g2}) and the first line of equation (\ref{eq:potb1g2b}) are directly transformed to each other via the duplication formula. On the mid-plane ($z=0$) and the symmetry axis ($R=0$), this potential simplifies to
\begin{equation*}\begin{split}
\frac{-\Phi(R,0)}{G\Lambda_0}
&=R_F(0,\varrho^2,1)=2R_F(\varrho,\varrho+\varrho^2,\varrho+1)
\\&=\frac2{1+\varrho}\mathbf K(\kappa_1)=\begin{cases}
\varrho^{-1}\mathbf K(\kappa_+)&(\varrho>1)\\
\mathbf K(\kappa_-)&(\varrho<1)\end{cases};
\\\frac{-\Phi(0,z)}{G\Lambda_0}&=\frac2{\!\sqrt{1+\zeta^2}}
\ln\frac{1+\zeta+\!\sqrt{1+\zeta^2}}{\!\sqrt{2\zeta}}.
\end{split}\end{equation*}
Here $\kappa_\pm$ and $\kappa_1$ are as defined in equation (\ref{eq:vcb1g2}).

The last case that we consider is for $(\beta,\gamma)=(1,\frac12)$ with
\begin{equation*}\begin{split}
\frac{\Sigma(R)}{\Lambda_0/R_0}
&=\frac18{_2F_1}\!\left(\stack{\frac32,\frac12}2;{1-\varrho^2}\right)
=\frac{R_D(0,\varrho^2;1)}{6\pi}
\\&=\frac1{2\pi}\begin{cases}
\dfrac{\varrho^2\mathbf E(\kappa_+)-\mathbf K(\kappa_+)}
{\varrho(\varrho^2-1)}&(\varrho>1)\smallskip\\
\dfrac{\mathbf K(\kappa_-)-\mathbf E(\kappa_-)}{1-\varrho^2}
&(\varrho<1)\end{cases}
\\&=\frac1{2\pi(1-\varrho)}\left[\frac{2\mathbf K(\kappa_1)}{1+\varrho}
-\mathbf E(\kappa_1)\right].
\end{split}\end{equation*}
Here the model with $\gamma=\frac12$ exhibits a linear growth of the mass and a flat rotation curve asymptotically. In particular,
\begin{equation*}\begin{split}
\frac{M(R)}{\Lambda_0R_0}&=(1+\varrho)\mathbf E(\kappa_1)
-\frac{2\varrho\mathbf K(\kappa_1)}{1+\varrho}-1
=\begin{cases}\varrho\mathbf E(\kappa_+)-1
\\\mathbf E(\kappa_-)-1\end{cases};
\\\frac{\varv_\mathrm c^2(R)}{G\Lambda_0}
&=\frac1{2\varrho}\,{_2F_1}\!\left(\stack{\frac32,1}2
;1-\frac1{\varrho^2}\right)
=\frac\varrho{1+\varrho}.
\end{split}\end{equation*}
As for the potential, we start with the zero point shifted one in equation (\ref{eq:g2potd}) with $(\beta,\gamma)=(1,\frac12)$. For analytic treatments however, it is better to transform the integral via changing the variable $t=x(x+\sigma)/(x+x_0)$, which then results in
\begin{equation*}
\frac{\Phi(\bm r)}{G\Lambda_0}
=\int_0^\infty\left[|\zeta|+\frac{\varrho^2}{2(x+x_0)}\right]
\frac{\dm x}{\!\sqrt{x^2(x+\sigma)^2+(x+x_0)^2}}.
\end{equation*}
Again changing the integration variable $x\to y=2x(x+\sigma)+2\!\sqrt{x^2(x+\sigma)^2+(x+x_0)^2}$ (note $2x_0=\sigma+|\zeta|$) leads this to
\begin{equation*}
\frac{\Phi(\bm r)}{G\Lambda_0}
=\int_{2x_0}^\infty\!\dm y\left[\frac{\varrho^2}{y(y+\varrho^2)}
+\frac{|\zeta|}{y+\varrho^2}\!\sqrt{\frac y{(y+\delta)(y+\alpha)}}\right].
\end{equation*}
which is easily reducible to a form over the canonical basis,
\begin{equation*}\begin{split}
\frac{\Phi(\bm r)}{G\Lambda_0}
&=\ln(1+\sigma-|\zeta|)+
2|\zeta|\,\Biggl[R_F(2x_0,2x_0+\delta,2x_0+\alpha)
\\&\phantom{=\ln(1+}
-\frac{\varrho^2}3R_J(2x_0,2x_0+\delta,2x_0+\alpha;2x_0+\varrho^2)\Biggr]
\\&=\ln(1+2x_0)+\frac{2|\zeta|}3
R_J(2x_0,\delta+2x_0,\alpha+2x_0;1+2x_0)
\\&=\ln\!\sqrt{p_-}+|\zeta|\left[
R_F(u,v,w)+\frac{1-\varrho^2}3R_J(u,v,w;p_-)\right]
\\&=\ln\!\sqrt{p_+}+\frac{|\zeta|}3R_J(a^2,b^2,c^2;1+a^2)
\\&=\ln\frac{\!\sqrt q}\sigma
+|\zeta|\,\Biggl[R_F(a^2,b^2,c^2)
\\&\phantom{=\ln\sqrt{(1+\sigma)^2+\zeta^2}}
-\frac{\varrho^2}3R_J(a^2,b^2,c^2;\varrho^2+a^2)\Biggr]
\end{split}\end{equation*}
where still $2x_0=\sigma+|\zeta|$ and also used are the short hands for
\begin{equation}\label{eq:shdf}\begin{split}
(u,v,w)&=[2(\sigma-\varrho),2(\sigma+\varrho),(1+\sigma)^2];
\\(a,b,c)&=\left(\frac{|\zeta|}{1+\sigma},\frac{\delta+\sigma}{1+\sigma},
\frac{\alpha+\sigma}{1+\sigma}\right);
\\p_\pm&=(1+\sigma)^2\pm\zeta^2;\quad q=\varrho^2(1+\sigma)^2+\zeta^2.
\end{split}\end{equation}
%
In terms of the Legendre bases, these are equivalent to
\begin{equation*}\begin{split}
\frac{\Phi(\bm r)}{G\Lambda_0}
&=\ln(1+\sigma-|\zeta|)+
\frac{2\xi^{1/2}}{\chi^{1/2}}
\frac{\mathrm F(\varphi,\kappa)-\delta\Pi(\varphi,1-\delta,\kappa)}
{\alpha^{1/2}}
\\&=\ln(1+\sigma+|\zeta|)+\frac{2\chi^{1/2}}{\xi^{1/2}}
\frac{\Pi(\varphi,1-\alpha^{-1},\kappa)-F(\varphi,\kappa)}{\alpha^{1/2}}
\\&=\ln\!\sqrt{p_-}+
\frac{(1-\varrho^2)\Pi(\varphi_1,n_1,\kappa_1)
+(\sigma^2-1)\mathrm F(\varphi_1,\kappa_1)}{\zeta\,\upsilon_+}
\\&=\ln\!\sqrt{p_+}+\frac{\chi^{1/2}}{\xi^{1/2}}
\frac{\Pi(\psi,1-\alpha^{-1},\kappa)-\mathrm F(\psi,\kappa)}{\alpha^{1/2}}
\\&=\ln\frac{\!\sqrt q}\sigma+\frac{\xi^{1/2}}{\chi^{1/2}}
\frac{\mathrm F(\psi,\kappa)-\delta\Pi(\psi,1-\delta,\kappa)}{\alpha^{1/2}}
\end{split}\end{equation*}
where $(\varphi,\kappa)$ is defined in equation (\ref{eq:b1g2kp}) and $\psi$ in equation (\ref{eq:b1g2ba}), while $\sin\varphi_1=\upsilon_+/(1+\sigma)$ and $(n_1,\kappa_1^2)=(\zeta^2/\upsilon_+^2,\upsilon_-^2/\upsilon_+^2)$ with $p_\pm$ and $q$ as in equation (\ref{eq:shdf}). All these expressions are formally equivalent but may have different computational stabilities. Finally the corresponding potential on the mid-plane and the symmetry axis simplify to
\begin{equation*}\begin{split}
\frac{\Phi(R,0)}{G\Lambda_0}&=\ln(1+\varrho);
\\\frac{\Phi(0,z)}{G\Lambda_0}&=
\frac{2\zeta}{\!\sqrt{1+\zeta^2}}
\ln\frac{1+\zeta+\!\sqrt{1+\zeta^2}}{\!\sqrt{2\zeta}}.
\end{split}\end{equation*}
In general, if $\beta$ is an integer and $\gamma$ is a half-integer, we have an algebraic expression for $\Phi(R,0)$ and $\varv_\mathrm c^2(R)$ but $\Sigma(R)$ requires elliptic integral basis to be written down. This contrasts to the case with a half-integer $\beta$ and an integer $\gamma$, for which $\Sigma(R)$ is algebraic but $\varv_\mathrm c^2(R)$ is an elliptic integral.

\section{Conclusion}

Equation (\ref{eq:vmeij1}) suggests that $\varv_\mathrm c^2$ due to $\Sigma$ of the form $G^{m,n}_{p,q}(\varrho^2)$ is also reducible to $G^{m',n'}_{p,q}(\varrho^2)$ (where $m+n=m'+n'$) if two among the upper/lower and front/back indices of the Me{\ij}er-$G$ (i.e., $a_i+\tfrac12$ or $b_j+\tfrac12$) can be canceled by the added fixed indices (viz., $1/2$ for the upper, and $1$ and $0$ for the lower front and back). Note that the same indices of the Me{\ij}er-$G$ can cancel diagonally with each other. Given that all $G^{m,n}_{2,2}$ reduce to the ${_2F_1}$-hypergeometric function, it follows that there exist 21 sets of two-parameter families of the models where both $\Sigma$ and $\varv_\mathrm c^2$ given by the Gauss-${_2F_1}$. However, for five sets of these models, the \citetalias{EdZ92} weight in equation (\ref{eq:vmeij1}) becomes $G^{0,0}_{p,q}$, indicating that proper weights for these models do not exist. The weights for the remaining cases consist of three cases of $G^{m,n}_{3,3}$ corresponding to $\Sigma\propto G^{m,n}_{2,2}(\varrho^2|\{a,\frac12\};\{b,0\})$ (but $m,n\ne2$); three cases of $G^{m-1,n-1}_{1,1}$ resulting in $\Sigma\propto G^{m,n}_{2,2}(\varrho^2|\{-\frac12,a\};\{b,0\})$ (but $m,n\ne0$); and other ten cases of $G^{m,n}_{2,2}$. This paper provides a full exploration of the models corresponding to the three $G^{m-1,n-1}_{1,1}$ cases. In other words, the models discussed here encompass all disks whose 1) surface density and mid-plane circular velocity are both expressible as a ${_2F_1}$-hypergeometric function (of $R^2/s^2$); and 2) \citetalias{EdZ92} weight $\Lambda$ exists and is expressed as one of the beta-related distributions: which are $h^{2c_1}(s^2-h^2)^{c_2}$, $h^{2c_1}(h^2-s^2)^{c_2}$, or $h^{2c_1}(s^2+h^2)^{c_2}$. We have shown that these include many common and interesting analytic disk models used in the literature. Many of these exhibit a fully ``analytic'' (i.e.\ not merely left as unresolved integrals) functional form for the potential. Since many ${_2F_1}$-functions are also reducible to elementary functions, this is not an exhaustive list of models with elementary functions as the \citetalias{EdZ92} weight, nor of models possessing a simple analytic expression for the full-space potential (see e.g., \cite{fdisk}). Nevertheless the current set of models, together with the general methodology utilizing the Mellin--Barnes integral and the Me{\ij}er-$G$ functions, serves as a robust starting point for exploring razor-thin disk models with tractable potential forms.


\clearpage
\appendix

\section{The differential inversion formula for the Stieltjes Transform}
\label{app:d_inv}

Analogous to Post's inversion formula \cite{Po30} for the Laplace transform, \citet{Wi38} provided an alternative differential inversion formula for the Stieltjes transform (eq.~\ref{eq:stiel}). This method relies solely on the behavior of the transformation on the same real interval as the integral support:
\begin{equation}\label{eq:wsif}
g(x)=\lim_{n\to\infty}\frac{(-x)^{n-1}}{n!(n-2)!}
\frac{\dm^{2n-1}[x^nf(x)]}{\dm x^{2n-1}},
\end{equation}
which however involves differentiations of infinite order. The factorial terms may be replaced using the Stirling approximation, such that $n!(n-2)!\to2\pi n^{2n-1}\rme^{-2n}$.
Similarly, there exist pure differential inversion formulae for the GST based solely on the real behavior of $f(x)$. In particular, for any pair of fixed integers $(k,\ell)$, equation (\ref{eq:stielt_g}) may be inverted via
\begin{multline}\label{eq:dinv_gsl}
g(x)=\lim_{n\to\infty}\left\lbrace(-1)^{n+k}Cx^{\eta+n+k}
\frac{\dm^{n+k}}{\dm x^{n+k}}\left[\frac1{x^\eta}
\frac{\dm^{n+\ell}(x^{\eta+n+\ell}f)}{\dm x^{n+\ell}}\right]\right\rbrace
\\=\lim_{n\to\infty}\left\lbrace(-1)^{n+k}C\frac{\dm^{n+\ell}}{\dm x^{n+\ell}}
\left[x^{\eta+2n+k+\ell}f^{(n+k)}(x)\right]\right\rbrace,
\end{multline}
where the constant $C$ is defined as
\begin{displaymath}\begin{split}
C&=\frac1{(1+\eta)_{n+k-1}(n+\ell)!}
=\frac{\Gamma(1+\eta)}{\Gamma(\eta+n+k)\Gamma(1+n+\ell)}
\\&\to\frac{\Gamma(1+\eta)}{2\pi}
\frac{\rme^{2n}}{n^{\eta+2n+k+\ell}}
\quad\text{(the Stirling approximation)}.
\end{split}\end{displaymath}
For $\eta=0$, this reduces to equation (\ref{eq:wsif}) with $(k,\ell)=(-1,0)$. While a rigorous proof is beyond the scope of this paper, we observe that for $f(x)=x^\alpha$, the formula implies
\begin{equation}\label{eq:p_inv}\begin{split}
g(x)&=\frac{A\Gamma(\eta+1)}{\Gamma(\alpha+\eta+1)\Gamma(-\alpha)}x^{\alpha+\eta};
\\A&=\lim_{n\to\infty}\frac{\Gamma(n+\ell+\alpha+\eta+1)\Gamma(n+k-\alpha)}
{\Gamma(n+k+\eta)\Gamma(n+\ell+1)}=1,
\end{split}\end{equation}
thanks to $(\dm/\dm x)^nx^\alpha=(-1)^n(-\alpha)_nx^{\alpha-n}$. Since equation (\ref{eq:stielt_g}) transforms $g(t)=t^\alpha$ to $f(x)=[\Gamma(\alpha+1)\Gamma(\eta-\alpha)/\Gamma(\eta+1)]x^{\alpha-\eta}$, this result is consistent with a valid inversion of the GST. Furthermore, applying equation (\ref{eq:p_inv}) to the integral definition of $\Sigma(R)$ in equation (\ref{eq:sigmeij}) -- thereby inverting the GST (with $\eta=1/2$) of $\Sigma(h^2)/4\pi\to\Sigma(R^2)$ in equation (\ref{eq:sigv}) -- allows one to recover the integral definition of $\Sigma(h)$ in equation (\ref{eq:vmeij1}).

The inversion formula of equation (\ref{eq:dinv_gsl}) ``generalizes'' to arbitrary real values of $k,\ell$ through the introduction of fractional derivatives using Weyl integrals \citep[chap.~XIII]{Er54}, but the resulting expression is not necessarily purely differential (which effectively shifts the order of the GST using a generalized Abel transform -- i.e.\ the Weyl or Riemann--Liouville integrals -- before applying the differential formula to the resulting GST of the new order). In particular, the formula of \cite[sect.~8]{Wi38} is recovered for $(k,\ell)=(-1-\eta,0)$ by identifying $(-1)^kf^{(n+k)}(x)={_\infty^-I}_x^{-k}f^{(n)}(x)$ for non-integral values of $k<0$ (where ${_\infty^-I}_x^\xi$ is defined as in equation \ref{eq:weyl}). On the other hand, choosing $(k,\ell)=(-\eta,1)$ yields equation (A4) of \cite{EdZ92} for integer values of $q=1+\eta$. Their statement (``This inversion ... given before'') notwithstanding, this particular case had appeared earlier in the literature; see Definition 8.1 and Theorem 8.3 of \cite{Wi38}.

\section{The ME{\IJ}ER-G functions}
\label{app:meijer}

The Me{\ij}er-G function is typically defined via
the Mellin--Barnes integral. Specifically, for a set of
non-negative integers $(m,n,p,q)$ satisfying $0\le m\le q$ \& $0\le n\le p$
(but $m+n\ne0$):
\begin{multline}\label{eq:meijer}
G^{m,n}_{p,q}\!\left(z\,\vrule
\stack{a_1,\dotsc,a_p}{b_1,\dotsc,b_q}\right)
\\=\frac1{2\pi\im}\!\int\limits_\mathcal C\!\frac{\dm t}{z^t}
\frac{\prod_{i=1}^m\Gamma(b_i+t)\cdot\prod_{j=1}^n\Gamma(1-a_j-t)}
{\prod_{i=n+1}^p\Gamma(a_i+t)\cdot\prod_{j=m+1}^q\Gamma(1-b_j-t)},
\end{multline}
which takes the form of an inverse Mellin transform (eq.~\ref{eq:mit}). That is to say, the integrand of equation (\ref{eq:meijer}) is the Mellin transform ($z\to t$) of the Me{\ij}er-G function. Note however that many standard references adopt an equivalent alternative definition in which the sign of the integration variable $t$ is reversed (e.g., \cite{Er53}, eq.~5.3.(1)). The integration path $\mathcal C$ is chosen to separate the poles of $\prod_{i=1}^m\Gamma(b_i+t)$ from those of $\prod_{j=1}^n\Gamma(1-a_j-t)$, starting and ending at infinity such that the integral converges. By the Cauchy residue theorem, this path essentially encircles either the poles of $\prod_{i=1}^m\Gamma(b_i+t)$ in the positive sense or those of $\prod_{j=1}^n\Gamma(1-a_j-t)$ in the negative sense. Detailed discussions on the canonical path and convergence criteria are available in standard reference such as \cite[\S~5.3]{Er53} or \cite[\S~16.17]{DLMF}. The definition implies that the Me{\ij}er-G function is well-defined only if no pole of $\prod_{i=1}^m\Gamma(b_i+t)$ coincides with a pole of $\prod_{j=1}^n\Gamma(1-a_j-t)$; that is, $a_j-b_i$ must not be a positive integer for any pair of $(i,j)$.

Although the Mellin transform is a powerful tool in various mathematical branches, it is not often emphasized in standard mathematical physics curricula. Consequently, the primary challenge for most physicists regarding the Me{\ij}er-G function lies in its definition itself. By contrast, traditional introductions to special functions usually prioritize power-series representations or differential equations. In order to approach the Me{\ij}er-G function via the power series, we recall that $\Gamma(x)$ possesses only simple poles at non-positive integers, with the residue at $x=-k$ being $(-1)^k/k!$. If no two values in the set $\set{b_j|j\in\mathbb N,1\le j\le m}$ differ by an integer, all the poles of $\prod_{i=1}^m\Gamma(b_i+t)$ are simple, allowing the function to be expressed as a sum of residues. Similarly if no two values in $\set{a_i|i\in\mathbb N,1\le i\le n}$ differ by an integer, a corresponding sum of residues for $\prod_{j=1}^n\Gamma(1-a_j-t)$ yields a representation of the Me{\ij}er-G function as a sum of hypergeometric power series. A canonical example of this procedure is the Me{\ij}er-G representation of a generalized hypergeometric function:
\begin{multline}\label{eq:mhgf}
G^{1,p}_{p,q+1}\!\left(z\,\vrule
\stack{a_1,\cdots,a_p}{0,b_1,\cdots,b_q}\right)
=\sum_{k=0}^\infty\frac{(-1)^k}{k!}
\frac{\prod_{j=1}^p\Gamma(1-a_j+k)}{\prod_{i=1}^q\Gamma(1-b_i+k)}z^k
\\=\frac{\prod_{j=1}^p\Gamma(1-a_j)}{\prod_{i=1}^q\Gamma(1-b_i)}\,
{_pF_q}\!\left(\stack{1-a_1,\cdots,1-a_p}{1-b_1,\cdots,1-b_q};-z\right)
\end{multline}
where $0\le p\le q+1$ (see also \cite{Er53}, eq.~5.3.(5)-(6)). This connection
extends to the associated differential equations: that is,
the Me{\ij}er-G functions $G^{m,n}_{p,q+1}[(-1)^{p-m-n}z\,|\{a_1,\cdots,a_p\},\{0,b_1,\cdots,b_q\}]$ with the same set of indices $\set{a_1,\cdots,a_p}$ and $\set{b_1,\cdots,b_q}$ constitute particular solutions to the same generalized hypergeometric differential equation satisfied by ${_pF_q}(\{1-a_1,\cdots,1-a_p\},\{1-b_1,\cdots,1-b_q\};-z)$. Furthermore, equation (\ref{eq:meijer}) also implies 
\begin{align}
z^\lambda G^{m,n}_{p,q}\!\left(z\,\vrule
\stack{a_1,\dotsc,a_p}{b_1,\dotsc,b_q}\right)
&=G^{m,n}_{p,q}\!\left(z\,\vrule
\stack{a_1+\lambda,\dotsc,a_p+\lambda}
{b_1+\lambda,\dotsc,b_q+\lambda}\right);
\label{eq:mp}\\\label{eq:mr}
G^{m,n}_{p,q}\!\left(\frac1z\,\vrule
\stack{a_1,\dotsc,a_p}{b_1,\dotsc,b_q}\right)
&=G^{n,m}_{q,p}\!\left(z\,\vrule
\stack{1-b_1,\dotsc,1-b_q}{1-a_1,\dotsc,1-a_p}\right),
\end{align}
which suggests that differential equations for more general cases are also reducible to the generalized hypergeometric equation. For explicit expression for these differential equations, please refer \citep[\S~16.8(ii) \& \S~16.21]{DLMF}. Here we simply note that $G^{m,n}_{p,q}(z)$ is associated with a linear homogeneous equation of degree $\max(p,q)$. For a more systematic treatment, \cite{BS13} provide a highly readable introduction to the Me{\ij}er-G function via ordinary differential equation.

\subsection{The Me{\ij}er--Mellin convolution theorem}

What makes the Me{\ij}er-G functions useful in analytic
representations of integral transformations is the fact that
the complete set of the Me{\ij}er-G function
is closed under the integral convolutions. In particular,
the convolution theorem on the Mellin transform,
\begin{equation}
\mathfrak M_{x\to s}\left[\int_0^\infty\!\frac{\dm y}y\,k\biggl(\frac xy\biggr)\,f(y)\right]
=\mathfrak M_{x\to s}[k(x)]\cdot\mathfrak M_{y\to s}[f(y)]
\end{equation}
indicates that the ``Me{\ij}er-G transformation'' of
any Me{\ij}er-G function is still a Me{\ij}er-G function; 
\begin{multline}\label{eq:mmc}
\int_0^\infty\!\frac{\dm y}y\,
G^{m,n}_{p,q}\!\left(\frac xy\,\vrule
\stack{a_1,\dotsc,a_p}{b_1,\dotsc,b_q}\right)
\,G^{r,s}_{u,v}\!\left(y\,\vrule
\stack{c_1,\dotsc,c_u}{d_1,\dotsc,d_v}\right)
\\=G^{m+r,n+s}_{p+u,q+v}\!\left(x\,\vrule
\stack{a_1,\dotsc,a_n,c_1,\dotsc,c_u,a_{n+1},\dotsc,a_p}
{b_1,\dotsc,b_m,d_1,\dotsc,d_v,b_{m+1},\dotsc,b_q}\right).
\end{multline}
Changing variables and utilizing equations (\ref{eq:mp})
and (\ref{eq:mr}), equation (\ref{eq:mmc}) is recast
in an equivalent symmetric form
\begin{multline*}
\int_0^\infty\!\dm z\,G^{m,n}_{p,q}\!\left(\alpha z\,\vrule
\stack{a_1,\dotsc,a_p}{b_1,\dotsc,b_q}\right)
\,G^{r,s}_{u,v}\!\left(\beta z\,\vrule
\stack{c_1,\dotsc,c_u}{d_1,\dotsc,d_v}\right)
\\=\frac1\beta\,G^{m+s,n+r}_{p+v,q+u}\!\left(\frac\alpha\beta\,\vrule
\stack{a_1,\dotsc,a_n,-d_1,\dotsc,-d_v,a_{n+1},\dotsc,a_p}
{b_1,\dotsc,b_m,-c_1,\dotsc,-c_u,b_{m+1},\dotsc,b_q}\right).
\end{multline*}

\subsection{Specific examples}
\label{app:mex}

Here we list the reduction of Me{\ij}er-G functions $G^{m,n}_{p,q}$
for $p,q\le2$ (but without any proof). First, if $p+q=1$,
the corresponding Me{\ij}er-G function is basically exponential;
\begin{equation*}
G^{1,0}_{0,1}\!\left(x\,\vrule\stack{}{\alpha}\right)=x^\alpha\exp(-x),
\end{equation*}
whereas $G^{m,n}_{1,1}$ results in a geometric series as in
\begin{equation*}\begin{split}
G^{1,0}_{1,1}\left(t\,\vrule\stack\gamma\alpha\right)
&=\frac{t^\alpha(1-t)^{\gamma-\alpha-1}}{\Gamma(\gamma-\alpha)}\Theta(1-|t|);
\\G^{1,1}_{1,1}\left(u\,\vrule\stack\beta\alpha\right)
&=\Gamma(1+\alpha-\beta)\frac{u^\alpha}{(u+1)^{1+\alpha-\beta}}.
\end{split}\end{equation*}
According to equation (\ref{eq:mhgf}), $G^{1,0}_{0,2}$ reduces to $_0F_1$,
which is also equivalent to some Bessel-type functions. In fact,
\begin{gather*}
J_a(z)=\biggl(\frac z2\biggr)^a
G^{1,0}_{0,2}\!\left(\frac{z^2}4\vrule\stack{}{0,-a}\right);\quad
I_a(z)=\biggl(\frac z2\biggr)^a
G^{1,0}_{0,2}\!\left(-\frac{z^2}4\vrule\stack{}{0,-a}\right);
\\
2K_a(z)=\biggl(\frac z2\biggr)^a
G^{2,0}_{0,2}\!\left(\frac{z^2}4\vrule\stack{}{0,-a}\right)
=\biggl(\frac2z\biggr)^a
G^{2,0}_{0,2}\!\left(\frac{z^2}4\vrule\stack{}{0,a}\right),
\end{gather*}
although $Y_a(z)$, at the simplest, is in the form of $G^{2,0}_{1,3}$.

The set of the Me{\ij}er-G functions $G^{m,n}_{1,2}$ encompasses
the confluent hypergeometric functions of both the first and second kinds,
i.e.\ $_1F_1$ (the Kummer-M) and $U$ (the Tricomi-U):
\begin{gather*}\begin{split}
\frac1{\Gamma(b)}{_1F_1}\!\left(\stack ab;x\right)
&=\Gamma(1-a)\,G^{1,0}_{1,2}\!\left(x\,\vrule\stack{1-a}{0,1-b}\right)
\\&=\frac{\rme^x}{\Gamma(b-a)}\,
G^{1,1}_{1,2}\left(x\,\vrule\stack{1+a-b}{0,1-b}\right);
\end{split}\\\begin{split}
U(a,b,x)&=\rme^x\,G^{2,0}_{1,2}\!\left(x\,\vrule\stack{1+a-b}{0,1-b}\right)
\\&=\frac1{\Gamma(a)\Gamma(1+a-b)}\,
G^{2,1}_{1,2}\!\left(x\,\vrule\stack{1-a}{0,1-b}\right).
\end{split}\end{gather*}
On the other hand, $G^{m,n}_{2,2}$ in general
reduces to the Gauss-$_2F_1$:
\begin{gather*}
G^{1,0}_{2,2}\!\left(x\,\vrule\stack{\gamma_1,\gamma_2}{\alpha,\delta}\right)
=\frac{x^\alpha
{_2F_1}(1+\alpha-\gamma_1,1+\alpha-\gamma_2;1+\alpha-\delta;-x)}
{\Gamma(\gamma_1-\alpha)\Gamma(\gamma_2-\alpha)\Gamma(1+\alpha-\delta)};
\\
G^{2,0}_{2,2}\!\left(x\,\vrule
\stack{\gamma_1,\gamma_2}{\alpha_1,\alpha_2}\right)
=\frac{(1-x)^{\mu-1}}{\Gamma(\mu)}x^{\alpha_1}
{_2F_1}\!\left(\stack{\gamma_1-\alpha_2,\gamma_2-\alpha_2}\mu;1-x\right)
\end{gather*}
for $|x|<1$ with $\mu=\gamma_1+\gamma_2-\alpha_1-\alpha_2$.
However both identically vanish for $|x|>1$. Here the second one
with $\alpha_1\leftrightarrow\alpha_2$ switched in the right-hand side is also valid
given the Euler-Pfaff linear transformation for the Gauss-$_2F_1$. By contrast,
equation (\ref{eq:mhgf}) indicates
\begin{multline*}
G^{1,2}_{2,2}\!\left(x\,\vrule\stack{\beta_1,\beta_2}{\alpha,\delta}\right)
=\frac{\Gamma(1+\alpha-\beta_1)\Gamma(1+\alpha-\beta_2)}
{\Gamma(1+\alpha-\delta)}\\\times x^\alpha
{_2F_1}\!\left(\stack{1+\alpha-\beta_1,1+\alpha-\beta_2}{1+\alpha-\delta}
;{-x}\right),
\end{multline*}
whereas (here $\mu=\beta_1+\beta_2-\alpha_1-\alpha_2$)
\begin{multline*}
G^{2,2}_{2,2}\left(x\,\vrule\stack{\beta_1,\beta_2}{\alpha_1,\alpha_2}\right)
=\prod_{i,j=1}^2\Gamma(1+\alpha_i-\beta_j)\\\times
\frac{x^{\alpha_1}}{\Gamma(2-\mu)}
{_2F_1}\!\left(\stack{1+\alpha_1-\beta_1,1+\alpha_1-\beta_2}{2-\mu}
;{1-x}\right),
\end{multline*}
or the same with $\alpha_1\leftrightarrow\alpha_2$. These are valid for the whole interval.
The last remaining case, excluding those cases that are reducible to one of the prior examples
through equation (\ref{eq:mr}), is found to be
\begin{multline*}
G^{1,1}_{2,2}\!\left(x\,\vrule\stack{\beta,\gamma}{\alpha,\delta}\right)
=\Gamma(1+\alpha-\beta)
\\\times\begin{cases}\displaystyle
\frac{x^\alpha{_2F_1}(1+\alpha-\beta,1+\alpha-\gamma;1+\alpha-\delta;x)}
{\Gamma(\gamma-\alpha)\Gamma(1+\alpha-\delta)}
&(x\le1)\smallskip\\\displaystyle
\frac{x^{\beta-1}{_2F_1}(1+\alpha-\beta,1+\delta-\beta;1+\gamma-\beta;x^{-1})}
{\Gamma(\beta-\delta)\Gamma(1+\gamma-\beta)}
&(x\ge1)\end{cases}.\end{multline*}

\section{The Carlson-R functions}
\label{app:carl}

The Carlson-R function \cite{Ca63} (see also \cite{DLMF}, \S~19.16)
is defined through (where $B=\sum_{j=1}^n\beta_j$)
\begin{multline}\label{eq:carldef}
R_{-\alpha}(\beta_1,\dotsc,\beta_n;x_1,\dotsc,x_n)=
\frac{\Gamma(B)}{\Gamma(\alpha)\Gamma(B-\alpha)}\\\times
\int_0^\infty\!\dm t\,t^{B-\alpha-1}
(t+x_1)^{-\beta_1}\dotsm(t+x_n)^{-\beta_n},
\end{multline}
which converges provided $B>\alpha>0$ (and all $x_i>0)$. Clearly 
the $R$-function is symmetric under any simultaneous permutation
of the variables $\bm x=(x_1,\dotsc,x_n)$ and indices
$\bm\beta=(\beta_1,\dotsc,\beta_n)$. The definition also indicates
that the $R_{-\alpha}(\bm\beta;\bm x)$ is a homogeneous function of
degree ``$-\alpha$'' regarding the variables $\bm x$; namely,
for any $\xi>0$,
\begin{equation}
R_{-\alpha}(\bm\beta;\xi\bm x)
=\xi^{-\alpha}R_{-\alpha}(\bm\beta;\bm x).
\end{equation}
In fact, the one-variable case simply reduces to
\begin{equation}
R_{-\alpha}(\beta;x)=x^{-\alpha}
;\quad
R_{-\alpha}(\beta;1)=1.
\end{equation}
In addition, changing the integration variable to $u=t^{-1}$ in
equation (\ref{eq:carldef}) results in the transformation given by
\begin{equation}\label{eq:carltr}
R_{-\alpha}(\bm\beta,\bm x)
=(\textstyle\prod_ix_i^{-\beta_i})\cdot
R_{-B+\alpha}(\bm\beta,\bm x^{-1}).
\end{equation}

The Carlson-R (more than one variable) is actually equivalent to
the multivariate Lauricella-${\rm F_D}$ hypergeometric series:
\begin{multline}\label{eq:laud}
F_D^{(n)}(a;b_1,\dotsc,b_n;c;x_1,\dotsc,x_n)
\\=\sum_{j_1=0}^\infty\dotsb\sum_{j_n=0}^\infty
\frac{(a)_{\sum_{i=1}^nj_i}}{(c)_{\sum_{i=1}^nj_i}}
\frac{(b_1)_{j_1}}{j_1!}\dotsm\frac{(b_n)_{j_n}}{j_n!}
x_1^{j_1}\dotsm x_n^{j_n},
\end{multline}
which absolutely converges if every $|x_i|<1$.
Let us observe the one-variable case $F^{(1)}_D$ is just
the Gauss ${_2F_1}$-hypergeometric series;
\begin{equation}
F_D^{(1)}(a;b;c;x)=\sum_{j=0}^\infty\frac{(a)_j(b)_j}{(c)_j}\frac{x^j}{j!}
={_2F_1}\!\left(\stack{a,b}c;x\right)
\end{equation}
In fact, the $F_D$-series is a generalization of
the ${_2F_1}$-function that is representable via
the one-dimensional integral
\begin{multline}\label{eq:laudi}
F_D^{(n)}(a;b_1,\dotsc,b_n;c;x_1,\dotsc,x_n)
=\frac{\Gamma(c)}{\Gamma(a)\Gamma(c-a)}\\\times
\int_0^1\!\dm u\,u^{a-1}(1-u)^{c-a-1}
(1-x_1u)^{-b_1}\dotsm(1-x_nu)^{-b_n},
\end{multline}
for $0<a<c$, which generalizes the Euler integral representation
of the ${_2F_1}$-series. Comparing to equation (\ref{eq:carldef}),
we then establish
\begin{multline}
F_D^{(n)}(a;b_1,\dotsc,b_n;c;x_1,\dotsc,x_n)
\\=R_{-a}(c-\textstyle\sum_{i=1}^nb_i,b_1,\dotsc,b_n;1,1-x_1,\dotsc,1-x_n),
\end{multline}
which may be taken as a definition for
the $R$-function as long as a convergent $F_D$-series exists
even if the integral representation does not converge.
Notably, $F_D^{(n)}(0;\bm b;c;\bm x)=1$ then implies $(B\ne0)$
\begin{equation}\label{eq:b9}
R_0(\bm\beta;\bm x)=1;\quad
R_{-B}(\bm\beta,\bm x)=\textstyle\prod_ix_i^{-\beta_i}.
\end{equation}
If $m$ is a non-negative integer (and $B\notin\set{0,1,\dotsc,m}$),
\begin{multline}\label{eq:Rmpoly}
R_m(\beta_1,\dotsc,\beta_n;x_1,\dotsc,x_n)
=\frac{m!}{(B)_m}\\\times\sum_{j_1+\dotsb+j_n=m}
\frac{(\beta_1)_{j_1}\dotsm(\beta_n)_{j_n}}{j_1!\dotsm j_n!}
x_1^{j_1}\dotsm x_n^{j_n}
\end{multline}
where the sum is over all sets of non-negative integer indices
$(j_1,\dotsc,j_n)$ with the constant sum $m$, is a homogeneous
degree-$m$ polynomial of the variables $(x_1,\dotsc,x_n)$.

Special cases of the Carlson-R reduce
the dimension for the domain of the variables:
e.g., if one of the indices is zero,
\begin{equation}
R_{-\alpha}(\bm\beta,0;\bm x,y)=R_{-\alpha}(\bm\beta;\bm x),
\end{equation}
while, if the two variables coincide,
\begin{multline}
R_{-\alpha}(\beta_1,\dotsc,\beta_n,\beta_{n+1};x_1,\dotsc,x_n,x_n)
\\=R_{-\alpha}(\beta_1,\dotsc,\beta_n+\beta_{n+1};x_1,\dotsc,x_n).
\end{multline}
These correspond not only to the trivial results on the $F_D$,
\begin{multline}
F_D^{(n+1)}(a;b_1,\dotsc,b_n,0;c;x_1,\dotsc,x_n,y)
\\=F_D^{(n+1)}(a;b_1,\dotsc,b_n,p;c;x_1,\dotsc,x_n,0)
\\=F_D^{(n)}(a;b_1,\dotsc,b_n;c;x_1,\dotsc,x_n)
\end{multline}
but also to
\begin{multline}
F_D^{(n+1)}(a;b_1,\dotsc,b_n,b_{n+1};c;x_1,\dotsc,x_n,x_n)
\\=F_D^{(n)}(a;b_1,\dotsc,b_n+b_{n+1};c;x_1,\dotsc,x_n),
\end{multline}
the proof of which based on the definition in
equation (\ref{eq:laud}) would involve the well-known combinatorial result:
%
\begin{equation}\label{eq:cvi}
\sum_{k=0}^n\binom nk(a)_k(b)_{n-k}=(a+b)_n.
\end{equation}
Finally, if one of the variables of the Carlson-R is zero, then
\begin{multline}
R_{-\alpha}(\beta_0,\beta_1,\dotsc,\beta_n;0,x_1,\dotsc,x_n)
\\=\frac{\Gamma(\sum_{i=0}^n\beta_i)\,\Gamma(\sum_{i=1}^n\beta_i-\alpha)}
{\Gamma(\sum_{i=1}^n\beta_i)\,\Gamma(\sum_{i=0}^n\beta_i-\alpha)}\,
R_{-\alpha}(\beta_1,\dotsc,\beta_n;x_1,\dotsc,x_n),
\end{multline}
which is valid provided both arguments of the two gamma functions
in the numerator in the last line are positive (i.e.\
$\sum_{i=0}^n\beta_i>0$ and $\sum_{i=1}^n\beta_i>\alpha$).
In terms of the $F_D$-series, this is equivalent to
\begin{multline}
F_D^{(n+1)}(a;b_1,\dotsc,b_n,p;c;x_1,\dotsc,x_n,1)
\\=\frac{\Gamma(c)\Gamma(c-p-a)}{\Gamma(c-p)\Gamma(c-a)}\,
F_D^{(n)}(a;b_1,\dotsc,b_n;c-p;x_1,\dotsc,x_n),
\end{multline}
which can be considered as a generalization of
the Gauss hypergeometric theorem (corresponding to the $n=0$ case):
\begin{equation}
{_2F_1}\!\left(\stack{a,b}c;1\right)
=\frac{\Gamma(c)\Gamma(c-b-a)}{\Gamma(c-b)\Gamma(c-a)}
\qquad\text{(if $c>a+b$)},
\end{equation}
which in itself is a generalization of equation (\ref{eq:cvi})
-- which results from the case when one of the upper indices
of the ${_2F_1}$-series is zero or a negative integer (and so the infinite sum
terminates after the finite number of terms).

\section{The Elliptic integrals}
\label{app:elint}

\subsection{The Carlson symmetric basis}

The newer basis for the elliptic integrals is 
the Carlson symmetric integrals (see \cite[\S~19]{DLMF}
or \cite[sect.~6.11]{NR}): i.e.\
\begin{subequations}\begin{equation}\label{eq:Rdef}\begin{split}
R_F(a,b,c)&=\frac12\!\int_0^\infty\!
\frac{\dm t}{\!\sqrt{(t+a)(t+b)(t+c)}};
\\R_J(a,b,c;\,&p)=\frac32\!\int_0^\infty\!
\frac{\dm t}{(t+p)\!\sqrt{(t+a)(t+b)(t+c)}},
\end{split}\end{equation}
which are the elliptic integrals of the first and third kinds, respectively.
The integrals converge provided that $a,b,c,p>0$ (or at most one among
$a,b,c$ being zero). If $p<0$ however, $R_J$ is still defined via
the Cauchy principal value (which is also transformable to an integral
with a positive argument; see eq.~\ref{eq:rjt2}). It is obvious that
both are symmetric under arbitrary permutation of indices $(a,b,c)$.
Also introduced is the elliptic integral of the second kind, viz.\
\begin{equation}
R_D(a,b;c)=R_J(a,b,c;c)
=-6\frac{\partial R_F(a,b,c)}{\partial c},
\end{equation}
which is symmetric in $a\leftrightarrow b$, but
%
there is an identity \citep[eq.~19.21.7]{DLMF}
allowing rearrangement of the third argument;
\begin{equation}\label{eq:rds}
\frac{(b-a)R_D(a,c;b)+(c-a)R_D(a,b;c)}3
=R_F(a,b,c)-\!\sqrt{\frac a{bc}},
\end{equation}\end{subequations}
which may be proven through the definitions of Carlson integrals
together with integrations by parts.
Unlike $R_F$, the partial derivatives of $R_D$ do not
introduce additional bases;
\begin{equation*}\begin{split}
\frac23(b-a)(c-a)\frac{\partial R_D}{\partial a}
&=R_F+\frac{2b-a-c}3R_D-\!\sqrt{\frac b{ac}};
\\\frac23(c-a)(c-b)\frac{\partial R_D}{\partial c}
&=R_F+\frac23(a+b-2c)\,R_D
-\!\sqrt{\frac{ab}{c^3}},
\end{split}\end{equation*}
where $R_F=R_F(a,b,c)$ and $R_D=R_D(a,b;c)$.

All three bases are the particular cases of the Carlson-R,
%
\begin{equation*}\begin{split}
R_F&(a,b,c)=R_{-\frac12}\bigl(\tfrac12,\tfrac12,\tfrac12;a,b,c\bigr);
\\R_D&(a,b;c)=R_{-\frac32}\bigl(\tfrac12,\tfrac12,\tfrac32;a,b,c\bigr);
\\R_J&(a,b,c;p)=R_{-\frac32}\bigl(\tfrac12,\tfrac12,\tfrac12,1;
a,b,c,p\bigr),
\end{split}\end{equation*}
and so $R_F(\xi\bm a)=\xi^{-1/2}R_F(\bm a)$,
$R_D(\xi\bm a)=\xi^{-3/2}R_D(\bm a)$, and
$R_J(\xi\bm a)=\xi^{-3/2}R_J(\bm a)$.
The integral is referred to as ``complete''
if one of the arguments is zero (provided it converges):
\begin{equation*}\begin{split}
R_F&(0,b,c)=\tfrac12\pi\,R_{-\frac12}\bigl(\tfrac12,\tfrac12;b,c\bigr);
\\R_D&(0,b;c)=\tfrac34\pi\,R_{-\frac32}\bigl(\tfrac12,\tfrac32;b,c\bigr);
\\R_J&(0,b,c;p)=\tfrac34\pi\,R_{-\frac32}\bigl(\tfrac12,\tfrac12,1;b,c,p\bigr).
\end{split}\end{equation*}
%
Note $\lim_{p\downarrow0^+}\!\sqrt p\,R_J(0,b,c;p)=(3\pi/2)(bc)^{-1/2}$;
this contrasts to the incomplete case $R_J(a,b,c;p)$ with $a,b,c>0$,
for which $p=0$ is the logarithmic singularity.
Following the discussion in Appendix~\ref{app:carl}, the complete
integrals $R_F(0,b,c)$ and $R_D(0,b,c)$ are reducible to
${_2F_1}$-functions, while $R_F(a,b,c)$ and $R_D(a,b;c)$
as well as $R_J(0,b,c;p)$ are also expressed as Appell-F$_1$
(equivalent to two-variable Lauricella-${\rm F_D}$).
In fact, it is also possible to express $R_J(0,b,c;p)$ in terms of
$R_F$ and $R_D$ (see \cite{DLMF}, eq.~19.21.6)
involving both complete and incomplete integrals.

There also exists a set of formulae \cite[\S~19.21(iii)]{DLMF}
to transform the last parameter of $R_J(a,b,c;p)$. For example,
\begin{subequations}\label{eq:rjt}\begin{equation}\begin{split}
R_J(0,b,c;p)
&=\frac32\!\int_0^\infty\!
\frac{u^{1/2}\dm u}{(pu+bc)\!\sqrt{(u+b)(u+c)}}
\\&=\frac1p\left[3R_F(0,b,c)
-\frac{bc}pR_J\biggl(0,b,c;\frac{bc}p\biggr)\right],
\end{split}\end{equation}
which reproduces equation (19.21.15) of \cite{DLMF}.
This interchanges the parameter between intervals,
$(0,b)\leftrightarrow(c,\infty)$ (assuming $0<b<c$),
while the parameter in $(b,c)$ is transformed within the interval.
If $p=b$ or $p=c$, then $bR_D(0,c;b)=3R_F(0,b,c)-cR_D(0,b;c)$,
which is equivalent to equation (\ref{eq:rds}) with $a=0$. An additional
distinct transformation applicable for the complete integral is given by
\begin{multline}\label{eq:rjt2}
R_J(0,b,c;p)=\frac{3\pi}2\frac{\Theta[p(c-p)(b-p)]}{\!\sqrt{p(c-p)(b-p)}}
\\-\frac1{c-p}\left[3R_F(0,b,c)
+\frac{c(c-b)}{c-p}R_J\biggl(0,b,c;\frac{c(b-p)}{c-p}\biggr)\right],
\end{multline}
which is proven from the identity [with $q=c(b-p)/(c-p)$]:
\begin{equation}
\int_0^\infty\!\frac{\dm x}{x^2+pq/c}
=\frac1{2(c-p)}\!\int_0^\infty\!\frac{(t^2+2ct+bc)\dm t}
{(t+p)(t+q)\!\sqrt{t(t+b)(t+c)}}
\end{equation}\end{subequations}
following the change of the integration variable, $x^2=t(t+b)/(t+c)$.
If $p$ lies between $b$ and $c$,
then the transformed parameter $q=c(b-p)/(c-p)$ is negative
(assuming $b,c>0$), which is understood to be the Cauchy principal value
of the integral defined in equation (\ref{eq:Rdef}). Also observe
that the composition of two transformations in equations (\ref{eq:rjt}) results
in equation (\ref{eq:rjt2}) with $b\leftrightarrow c$.

\subsection{The Legendre canonical basis}
\label{app:lei}

Traditional approaches to elliptic integrals utilize the canonical forms established by Legendre. Unfortunately, the Legendre canonical set is notorious for a confusing variety of notations found in the literature. Here we primarily follow that of the {\sl Digital Library of Mathematical Functions} (DLMF) \cite{DLMF}, which is consistent with \cite[sect.~6.11]{NR} and \cite{GR}. This convention, however, differs from \cite[chap.~17]{AS}, where the elliptic parameter ($m=k^2$) or the modular angle ($\alpha=\sin^{-1}k$)
is used as the argument instead (cf.\ the Wolfram research functions site\footnote{\url{http://functions.wolfram.com}}). The notation for the integral of the third kind $\Pi(\phi,\omega^2,k)$ also follows the \citetalias{DLMF}. This not only differs from \cite{AS}, specifically their
$\Pi(n;\phi\backslash\sin^{-1}k)$ in eq.~17.2.14 or $\Pi(n;\mathrm F(\phi,k)|k^2)$ in eq.~17.2.15, but also from \cite{Er53} and \cite{NR}, which adopt the opposite sign convention for the parameter $n\to-n$. While the definition used in the {\sl GNU Scientific Library} (GSL) is consistent with \cite{Er53} and \cite{NR}, but the adopted order of arguments $n$ and $k$ is switched. Assuming $0\le k<1$, the complete integral for $\mathbf\Pi(\omega^2,k)=\Pi(\pi/2,\omega^2,k)$ converges for $\omega^2<1$, while it diverges as $\omega^2\uparrow1^-$. Technically, $\mathbf\Pi(\omega^2,k)$ possesses a branch cut along the real axis for $\omega^2\in[1,\infty)$, where the Cauchy principal value is typically chosen for the principal branch. Note that the sign convention for the parameter $\chi=-\omega^2$ adopted by \cite{Er53} and \cite{NR} places this branch cut along the negative real line instead. In this paper, we also adopt the nonstandard font choices 
to facilitate distinctions between the complete ($\mathbf K$, $\mathbf E$ and $\mathbf\Pi$) and incomplete ($\mathrm F$, $\mathrm E$ and $\Pi$) elliptic integrals.

Let us define Legendre's elliptic integral of the first kind;
\begin{align}
\mathrm F(\phi,k)&=
\int_0^\phi\frac{\dm\theta}{\!\sqrt{1-k^2\sin^2\!\theta}}
=\int_0^{\sin\phi}\!\frac{\dm t}{\!\sqrt{(1-t^2)(1-k^2t^2)}}
\nonumber\\&=\sin\phi\,F_1\bigl(\tfrac12;\tfrac12,\tfrac12;\tfrac32;
k^2\sin^2\!\phi,\sin^2\!\phi\bigr)
\nonumber\\&=\sin\phi\,R_{-\frac12}\bigl(\tfrac12,\tfrac12,\tfrac12;
\cos^2\!\phi,1-k^2\sin^2\!\phi,1\bigr)
\nonumber\\&=\sin\phi\,R_F\bigl(\cos^2\!\phi,1-k^2\sin^2\!\phi,1\bigr);
\end{align}
for $0\le\phi\le\pi/2$ and $k^2\sin^2\phi<1$,
and the second kind;
\begin{equation}\begin{split}
\mathrm E(\phi,k)&=
\int_0^\phi\!\sqrt{1-k^2\sin^2\!\theta}\,\dm\theta
=\int_0^{\sin\phi}\!\sqrt{\frac{1-k^2t^2}{1-t^2}}\,\dm t
\\&=\sin\phi\,F_1\bigl(\tfrac12;-\tfrac12,\tfrac12;\tfrac32;
k^2\sin^2\!\phi,\sin^2\!\phi\bigr)
\\&=\sin\phi\,R_{-\frac12}\bigl(\tfrac12,-\tfrac12,\tfrac32;
\cos^2\!\phi,1-k^2\sin^2\!\phi,1\bigr).
\end{split}\end{equation}
It is easy to find that $\mathrm F(0,k)=\mathrm E(0,k)=0$
and $\mathrm F(\phi,0)=\mathrm E(\phi,0)=\phi$,
while the integrals with $\phi=\pi/2$ are referred to as ``complete'';
\begin{align}\label{eq:d6}
\mathbf K(k)&
=\int_0^\frac\pi2\!\frac{\dm\theta}{\!\sqrt{1-k^2\sin^2\!\theta}}
=R_F(0,1-k^2,1);
\\\nonumber\mathbf E(k)&
=\int_0^\frac\pi2\!\dm\theta\sqrt{1-k^2\sin^2\!\theta}
=(1-k^2)
\int_0^\frac\pi2\!\frac{\dm\hat\theta}{(1-k^2\sin^2\!\hat\theta)^{3/2}}.
\end{align}
Then $\mathbf K(0)=\mathbf E(0)=\pi/2$ and $\mathbf E(1)=1$,
whereas $\mathbf K(k)$ diverges logarithmically as $k\uparrow1^-$.
The last part of equation (\ref{eq:d6})
follows the change of the variable:
$\sin^2\hat\theta=(1-\sin^2\!\theta)/(1-k^2\sin^2\!\theta)$,
which enables us to establish that
\begin{equation}
k\frac{\dm\mathbf K(k)}{\dm k}=
\frac{\mathbf E(k)}{1-k^2}-\mathbf K(k),\quad
k\frac{\dm\mathbf E(k)}{\dm k}=
\mathbf E(k)-\mathbf K(k).
\end{equation}
In addition, the ``incomplete'' integrals with $k\sin\phi=1$ are reducible to the
complete integrals; that is, for $0<x<1$,
\begin{equation}\begin{split}
\mathrm F(\sin^{-1}\!x,x^{-1})&=x\mathbf K(x);
\\\mathrm E(\sin^{-1}\!x,x^{-1})&=x^{-1}\mathbf E(x)-(x^{-1}-x)\mathbf K(x).
\end{split}\end{equation}
with the inverse transformations of
$\mathbf K(k)=k^{-1}\mathrm F(\sin^{-1}\!k,k^{-1})$ and
$\mathbf E(k)=k\,\mathrm E(\sin^{-1}\!k,k^{-1})
+k^{-1}(1-k^2)\mathrm F(\sin^{-1}\!k,k^{-1})$,
Both complete integrals also reduce to the hypergeometric series;
\begin{equation}\begin{split}
\frac2\pi\mathbf K(k)&={_2F_1}\!\left(\stack{\frac12,\frac12}{1};k^2\right);\\
\frac2\pi\mathbf E(k)&={_2F_1}\!\left(\stack{-\frac12,\frac12}{1};k^2\right)
=(1-k^2)\,{_2F_1}\!\left(\stack{\frac32,\frac12}{1};k^2\right).
\end{split}\end{equation}
The Euler--Pfaff transform of the ${_2F_1}$-series further indicates
\begin{equation}
\!\sqrt{1-k^2}\,\mathbf K(k)=\mathbf K(k_1);\quad
\mathbf E(k)=\!\sqrt{1-k^2}\,\mathbf E(k_1),
\end{equation}
where $k_1^2=k^2/(k^2-1)$.
Note Carlson's $R_F$ encompasses
Legendre's $\mathrm F$ and $\mathbf K$. As for Carlson's $R_D$,
we observe $3R_{-\frac12}\bigl(\tfrac12,-\tfrac12,\tfrac32;a,b,c\bigr)
=3R_F(a,b,c)+(b-c)R_D(a,b;c)$ and 
\begin{align}
&3[\mathrm F(\phi,k)-\mathrm E(\phi,k)]
=k^2\sin^3\!\phi\,R_D(\cos^2\!\phi,1-k^2\sin^2\!\phi;1);
\nonumber\\&k^2R_D(0,1-k^2;1)=3[\mathbf K(k)-\mathbf E(k)];
\nonumber\\&3\mathbf E(k)=3R_F(0,1-k^2,1)-k^2R_D(0,1-k^2;1).
\end{align}

The traditional alternative for the Carlson-$R_J$
is the Legendre integral of the third kind, which we define to be
\begin{align}
\Pi&(\phi,\omega^2,k)=\int_0^\phi\!\frac{\dm\theta}
{(1-\omega^2\sin^2\theta)\!\sqrt{1-k^2\sin^2\theta}}
\\\nonumber&=\sin\phi\,F^{(3)}_D
\bigl(\tfrac12;\tfrac12,\tfrac12,1;\tfrac32;
k^2\sin^2\phi,\sin^2\phi,\omega^2\sin^2\phi\bigr)
\\&=\sin\phi\,R_{-\frac12}
\bigl(\tfrac12,\tfrac12,-\tfrac12,1;
\cos^2\phi,1-k^2\sin^2\phi,1,1-\omega^2\sin^2\phi\bigr)
\nonumber\\&=\sin\phi\,R_F\bigl(\cos^2\phi,1-k^2\sin^2\phi,1\bigr)
\nonumber\\&\phantom{=}\quad
+\tfrac13\omega^2\sin^3\phi\,R_J(\cos^2\phi,1-k^2\sin^2\phi,1;1-\omega^2\sin^2\phi),
\nonumber\end{align}
where $F^{(3)}_D$ is the three-variable Lauricella-${\rm F_D}$
(see eqs.~\ref{eq:laud} and \ref{eq:laudi}). The last equality
is also recast to be
\begin{align}
3[\Pi&(\phi,\omega^2,k)-\mathrm F(\phi,k)]
\\\nonumber
&=\omega^2\sin^3\phi\,F^{(3)}_D\bigl(\tfrac32;\tfrac12,\tfrac12,1
;\tfrac52;k^2\sin^2\phi,\sin^2\phi,\omega^2\sin^2\phi\bigr)
\\\nonumber
&=\omega^2\sin^3\!\phi\,R_J(\cos^2\phi,1-k^2\sin^2\phi,1;1-\omega^2\sin^2\phi). 
\end{align}
While it is less common, one can also define the complete Legendre
integral of the third kind, namely
\begin{subequations}\begin{gather}\begin{split}
\mathbf\Pi(\omega^2,k)&=\Pi(\pi/2,\omega^2,k)
=\tfrac12\pi\,F_1\bigl(\tfrac12;\tfrac12,1;1;k^2,\omega^2\bigr)
\\&=\tfrac12\pi\,R_{-\frac12}\bigl(\tfrac12,{-\tfrac12},1;1-k^2,1,1-\omega^2\bigr);
\end{split}\\\begin{split}
3[\mathbf\Pi&(\omega^2,k)-\mathbf K(k)]
=\tfrac34\pi\,\omega^2F_1\bigl(\tfrac32;\tfrac12,1;2;k^2,\omega^2\bigr)
\\&=\tfrac34\pi\,
\omega^2R_{-\frac32}\bigl(\tfrac12,\tfrac12,1;1-k^2,1,1-\omega^2\bigr).
\end{split}\end{gather}
This is also related to the Carlson bases through
$\tfrac\pi2 R_{-\frac12}\bigl(\tfrac12,{-\tfrac12},1;a,b,c\bigr)
=R_F(0,a,b)\!+\!\tfrac13(b-c)R_J(0,a,b;c)$. That is to say,
\begin{align}
\mathbf\Pi&(\omega^2,k)=
R_F(0,1-k^2,1)+\tfrac13\omega^2R_J(0,1-k^2,1;1-\omega^2);
\nonumber\\\omega^2&R_J(0,1-k^2,1;1-\omega^2)=3[\mathbf\Pi(\omega^2,k)-\mathbf K(k)].
\end{align}\end{subequations}
It is fairly straightforward to establish that
\begin{equation}\begin{split}
\mathbf\Pi&(0,k)
=\int_0^{\frac\pi2}\!\frac{\dm\theta}{(1-k^2\sin^2\theta)^{1/2}}
=\mathbf K(k);
\\\mathbf\Pi&(\omega^2,0)
=\int_0^{\frac\pi2}\!\frac{\dm\theta}{1-\omega^2\sin^2\theta}
=\frac\pi{2\!\sqrt{1-\omega^2}};
\\\mathbf\Pi&(k^2,k)
=\int_0^{\frac\pi2}\!\frac{\dm\theta}{(1-k^2\sin^2\theta)^{3/2}}
=\frac{\mathbf E(k)}{1-k^2},
\end{split}\end{equation}
for $0\le k^2<1$ and $\omega^2<1$. In terms of the Legendre forms,
the formulae in equations (\ref{eq:rjt}) are equivalent to
\begin{align}
\mathbf\Pi(\omega^2&,k)
=\frac{1-k^2}X\mathbf\Pi\biggl(-\frac{\omega^2-k^2}{1-\omega^2},k\biggr)
-\frac{k^2\mathbf K(k)}{\omega^2-k^2}
\nonumber\\&=\frac\pi2\frac{\Theta(X)}{\!\sqrt X}
-\mathbf\Pi\biggl(\frac{k^2}{\omega^2},k\biggr)+\mathbf K(k)
\label{eq:pit}\\\nonumber&=\frac\pi2\frac{\Theta(X)}{\!\sqrt X}
-\frac{1-k^2}X\mathbf\Pi\biggl(-\frac{k^2(1-\omega^2)}{\omega^2-k^2},k\biggr)
+\frac{\mathbf K(k)}{1-\omega^2};
\end{align}
where $X=(1-\omega^2)(\omega^2-k^2)/\omega^2$. Here if $1>\omega^2=\pm k>-1$,
then $X=(1\mp k)^2$, and so follows the second line that,
\begin{equation}\label{eq:Pik}
2\mathbf\Pi(\pm k,k)=\frac\pi{2(1\mp k)}+\mathbf K(k).
\end{equation}
for $0\le k^2<1$. If $\omega^2=1+\alpha^2>1$, equation (\ref{eq:pit})
indicates 
\begin{equation}
\mathbf\Pi(1+\alpha^2,k)
=\mathbf K(k)-\mathbf\Pi\biggl(\frac{k^2}{1+\alpha^2},k\biggr),
\end{equation}
which is consistent with the Cauchy principal value of the integral.

\onecolumngrid
\clearpage
\twocolumngrid
\pagestyle{empty}
\label{lastpage}
\small\tableofcontents


\vfill
\begin{thebibliography}{}
\renewcommand\baselinestretch0
\footnotesize

\bibitem[Evans \& de~Zeeuw(1992)]{EdZ92}
N.W.~Evans, \and P.T.~de~Zeeuw, \mnras, {\bf 257}, 152 (1992).

\bibitem[Qian(1992)]{Qi92}
E.~Qian, \mnras, {\bf 257}, 581 (1992).

\bibitem[An(2013)]{An13}
J.~An, \mnras, {\bf 435}, 3045 (2013).

\bibitem[Mestel(1963)]{Me63}
L.~Mestel, \mnras, {\bf 126}, 553 (1963).

\bibitem[Routh(1892)]{Ro92}
E.J.~Routh, {\it A Treatise on Analytical Statics, vol.~2.}
(Cambridge Univ.\ Press, Cambridge, 1892). 

\bibitem[Kuzmin(1953)]{Ku53}
G.G.~Kuzmin, Tartu~Astron.~Obs.~Teated, {\bf 1} (1953).

\bibitem[Kuzmin(1956)]{Ku56}
G.G.~Kuzmin, Astron.~Zh., {\bf 33}, 27 (1956).

\bibitem[Toomre(1963)]{To63}
A.~Toomre, \apj, {\bf 138}, 385 (1963).

\bibitem[Lynden-Bell(1989)]{Ly89}
D.~Lynden-Bell, \mnras, {\bf 237}, 1099 (1989).

\bibitem[Erd\'elyi et al.(1954)]{Er54}
A.~Erd\'elyi, W.~Magnus, F.~Oberhettinger, \and F.G.~Tricomi,
{\it Tables of Integral Transforms (in 2 volumes), Bateman Manuscript Project}
(McGraw-Hill, New York NY, 1954).

\bibitem[Widder(1938)]{Wi38}
D.V.~Widder, Trans.~American~Math.~Soc., {\bf 43}, 7 (1938).

\bibitem[Schwarz(2005)]{Sc05}
J.H.~Schwarz, J.~Math.~Phys., {\bf 46}, 013501 (2005).

\bibitem[Whittaker \& Watson(1927)]{WW27}
E.T.~Whittaker, \and G.N.~Watson, {\it A Course of Modern Analysis, 4th edn.}
(Cambridge Univ.\ Press, Cambridge, 1927). 

\bibitem[An(2026)]{fdisk}
J.~An, ``A family of finite-radius razor-thin disk models,'' preprint

\bibitem[Erd\'elyi et al.(1953)]{Er53}
A.~Erd\'elyi, W.~Magnus, F.~Oberhettinger, \and F.G.~Tricomi,
{\it Higher Transcendental Functions (in 3 volumes), Bateman Manuscript Project}
(McGraw-Hill, New York NY, 1953).

\bibitem[Olver et al.(2010)]{DLMF}
F.W.J.~Olver, D.W.~Lozier, R.F~Boisvert, \and C.W.~Clark,
{\it NIST Handbook of Mathematical Functions} (Cambridge Univ.\ Press, Cambridge, 2010).

\bibitem[An(2011)]{An11}
J.~An, \apj, {\bf 736}, 151 (2011).

\bibitem[An et al.(2012)An, Van~Hese \& Baes]{AvHB}
J.~An, E.~Van~Hese, E., \and M.~Baes, \mnras, {\bf 422}, 652 (2012).

\bibitem[Lynden-Bell(2003)]{Ly03}
D.~Lynden-Bell, \mnras, {\bf 338}, 208 (2003).

\bibitem[Agarwal(1965)]{Ag65}
R.P.~Agarwal, Proc.~Natl.~Inst.~Sci.~India, Part~A, {\bf 13}, 536 (1965).

\bibitem[Sharma(1965)]{Sh65}
B.L.~Sharma, Ann.~Soc.~Sci.~Bruxelles, Series~I, {\bf 79}, 26 (1965).

\bibitem[Mittal \& Gupta(1972)]{MG72}
P.K.~Mittal, \and K.C.~Gupta, Proc.~Indian~Acad.~Sci., Section~A, {\bf 75}, 117 (1972).

\bibitem[Kalnajs(1976)]{Ka76}
A.~Kalnajs, \apj, {\bf 205}, 751 (1976).


\bibitem[H\'enon(1959)]{He59}
M.~H\'enon, Ann.~Astrophys., {\bf 22}, 126 (1959).

\bibitem[Post(1930)]{Po30}
E.~Post, Trans.~American~Math.~Soc., {\bf 32}, 723 (1930).

\bibitem[Beals \& Szmigielski(2013)]{BS13}
R.~Beals, \and J.~Szmigielski, Not.~American~Math.~Soc., {\bf 60}, 866 (2013).

\bibitem[Carlson(1963)]{Ca63}
B.C.~Carlson, J.~Math.~Anal.~Appl., {\bf 7}, 452 (1963).

\bibitem[Press et al.(1992)]{NR}
W.H.~Press, S.A.~Teukolsky, W.T.~Vetterling, \and B.P.~Flannery,
{\it Numerical Recipes in {\sc Fortran~77}, 2nd edn.} (Cambridge Univ.\ Press, Cambridge, 1992).

\bibitem[Gradshteyn \& Ryzhik(2007)]{GR}
I.S.~Gradshteyn, \and I.M.~Ryzhik,
{\it Table of Integrals, Series, and Products 
(A.~Jeffrey \& D.~Zwillinger eds.), 7th edn.}
(Elsevier Academic Press, Burlington MA, 2007).

\bibitem[Abramowitz \& Stegun(1964)]{AS}
M.~Abramowitz, \and I.A.~Stegun,
{\it Handbook of Mathematical Functions, Applied Mathematics Series, vol~55}
(National Bureau of Standards, Washington DC, 1964).

\end{thebibliography}
\end{document}